\documentclass[12pt]{article}

\title{\vspace{5mm} \textbf{
Perturbative path-integral of string field and the $A_{\infty }$ structure of the BV master equation 
} \vspace{5mm} }
\author{\Large Toru Masuda,${}^{b}$ \hspace{3mm} Hiroaki Matsunaga\,${}^{a,b,c,}$\footnote{Current affiliation: hiroaki.matsunaga@omu.ac.jp} \vspace{5mm} } 
\date{
\textit{${}^{a}$Mathematical Institute, 
Faculty of Mathematics and Physics, \\ 
Charles University Prague, \\[1mm] 
Sokolovska 83, 18675 Prague 8, Czech Republic \\[4mm] 
${}^{b}$CEICO, Institute of Physics, 
Czech Academy of Sciences, \\[1mm] 
Na Slovance 2, 18221 Prague 8, Czech Republic   \\[4mm] 
${}^{c}$Yukawa Institute for Theoretical Physics, Kyoto University, \\[1mm] 
Kitashirakawa Oiwake-cho, Sakyo-Ku, Kyoto 606-8502, Japan \\[4mm]
${}^{\ast }$Osaka Metropolitan University College of Technology, \\[1mm]
Saiwai-cho, Neyagawa, Osaka 572-8572, Japan
} \\ 
}

\usepackage{geometry} 
\usepackage[dvips]{graphicx}
\usepackage{color, upgreek, mathrsfs, enumitem}
\usepackage{amsmath,amscd, amssymb, amsfonts, amsthm}
\usepackage{url}
\usepackage{cite}

\makeatletter
  
  \@addtoreset{equation}{section}
\makeatother

\newcommand{\no}{\nonumber\\}

\newcommand{\la}{\big{\langle }}
\newcommand{\ra}{\big{\rangle }}
\newcommand{\La}{\Big{\langle }}
\newcommand{\Ra}{\Big{\rangle }}

\newcommand{\cD}{\mathcal{D}}

\newcommand{\cH}{\mathcal{H}}

\allowdisplaybreaks[3] 

\begin{document}
\maketitle 

\begin{abstract} 
The perturbative path-integral gives a morphism of the (quantum) $A_{\infty }$ structure intrinsic to each quantum field theory, 
which we show explicitly on the basis of the homological perturbation. 
As is known, 
in the BV formalism, 
any effective action also solves the BV master equation, 
which implies that the path-integral can be understood as a morphism of the BV differential.  
Since each solution of the BV master equation is in one-to-one correspondence with a quantum $A_{\infty }$ structure, 
the path-integral preserves this intrinsic $A_{\infty }$ structure of quantum field theory, 
where $A_{\infty }$ reduces to $L_{\infty }$ whenever multiplications of space-time fields are graded commutative. 
We apply these ideas to string field theory and (re-)derive some quantities based on the perturbative path-integral, 
such as effective theories with finite $\alpha ^{\prime }$, 
reduction of gauge and unphysical degrees, 
S-matrix and gauge invariant observables.  
\end{abstract}

\thispagestyle{empty}\addtocounter{page}{-1} 

\clearpage

\tableofcontents 

\section{Introduction} 

In quantum theory, 
partition functions or expectation values of observables are central objects. 
For a Lagrangian field theory, 
the path-integral provides these objects, 
though 
how to integrate is obscure except for free theories. 
The perturbative path-integral is a standard technique that enables us to treat interacting fields in terms of free theories. 
In this paper, 
we show explicitly that the perturbative path-integral can be regarded as a morphism of the (quantum) $A_{\infty }$ structure intrinsic to each quantum field theory. 
Such a perspective provides simple explanations of some algebraic properties of the quantities based on the perturbative path-integral, 
which will be useful for 
calculating the scattering amplitudes, 
deriving effective theories, 
fixing gauge degrees, 
studying flows of exact renormalization group and so on.

\vspace{2mm}

Homotopy algebras, 
such as quantum $A_{\infty }$ or $L_{\infty }$, 
arise naturally in the context of the ordinary Lagrangian description of quantum field theory. 
As is mentioned at the end of section 6, 
they describe not only the gauge invariance of Lagrangian but also the Feynman graph expansion. 
%
Thus, 
theoretical physicists already know some of these structures, 
albeit implicitly, 
even for the theory without gauge degrees.\footnote{String field theories will be typical examples revealing these explicitly. } 
%
%
%
%
%
The Batalin-Vilkovisky (BV) formalism makes these structures visible and provides the translation between Lagrangian field theories and homotopy algebras \cite{Zwiebach:1992ie}. 
The BV formalism is 
one of the most powerful and general frameworks for quantization of gauge theories, 
which is based on the homological perturbation \cite{Batalin:1981jr, Batalin:1984jr, Henneaux}. 
For a given Lagrangian field theory, 
we can defines a complex with an appropriate BV differential by solving the BV master equation, 
which is one equivalent description of a quantum $A_{\infty }$ algebra \cite{Barannikov:2017, Barannikov:2010np, Doubek:2013dpa}. 
This $A_{\infty }$ algebra reduces to an $L_{\infty }$ algebra whenever multiplications of space-time fields are graded commutative. 
Since the BV formalism assigns a homotopy algebra to each quantum field theory, 
we can extract the intrinsic $A_{\infty }$ structure 
explicitly by casting the BV master action into the homotopy Maurer-Cartan form \cite{Schwarz:1992nx, Alexandrov:1995kv, Jurco:2018sby, Macrelli:2019afx, Jurco:2019yfd}.\footnote{The original action is recovered by the homotopy Maurer-Cartan action by setting antifields to zero, 
whose $A_{\infty }$ structure is just a piece of the full (quantum) $A_{\infty }$ structure of the BV master action. }  

\vspace{2mm}

As is well-known, 
in the BV formalism, 
any effective action also solves the BV master equation. 
This fact implies that the path-integral can be understood as a morphism of the BV differential. 
Since any solution of the BV master equation is in one-to-one correspondence with a quantum $A_{\infty }$ structure,\footnote{
In this paper, we consider Lagrangians field theory describing fluctuations around the perturbative vacuum where its free theory is solved, 
namely, the standard perturbation theory based on the Feynman graph expansion: 
Our $A_{\infty }$ structure arises from the expansion of a given BV master action around a Gaussian critical point. } 
the path-integral preserves this intrinsic $A_{\infty }$ structure of quantum field theory. 
Although these properties may valid for non-perturbative path-integral, 
in this paper, 
we consider the perturbative path-integral. 
We first show that the perturbative path-integral can be performed as a result of the homological perturbation for the intrinsic $A_{\infty }$ structure and thus it gives a morphism of this (quantum) $A_{\infty }$ structure in any BV-quantizable quantum field theory. 
Then, 
we apply these ideas to string field theory and consider some quantities based on the perturbative path-integral of string fields. 
As a result of the homological perturbation, 
we derive effective theory with finite $\alpha '$ \cite{Sen:2016qap }, 
the light-cone reduction \cite{Matsunaga:2019fnc, EM}, 
and string S-matrix 
in a simple way. 
In addition to these (re-)derivations, 
we explain that this approach may enable us to use unconventional pieces of perturbative calculus. 
We discuss the open string $S$-matrix based on unconventional propagators whose $4$-point amplitude reproduces the gauge invariant quantity given by \cite{Masuda:2019rgv} directly.

\vspace{2mm}

This paper is organized as follows. 
In section 2, 
we explain our settings and several important facts that we need later, 
in which the relation between Lagrangian's $A_{\infty }$ and $L_{\infty }$ is explained. 
In section 3, 
after presenting the relation between the BV master equation and the quantum $A_{\infty }$ structure, 
we show explicitly that the homological perturbation indeed performs the perturbative path-integral in terms of the BV-BRST cohomology. 
%
This result would be known, as mentioned at the end of section 6, except for incidental details.
The quantum $A_{\infty }$ structure of effective theory and the classical limit are also discussed. 
In section 4, 
we translate the results based on the BV formalism into corresponding results based on the (quantum) $A_{\infty }$ structure. 
We show that when the original BV master action includes source terms, 
its effective theory must have a weak $A_{\infty }$ structure. 
In section 5, 
we apply these results to string field theory. 
We (re-)derive effective theories with finite $\alpha '$, 
reduction of gauge degrees, 
string amplitudes, 
and gauge invariant quantities discussed in \cite{Masuda:2019rgv} in a systematic way. 
In section 6, 
we conclude with summary and mentioning earlier works. 
In appendix, 
we explain the grading of our $A_{\infty }$ structure and a basis of the symplectic pairing in the BV formalism that makes manifest Lagrangian's homotopy algebraic structure.

\subsubsection*{Remarks}

In this paper, 
we start from \textit{a given integrand} $e^{S[\phi ]}$ of the perturbative path-integral $\int \mathcal{D}[\phi ] \, e^{S [\phi ]}$ and explain that the flow of $\int \mathcal{D} [\phi '' ]$ for any $\phi = \phi ' + \phi ''$ can be described in terms of the BV-BRST cohomology (in section 3) and Lagrangian's homotopy algebra (in section 4) explicitly. 
Theoretical physicists know that the physical configurations or states are described by both of the path-integral and the BRST cohomology, 
and several experts know that the method of the BV-BRST cohomology is equivalent to the approach based on Lagrangian's homotopy algebra. 
In this sense, 
this statement itself is not new, 
although it would be implicit and indirect. 
An explicit and direct connection between the path-integral, 
the BV-BRST cohomology and Lagrangian's homotopy algebraic structure has not been presented enough and would be worth elucidating.  
The authors believe that contents of subsection 3.3 (and the relevant part in 3.5) and subsection 4.7 include new contributions to the development of the homotopy algebraic approach to QFTs, 
and that subsection 5.2---a quantum extension of the light-cone reduction---and the last part of subsection 5.3 are new results in the study of strings.

\vspace{2mm} 

Although it is just a reformulation for the well-known technique of Feynman graph expansion,  
we believe that there are some advantages: 
It tells us how usual perturbative calculations can be mathematically rigorous ones; 
It rewrites the perturbative path-integral into a simple recursion relation; 
It makes each step of such calculations manifest, exact and systematic, 
so that any oversight -- what we could know from the usual perturbative calculations -- would be detected after the reformulation.  
In addition, 
it shows that for any integrands $e^{S[\phi ]}$, 
the path-integral eliminating gauge and unphysical degrees, 
the flow of $\int \mathcal{D}[ \phi _{\mathrm{gauge}} ] \mathcal{D}[ \phi _{\mathrm{unphys}}] $ for any $\phi = \phi _{\mathrm{phys} } + \phi _{\mathrm{gauge}} + \phi _{\mathrm{unphys}}$, 
and the path-integral eliminating off-shell fields, 
the flow of $\int \mathcal{D}[ \phi _{\mathrm{off \, shell}} ] $ for $\phi = \phi _{\mathrm{on \,shell} } + \phi _{\mathrm{off \, shell}} $, 
are performed in a consistent way, 
as physicists know without proofs. 
In order to enjoy such advantages, 
we need rather \textit{explicit} derivations, 
which is our motivation.

\vspace{2mm} 

Please note that how to prepare a ``consistent'' field theory $e^{S[\phi ]}$ is up to readers: 
The results can be concrete (or formal) ones as much as given initial data. 
It would be preferable to consider any renormalized (effective) field theories that solves the BV master equation or string field theories as given integrands $e^{S[\phi ]}$ of our consideration.  

\section{Preliminaries} 

In this section, 
we explain our settings and notation and present several mathematical facts. 

\vspace{2mm} 

Let us explain field theories of our consideration. 
In this paper, 
we consider field theories $S_{\mathrm{cl}} [\psi _{\mathrm{cl}}  ] $ that consist of the kinetic term $S_{\mathrm{cl \, free}} [\psi _{\mathrm{cl}} ]$ and the interacting terms $S_{\mathrm{cl \, int}} [\psi _{\mathrm{cl}} ]$ as follows 
\begin{align}
\label{classical physical action}
S_{\mathrm{cl}} [ \psi _{\mathrm{cl} }]
\equiv - \underbrace{ \frac{1}{2} \int dx \, 
\psi _{\mathrm{cl}} \, \mu ^{\mathrm{cl}}_{1} ( \psi _{\mathrm{cl}} ) 
}_{S_{\mathrm{cl \, free}} [\psi _{\mathrm{cl}}  ] }
-
\underbrace{ \sum_{n} \frac{1}{n+1} \int dx \, 
\psi _{\mathrm{cl}} \, \mu ^{\mathrm{cl}}_{n} ( \underbrace{\psi _{\mathrm{cl}}  , ... , \psi _{\mathrm{cl}}  }_{n} ) }_{S_{\mathrm{cl \, int}} [\psi _{\mathrm{cl}} ] } 
\, , 
\end{align} 
where the $\psi _{\mathrm{cl}} $ denotes classical fields, 
the $\mu _{1}^{\mathrm{cl}}$ is kinetic operators and the $\mu _{n}^{\mathrm{cl} } $ are the $(n+1)$-point vertices. 
We assume that the classical action (\ref{classical physical action}) consists of the physical degrees only or is consistently gauge-fixable. 
We also assume that  the free part $S_{\mathrm{cl \, free} } [\psi _{\mathrm{cl}}  ]$ can have a non-degenerate Hessian, 
so that the Gaussian integral can be normalized as usual 
\begin{align}
\label{Gaussian}
1 = \int \cD [\psi _{\mathrm{cl}} ] \, e^{S_{\mathrm{cl \, free}} [ \psi _{\mathrm{cl}}  ] } \, , 
\end{align}
which requires that the free part $S_{\mathrm{cl \, free} } [\psi _{\mathrm{cl}}  ]$ is solved and the value of $\sqrt{\mathrm{det}\, (\mu ^{\mathrm{cl}}_{1})^{-1} }$ is given. 


\subsection{Perturbative path-integral} 

In section 3 and 4, 
we explain that the perturbative path-integral can be performed in terms of the BV-BRST cohomology or Lagrangian's homotopy algebra. 
We explain the word ``the perturbative path-integral'' below.  
In field theories, 
the expectation value of observables $\langle ... \rangle _{J}$ is described by using the path-integral, 
\begin{align}
\label{non-perturbative}
Z_{J}^{-1} 
\int \mathcal{D} [\psi _{\mathrm{cl}}  ] \, \big{(} \, ... \, \big{)} \, e^{S[\psi _{\mathrm{cl}} ] + J\psi _{\mathrm{cl}} } \, , 
\hspace{5mm} 
Z_{J}  = \int \cD [\psi _{\mathrm{cl}}  ] \, e^{S [ \psi _{\mathrm{cl}}  ] + J \psi _{\mathrm{cl}}  } \, , 
\end{align}
where $Z_{J}$ denotes the partition function.\footnote{We set $\hbar = 1$ for convenience. 
If necessary, 
we write $\hbar $ explicitly, 
such as $\exp ( \hbar ^{-1} S )$. } 
Although the non-perturbative path-integral of interacting fields is a deep question, 
we can perform it for free theories since free actions are at most quadratic. 
The perturbative expansion enables us to perform the path-integral of interacting fields formally, 
which is a standard procedure in a Lagrangian field theory. 
In terms of the free theory, 
which should be well solved, 
we can represent (\ref{non-perturbative}) as the following expectation values 
\begin{align}
\label{free rep}
 \la \, ... \, 
 e^{S_{\mathrm{cl \, int}} [\psi _{\mathrm{cl} }] } \, ... \, \ra _{\mathrm{free}, \, J} 
= Z_{J}^{-1}  
\int \mathcal{D} [\psi _{\mathrm{cl}}  ] \, \big{(} \, ... \, 
e^{S_{\mathrm{cl \, int}} [\psi _{\mathrm{cl} } ]} \, ... \, \big{)} \, e^{S_{\mathrm{cl \, free}} [\psi _{\mathrm{cl}} ] + J\psi _{\mathrm{cl}} } 
\, . 
\end{align}
The partition function $Z_{J}$ can be represented as $Z_{J} = \langle \, e^{S_{\mathrm{cl \, int}} [\psi _{\mathrm{cl}}  ] } \, \rangle _{\mathrm{free}, \, J}$. 
We then consider to \textit{replace} a given functional of $\psi _{\mathrm{cl}} $ by using a formal power series of $\psi _{\mathrm{cl} }$, 
for which we write $F [\psi _{\mathrm{cl}} ] $, 
and \textit{replace} 
the expectation value of a given functional by $\langle \, F[\psi _{\mathrm{cl} } ] \, \rangle _{\mathrm{free}, \, J}$ formally.  
This type of integral reduces to the Gaussian integral (\ref{Gaussian}) because of $F [\psi ] \, e^{J \psi } = F[\partial _{J} ] \, e^{J \psi }$.  
Hence, 
whenever the free theory is well solved, 
we can perform the \textit{perturbative} path-integral of (\ref{free rep}) as follows 
\begin{align} 
\label{J rep}
 \la \, ... \, F' [ \psi _{\mathrm{cl}} ] \, ...\, \ra _{\mathrm{free} , \, J} 
& \equiv  \Big{(} \, ...\, F' [ \, \partial _{J} ] \, ... \, 
\Big{)} \, 
e^{\frac{1}{2} J \, \mu _{1}^{-1} \, J } \, , 
\end{align} 
where $F' [\psi _{\mathrm{cl}} ] $ is a formal power series of $\psi _{\mathrm{cl} }$. 
The expectation value $\langle ... \rangle _{J} \equiv \langle ... e^{S_{\mathrm{int} } [\psi _{\mathrm{cl} } ] } ... \rangle _{\mathrm{free}, \, J}$ is always defined by the perturbative path-integral (\ref{J rep}) in the rest of this paper. 
The Feynman graph expansion of $F [\psi _{\mathrm{cl} }]$ is an alternative representation of (\ref{J rep}) with $F' [\psi _{\mathrm{cl}} ] = F[\psi _{\mathrm{cl} }] \, e^{S_{\mathrm{int} } [\psi _{\mathrm{cl} } ] }$. 
As is well-known, 
by adding the source term $e^{J \, \psi _{\mathrm{ev} }}$, 
(\ref{J rep}) can be cast as 
\begin{align} 
\label{Feynman rep}
 \la \, ... \, F[ \psi _{\mathrm{cl}} ] \, ...\, \ra _{J}  
 \equiv  e^{ \frac{1}{2} \partial _{\psi _{\mathrm{ev} } }  \mu _{1}^{-1} \partial _{\psi _{\mathrm{ev} }}  } 
\Big{[} ( \, ... \, F [\psi _{\mathrm{ev} } ] \, ... \, e^{S_{\mathrm{cl \, int} } [\psi _{\mathrm{ev} } ] }  \, ) \, 
e^{ J \psi _{\mathrm{ev} } } \, \Big{]} _{\psi _{\mathrm{ev} } = 0 } \,  . 
\end{align}
In this paper, 
the word ``the perturbative path-integral'' always mean (\ref{J rep}) or (\ref{Feynman rep}). 
Note that fields $\psi _{\mathrm{cl} }$ are integrated by using (\ref{Gaussian}) in both representations (\ref{J rep}) and (\ref{Feynman rep}).

\subsection{The BV master equation} 

The BV formalism is one of the most general and systematic prescription to quantize gauge theories, 
which enable us to treat open or redundant gauge algebras \cite{Batalin:1981jr, Batalin:1984jr, Henneaux}. 
As is known, 
a gauge-fixing is necessary to perform the path-integral for a given gauge theory, 
to which we can apply the BV formalism even if ordinary methods such as fixing-by-hand, 
deriving the Dirac bracket, 
brute-force computations and the BRST procedure do not work. 
To carry out (\ref{non-perturbative}) or (\ref{Gaussian}), 
the action must provides a regular Hessian. 
In addition, 
symmetries proportional to the equations of motion are redundant and must be taken into account. 
We thus introduce antifields $\psi ^{\ast }_{\mathrm{cl}}$, 
ghost fields $c$, 
antifields for ghosts $c^{\ast }$ and pairs of higher fields-antifields as much as needed, 
\begin{align} 
\label{example of BV fields}
S_{\mathrm{cl} }[\psi _{\mathrm{cl}}  ] \,\, \longrightarrow \,\, 
S [\psi ] = S_{\mathrm{cl} } [\psi _{\mathrm{cl}} ] 
+  \underbrace{ \psi ^{\ast }_{\mathrm{cl }} ( S , \psi ) }_{\mathrm{ghost\,\,terms}}
+ \underbrace{ c^{\ast }  \, ( S , c ) + \cdots }_{\mathrm{higher\,\,ghost\,\, terms}} \, . 
\end{align} 
We write $\psi $ for the sum of all fields and antifields.  
This extended action $S[\psi ]$ is called a BV master action, 
which can provide a regular Hessian, 
and enables us to perform the path-integral of gauge theory. 
The BV master action $S[\psi ]$ is a solution of the BV master equation 
\begin{align}
\label{quantum BV master equation}
\hbar \, \Delta \, e^{S[\psi ] }
= 
\Big[ \, 
\hbar \, \Delta S [\psi ]
+ \frac{1}{2} \big{(} \, S [\psi ] , \, S[\psi ] \, \big{)} 
\, \Big] \, e^{S[\psi ] } 
= 0  
\end{align}
and must satisfy the initial condition $S[\psi ] |_{\psi ^{+} = 0} = S_{\mathrm{cl}} [\psi _{\mathrm{cl}} ]$ where $\psi ^{+}$ denotes all of the antifields. 
The BV master equation guarantees that the theory is independent of gauge-fixing conditions and has no gauge anomaly arising from the measure factor of the path-integral. 
A gauge-fixing is carried out by choosing appropriate gauge-fixing fermions, 
which determines a Lagrangian submanifold. 
Note that the original action itself becomes the BV maser action whenever field theories have no gauge degree: 
The antifields are nothing but a useful tool. 

\vspace{2mm} 
In the context of the BV formalism, 
De\,Witt's notation is often adopted: 
The repeated labels imply not only a sum over discrete indices but also give an integration over continuous variables. 
E.g. $A_{a} \, A^{a}  = \int dx \, A_{n \, \mu }(x) A_{n}^{\, \, \mu } (x) = \mathrm{tr}  \int (\star A) \wedge A $ for Yang-Mills fields $A = A_{\mu } (x) dx^{\mu }$ and $\phi _{a}^{+} \, \phi ^{a} = \int dx \, \phi _{n}^{+}(x) \phi _{n}(x) $ for the $O(N)$ scalar fields $\{ \phi ^{a} \} _{a} = \{ \phi _{n}(x) \}_{n , x}$. 
We write $\psi _{g}$ and $\psi ^{\ast }_{g}$ for fields and antifields having space-time ghost number $g$ and $-g-1$ respectively: 
$\psi _{0} \equiv \psi _{\mathrm{cl}}$, 
$\psi ^{\ast }_{0} \equiv \psi ^{\ast }_{\mathrm{cl}}$, 
$\psi _{1} \equiv c$ and $\psi ^{\ast }_{1} \equiv c^{\ast }$ in (\ref{example of BV fields}) for example. 
We also use $\psi _{-1-g} = \psi _{g}^{\ast } $ for brevity. 
The BV Laplacian $\Delta $ is an odd second-order functional derivative, 
which is given by 
\begin{align}
\Delta  \equiv \sum_{g} (-)^{g} \frac{\partial }{\partial \psi _{g} }  \frac{\partial }{\partial \psi _{g}^{\ast }} 
= \frac{\partial }{\partial \psi _{\mathrm{cl} } }  \frac{\partial }{\partial \psi ^{\ast }_{\mathrm{cl} } } 
- \frac{\partial }{\partial c } \, \frac{\partial }{\partial c^{\ast } } 
+ \cdots \, . 
\end{align}
It is a fundamental object in the BV formalism and has geometrical meaning \cite{Schwarz:1992nx, Alexandrov:1995kv}. 
The BV bracket is defined by $ (-)^{F} ( F , G ) \equiv \Delta (FG) - (\Delta F) G - (-)^{F} F (\Delta G )$, 
where $F$ and $G$ are any functionals of fields and antifields.  
The BV bracket can be cast as 
\begin{align} 
\label{antibracket}
\big{(} \, F \, , \, G \, \big{)}  
= \sum_{g} \bigg{[} \, 
\frac{\partial _{r} F}{\partial \psi _{g} } \, \frac{\partial G}{\partial \psi ^{\ast }_{g} } 
-  \frac{\partial _{r} F}{\partial \psi ^{\ast }_{g} } \, \frac{\partial G}{\partial \psi _{g} } \, 
\bigg{]} \, . 
\end{align}
Note that $\partial _{r}$ denotes the right derivative and it satisfies  
$\frac{\partial }{\partial \psi _{g} } F = (-)^{g ( F +1 )} \frac{\partial _{r} }{\partial \psi _{g} } F$\,. 

\vspace{2mm}

\subsubsection*{BV master action}

We write $\{ \psi ^{a} \}_{a}$ for all of the field contents of fields and antifields, 
which are determined by solving the BV master equation.  
The contracted labels of $\{ \psi ^{a} \}_{a}$ run over all different types of fields.\footnote{ 
E.g. $\frac{1}{2} \int  \psi _{a}^{+} \, \psi ^{a} = \int dx [ A^{+}_{\mu } (x) \, A^{\mu } (x) + 
c^{+} (x) c(x) + 
\psi ^{+}_{\alpha } (x) \psi ^{\alpha } (x) + 
\bar{\psi }^{+}_{\alpha }(x)  \bar{\psi }^{\alpha } (x) ]$ for the field contents $\{ \psi ^{a} \} _{a} = \{ A^{\mu } (x) , c(x) , \psi ^{\alpha } (x) , \bar{\psi }^{\alpha } (x) , \, A^{+}_{\mu } (x) , c^{+} (x) , \psi ^{+}_{\alpha } (x) , \bar{\psi }^{+}_{\alpha } (x) \}_{\mu , \alpha , x}$ in QED. } 
In other words, 
the $a$-label of $\psi ^{a}$ distinguishes not only indices of a given field, 
such as the Lorentz indices, 
spinor indices, 
labels of symmetry generators or space-time points, 
but also species of fields themselves. 
Hence, 
we can always rewrite a given action into the form of 
\begin{align}
\label{L contracted form}
S [\psi ] = \frac{1}{2} \int 
\mu _{ab} \, \psi ^{b} \, \psi ^{a} 
+ \sum _{n > 2} \frac{1}{n!} \int 
\mu ^{\mathsf{sym}}_{a_{1} ... a_{n} } \, \psi ^{a_{n} } \, ... \, \psi ^{a_{1} }  
\end{align}
for every BV-quantizable field theory.\footnote{$S[\psi ]$ may have the $\hbar $ dependence $S[\psi ] = S_{\mathrm{cl}} [\psi _{\mathrm{cl} }] + \sum_{g} \hbar^{g} S_{\hbar ^{g} } [\psi ]$ when $(S_{\mathrm{cl}} , S_{\mathrm{cl}} ) =0 $ but $\Delta S_{\mathrm{cl}} \not=0$. } 
The $\mu _{ab}$ are kinetic operators between $\psi ^{a}$ and $\psi ^{b}$ and the $\mu ^{\mathsf{sym}}_{a_{1} ... a_{n}}$ are $n$-point vertices at which $\psi ^{a_{1}} , ... , \psi^{a_{n}}$ interact. 

\vspace{2mm} 

We write $(-)^{|\psi ^{a} |}$ for the sign of the total grading\footnote{This is the total ghost number (the total ghost number plus one) for the field contents starting with Grassmann even (odd) classical fields. } $|\psi ^{a}|$ of a given field content $\psi ^{a}$, 
where $|\psi ^{a} |$ is often abbreviated to $|a|$ for simplicity. 
The action $S [\psi ] \in \mathbb{R}$ and the coefficients $\{ \mu _{a_{1} .. a_{n}}^{\mathsf{sym} } \}_{n}$ of (\ref{L contracted form}) are of neutral total grading, 
for which we write $|S[\psi ] | = 0$ and $|\mu ^{\mathsf{sym}}_{a_{1} ... a_{n}} | = 0$ respectively. 
We can confirm the condition $\sum _{k=1}^{n}|\psi ^{a_{k}}| = 0$ on (\ref{L contracted form}). 
The BV Laplacian and the BV bracket can be expressed as  
\begin{align}
\hbar \, \Delta \, A = \frac{\hbar }{2} (-)^{ |\psi ^{a}| }  \frac{\overset{\rightarrow }{\delta } }{\delta \psi ^{a} }
 \omega ^{ab} \frac{\overset{\rightarrow }{\delta } }{\delta \psi ^{b} } \, A \, , \,
 \hspace{5mm} 
 \big{(} \, A \, , \, B \, \big{)} = \frac{\overset{\leftarrow }{\delta } A}{\delta \psi ^{a} } \, \omega ^{ab} \, \frac{\overset{\rightarrow }{\delta } B}{\delta \psi ^{b} } \, , 
\end{align}
in which the BV Poisson structure $\omega ^{ab}$ naturally appears.  
The components $\omega ^{ab}$ assign degree $1$ to inputs and satisfies $\omega ^{ab} = - (-)^{(|a|+1)(|b|+1)} \omega ^{ba}$ with $|a| + |b| = 1$, 
so that $\frac{\overset{\leftarrow }{\delta } }{\delta \psi ^{a} } \, \omega ^{ab} \, \frac{\overset{\rightarrow }{\delta } }{\delta \psi ^{b} }$ 
is zero unless $|\psi ^{a}| = - |\psi ^{b} | -1$ since $\omega ^{ab}$ assigns degree $1$.  
Note that as the $\mu _{a_{1} ... a_{n}}$ of (\ref{L contracted form}) satisfy $|\mu _{a_{1} ... a_{n}}| = 0$\,, 
the $\omega ^{ab}$ satisfy $|\omega ^{ab}| = 0$: 
They are just components.

\subsection{Lagrangian's $L_{\infty }$ structure}

For each (ordinary) field theory solving the BV master equation, 
we can always find a set of algebraic relations among its vertices,     
\begin{align}
\label{Zwiebach quantum L}
 \frac{\hbar }{2} (-)^{\varphi ^{a} } 
\omega ^{ab} \, \mu ^{\mathsf{sym}}_{ba a_{1} ... a_{n}}
+ \sum_{m=0}^{n-1}  
\sum_{\sigma \in \mathbb{S}_{n ; m} } \sigma \circ 
\mu ^{\mathsf{sym}}_{a_{1} ... a_{m} a} \, \omega ^{ab} \, \mu ^{\mathsf{sym}}_{b a_{m+1} ... a_{n} } 
= 0 \,  
\end{align}
for each fixed $n \in \mathbb{N}$, 
which is Lagrangian's (quantum) $L_{\infty }$ structure. 
The permutation $\sigma \in \mathbb{S}_{n ; m}$ denotes a $(m,n-m)$-shuffle of the $\{ a_{1} , ... , a_{m} ; a_{m+1} , ... , a_{n} \}$ indices: 
For $0 \leq m \leq n$, 
a permutation $\sigma \in \mathbb{S}_{n}$ of $1 < ... < n$ is called a $(m,n-m)$-shuffle if $\sigma (k) < \sigma (k+1)$ when $k \not= m$, 
where $<$ denotes the ordering of $\{ 1 , ... , n \}$ or $\{ \sigma (1) , ... , \sigma (n) \}$.  
In most cases, 
homotopy algebras are written as a set of algebraic relations between multilinear maps $\{ \mu _{n} \}_{n \in \mathbb{N}} $ acting on a graded vector space \cite{Zwiebach:1992ie, Markl:1997bj}: 
A set of the relations (\ref{Zwiebach quantum L}) is nothing but its component expression. 
 
\subsubsection*{Graded commutativity} 

We extract the set of vertices $\{ \mu ^{\mathsf{sym}}_{a_{0} ... a_{n}} \}_{a,n}$ from a given Lagrangian.\footnote{
It is a ``field-theoretically'' unique set. 
All possible field redefinitions and addition of auxiliary fields are included in the BV canonical transformations, 
which are BV-BRST exact shifts of $S[\varphi ]$ preserving the BV master equation. 
They are nothing else but homotopy algebra morphisms preserving a given cohomology and quasi-isomorphisms of homotopy algebras discussed in \cite{Kajiura:2003ax, Kajiura:2001ng, Jurco:2018sby} describe the field theoretical equivalence as well. } 
What we need to assume to obtain (\ref{Zwiebach quantum L}) from a given (\ref{L contracted form}) is the graded commutativity of vertices. 
We show that every field theory (\ref{L contracted form}) solving the BV master equation has a quantum $L_{\infty }$ structure whenever the the vertices are graded commutative with respect to all inputs 
\begin{align} 
\label{commutative vertices}
\mu ^{\mathsf{sym}}_{... a b ... } = (-)^{|\psi ^{a}| |\psi ^{b}| } 
\mu ^{\mathsf{sym}}_{ ... b a ... }  \, . 
\end{align}

\vspace{2mm} 

Ordinary field theories satisfy this requirement of (\ref{commutative vertices}).  
When we assume that the Lagrangian (\ref{L contracted form}) has the graded symmetric vertices (\ref{commutative vertices}), 
then, 
the cyclic property\footnote{
The requirement of cyclic property (\ref{cyclic vertices}) is weaker than that of graded commutativity (\ref{commutative vertices}).
Actually, 
the requirement (\ref{cyclic vertices}) is enough to extract a homotopy algebra from a given Lagrangian. } is automatic, 
\begin{align}
\label{cyclic vertices}
\mu ^{\mathsf{sym}}_{a_{0} a_{1} ... a_{n} } 
= (-)^{|\psi ^{a_{0}}| ( |\psi ^{a_{1}}| + \cdots + |\psi ^{a_{n}}| ) } 
\mu ^{\mathsf{sym}}_{a_{1} ... a_{n} a_{0} } 
\, . 
\end{align}
The upshot of the graded commutativity\footnote{
It is the same as the requirement that components of given fields are graded symmetric 
\begin{align*}
\int ... \, \psi ^{a} \, \psi ^{b} ... 
& = (-)^{|\psi ^{a}| |\psi ^{b}|} 
\int ... \, \psi ^{b} \, \psi ^{a} ... 
\, . 
\end{align*} 
In the expression of (\ref{L contracted form}), 
inputs of vertices are prepared to be these components of fields. } 
is that the Lagrangian of (\ref{L contracted form}) is permutation invariant with respect to its field contents 
\begin{align} 
\mu ^{\mathsf{sym}}_{a_{1} ... a_{n} } \psi ^{a_{n}} ... \, \psi ^{a_{1} } 
= (-) ^{|\sigma (\varphi )|}\mu ^{\mathsf{sym}}_{a_{1} ... a_{n} } \psi ^{a_{\sigma (n)}} ... \, \psi ^{a_{\sigma (1)} }  \, . 
\end{align} 

\subsubsection*{The equivalence of BV and homotopy algebra}

Under the assumption of the graded commutativity, 
Lagrangian's $L_{\infty }$ algebra is equivalent to imposing the BV master equation \cite{Jurco:2018sby, Doubek:2017naz}. 
While the BV Laplacian acts on the action as 
\begin{align}
\hbar \, 
\Delta \, S  
& = - \sum_{n \geq 1} \frac{1}{n!} \int 
\frac{\hbar }{2} (-)^{|\psi ^{a} |}  
\omega ^{ab} \, \mu ^{\mathsf{sym}}_{ba a_{1} ... a_{n} } 
\psi ^{a_{n}} ... \, \psi ^{a_{1}} \, , 
\end{align}
the BV-BRST transformation or homological vector field acts on the action as  
\begin{align}
\label{Zwiebach (S,S)}
\big{(} \, S  \, , \, S \, \big{)} 
& = \sum_{k,l} \frac{1}{k! l! } \int \, 
\psi ^{a_{k} } ... \, \psi ^{a_{1}} 
\, \mu ^{\mathsf{sym}}_{a a_{1} ... a_{k} } \, \omega ^{ab} \, \mu ^{\mathsf{sym}}_{b_{1} ... b_{l} b} 
\, \psi ^{b_{l}} ... \, \psi ^{b_{1} } 
\no 
& = - \sum _{n \geq 1} \frac{1}{n!} \int \,  \sum_{m=1}^{n} \frac{n!}{m! (n-m)!} 
\mu ^{\mathsf{sym}}_{a_{1} ... a_{m} b} \, \omega ^{ba} \, \mu ^{\mathsf{sym}}_{a a_{m+1} ... a_{n} }
\psi ^{a_{n} } ... \, \psi ^{a_{1}} 
\end{align}
We used the equality $\sum_{k,l} \frac{1}{k! l!} F_{k} G_{l} = \sum_{n\geq 0} \frac{1}{n!} \sum_{m=1}^{n} \frac{n!}{m! (n-m)!} F_{m} G_{n-m}$. 
We notice that since each vertex is permutation invariant, 
the equality (\ref{Zwiebach (S,S)}) can be expressed as follows 
\begin{align}  
\big{(} \, S \, , \, S \, \big{)} 
& = - \sum _{n \geq 1} \frac{1}{n!}  \int 
\sum_{m=1}^{n}  
\sum_{\sigma \in \mathbb{S}_{n ; m} } 
\mu ^{\mathsf{sym}}_{a_{\sigma(1)} ... a_{\sigma (m)} a} \, \omega ^{ab} \, \mu ^{\mathsf{sym}}_{b a_{\sigma (m+1)} ... a_{\sigma (n)} } 
\psi ^{a_{\sigma (n)} } ... \, \psi ^{a_{\sigma (1)}} \, . 
\end{align}
For $0 \leq m \leq n$, 
we write $\sigma \in \mathbb{S}_{n;m}$ for a permutation $\sigma \in \mathbb{S}_{n}$ of $\{ 1 , ... , n \}$ such that the order of $\sigma (1) < ... < \sigma (m)$ and $\sigma (m+1) < ... < \sigma (n)$ are preserved, 
namely, 
a $(m,n-m)$-shuffle. 
Hence, 
the BV master equation indeed gives Lagrangian's quantum $L_{\infty }$ relations (\ref{Zwiebach quantum L}) whenever the vertices are graded symmetric (\ref{commutative vertices}). 
See \cite{Schwarz:1992nx, Alexandrov:1995kv, Kajiura:2003ax, Kajiura:2001ng, Macrelli:2019afx} for the case of cyclic $L_{\infty }$ algebras. 

\subsection{Lagrangian's $A_{\infty }$ structure} 

We can always \textit{weaken} the requirement of the graded commutative vertices (\ref{commutative vertices}), 
from which Lagrangian's $A_{\infty }$ structure arises.  
We rewrite a given action into the form of the cyclic weight 
\begin{align}
\label{contracted form of S}
S [\varphi ] = \frac{1}{2} \int \, 
\mu _{ab} \, \psi ^{b} \, \psi ^{a} 
+ \sum _{n >1} \frac{1}{n+1} \int \, 
\mu _{a_{0} a_{1} ... a_{n} } \, \psi ^{a_{n} } \, ... \, \psi ^{a_{1} } \, \psi ^{a_{0} } 
\, . 
\end{align}
Notice that solving the BV master equation is nothing but a systematic prescription to determine a set of the above coefficients $\{ \mu _{a_{0} ... a_{n}} \}_{a}$. 
Then, 
instead of (\ref{commutative vertices}), 
we assume the cyclic property of the vertices only\footnote{It is the same as the requirement that all of given field contents $\varphi = \{ \varphi ^{a} \}_{a}$ satisfy the cyclic property 
\begin{align*} 
\int \, 
\varphi ^{a_{0} } \, \psi ^{a_{1}} \cdots \psi ^{a_{n} } 
= (-)^{|\psi ^{a_{0}} | ( | \psi ^{a_{1}}| + \cdots + | \psi ^{a_{n}} | ) } 
\int \, 
\psi ^{a_{1}} \cdots \psi ^{a_{n} } \, \psi ^{a_{0} } 
\, . 
\end{align*}
The integral symbol denotes not only the space-time (or momentum) integral but also the trace if any. }   
\begin{align}
\label{cyclic component} 
\mu _{a_{0} a_{1} ... a_{n} } = (-)^{| \psi ^{a_{0}} | ( | \psi ^{a_{1}} | + \cdots + | \psi ^{a_{n}} | ) } 
\mu _{a_{1} ... a_{n} a_{0} } 
\, . 
\end{align}
The (quantum) $A_{\infty }$ structure of quantum field theory can be extracted from the BV master equation imposed on (\ref{contracted form of S}) under the requirement of (\ref{cyclic component}) in the same way as $L_{\infty }$, 
which we explain in the next section.  
For each fixed $n \in \mathbb{N}$, 
a quantum $A_{\infty }$ relation is 
\begin{align}
\sum_{s,t} 
\bigg{[} 
 \frac{\hbar }{2} (-)^{\psi ^{a_{t}} } 
\omega ^{a_{t} a_{s} } \, \mu _{a_{1} ...a_{s} ... a_{t} ... a_{n}}
+ \sum_{m=0}^{n-1}  
\mu _{a_{1} ... a_{s} ... a_{m}} \, \omega ^{a_{s}a_{t}} \, \mu _{a_{m+1} ... a_{t} ... a_{n} } 
\, \bigg{]} 
= 0 \,  . 
\end{align}

\vspace{2mm} 

Lagrangian's $A_{\infty }$ can be always found whenever Lagrangian's $L_{\infty }$ structure exists \cite{Kajiura:2003ax, Kajiura:2001ng, Jurco:2019yfd}. 
Note however that whenever the vertices of (\ref{contracted form of S}) are graded commutative $\mu _{... st ...} = (-)^{|\psi _{s} || \psi _{t} |} \mu _{... ts ...}$ as usual, 
the description based on (\ref{contracted form of S}) with $A_{\infty }$ becomes \textit{redundant} -- all permuted vertices $\mu _{a_{\sigma (1)} ... a_{\sigma (n) }}$ reproduce the same contributions as the original vertex $\mu _{a_{1} ... a_{n} }$ -- and then Lagrangian's $A_{\infty }$ structure automatically reduces to the $L_{\infty }$ structure.  
As usual, 
then, 
it is economical and physically natural to use the vertices $\{ \mu ^{\mathsf{sym}}_{a_{0} ... a_{n}} \} _{a}$ defined by the sum of all possible permutation 
\begin{align}
\mu ^{\mathsf{sym} }_{a_{0} ... a_{n} } \equiv \sum_{\sigma \in \mathbb{S}_{n+1} } \mu _{a_{\sigma (0)} ... a_{\sigma (n) } } \, . 
\end{align}
The action (\ref{contracted form of S}) of the cyclic weight is the same as the action (\ref{L contracted form}) of the symmetric wight then, 
so that two Feynman graphs based on $A_{\infty }$ and $L_{\infty }$ are the same then.  

\vspace{2mm} 

The form (\ref{contracted form of S}) is nothing but the component expression of the action (\ref{classical physical action}).

\subsection{(Quantum) $A_{\infty }$ reduces to (quantum) $L_{\infty }$}

To see the relation between Lagrangian's $A_{\infty }$ and $L_{\infty }$ structures directly, 
it is helpful to consider the state space of fields and to regard given vertices as multilinear maps on such a vector space. 
Let $\mathcal{\widehat{H}}$ be the state space of fields and $\mathcal{T} (\mathcal{\widehat{H}} )$ be the tensor algebra. 
We write $\{ \hat{e}_{a} \}_{a}$ for a basis of $\mathcal{\widehat{H}}$ explicitly and write $\psi = \psi ^{a} \, \hat{e}_{a}$ for a vector of the state space. 
See appendix A for the basis. 

\vspace{2mm} 

We notice that given vertices of the form $\mu _{a_{0} a_{1} ... a_{n}} \, \psi ^{a_{1}} \cdots \, \psi ^{a_{n}}$ can be regarded as products or multiplications of fields $\psi _{1}, ... , \psi _{n}$ that belong to the state space. 
In general, 
we can regard the products of given fields as values of multilinear maps $\mu _{n}$ acting on the tensor product $\psi _{1} \otimes \cdots \otimes \psi _{n}$, 
which may be non-commutative in general. 
We write 
\begin{align}
\label{multilinear mu}
\mu _{n} ( \psi _{1} , ... , \psi _{n} ) 
 \equiv \mu _{n} \, ( \psi _{1} \otimes \cdots \otimes \psi _{n} ) \, . 
\end{align}
%
%
%
For instance, 
the vertices appearing in the action (\ref{classical physical action}) can be specified by these multilinear maps. 
%
%
In most cases, 
these multilinear maps acting on fields associate with algebraic relations, 
such as coupling constant, 
delta functions of momentum conservation, 
contractions of indices, 
cut-off functions, 
space-time differentials or structure constants of Lie algebras. 
%
The $A_{\infty }$ or $L_{\infty }$ structure we consider in this paper is a special combination of such algebraic relations,\footnote{As we see, the BV formalism uniquely specifies such a set of algebraic relations for a give field theory. } 
$\mu = \{ \mu _{n} \}_{n}$, 
%
that can be identified with properties of multilinear maps $\{ \mu _{n} \}_{n}$ acting on the tensor algebra. 

\vspace{2mm} 

When we consider ordinary quantum field theory,\footnote{A quantum field theory of particles that is not based on the non-commutative geometry. } 
because of the graded commutativity of fields $\psi _{1} \cdot \psi _{2} = (-)^{\psi _{1} \psi _{2}} \psi _{2} \cdot \psi _{1}$, 
it is economical to consider the symmetric tensor algebra $\mathcal{S} (\mathcal{\widehat{H}} )$ instead of $\mathcal{T} (\mathcal{\widehat{H}})$.  
%
The symmetrized tensor product, 
%
%
%
%
\begin{align}
\label{symmetrized tensor product}
\psi _{1} \wedge \cdots \wedge \psi _{n} \equiv \sum_{\sigma \in \mathbb{S} } (-)^{\sigma (\psi ) } 
\psi _{\sigma (1) } \otimes \cdots \otimes \psi _{\sigma (n) } \, , 
\end{align}
is a natural product of the symmetric tensor algebra $\mathcal{S} (\widehat{\mathcal{H}} )$. 
Then, 
instead of (\ref{multilinear mu}), 
it is reasonable to consider the values of multilinear maps $\{ \mu _{n} \}_{n}$ acting on the symmetrized tensor product 
\begin{align} 
\mu ^{\mathsf{sym}}_{n} (\psi _{1} , ... , \psi _{n} ) \equiv \mu _{n} (\psi _{1} \wedge \cdots \wedge \psi _{n} ) \, . 
\end{align} 
This kind of $\mu ^{\mathsf{sym} }(\psi , ... , \psi )$ gives natural algebraic structures appearing in \textit{commutative} quantum field theory. 
%
Note that $\mu$ and $\mu ^{\mathsf{sym}}$ are distinguished by just input states: 
the commutativity of space-time fields can be identified with a property of inputs. 
When we consider \textit{commutative} quantum field theory, 
we can obtain $\mu ^{\mathsf{sym}} (\psi , ... , \psi )$ from a given $\mu (\psi , ... , \psi )$ as follows  
\begin{align}
\label{A to L}
\mu _{n} (\psi _{1} , ... , \psi _{n} )
= \frac{1}{n!} \, \mu ^{\mathsf{sym}}_{n} (\psi _{1} , ... , \psi _{n} ) \, .
\end{align} 
The factor $n !$ comes from the symmetrization of the tensor product. 
In this paper, 
we thus consider properties of algebraic structures $\mu (\psi , ... , \psi )$ that do not depend on the graded commutativity of space-time fields. 
As (\ref{A to L}), 
our $\mu (\psi, ... , \psi )$ reduces to $\mu ^{\mathsf{sym} }(\psi , ... , \psi )$ automatically whenever we consider ordinary quantum field theory. 
Actually, 
the relation of $\mu $ and $\mu ^{\mathsf{sym} }$ is nothing but that of (quantum) $A_{\infty }$ and $L_{\infty }$\,. 
The quantum $A_{\infty }$ structure appearing in this paper can be always switched to the quantum $L_{\infty }$ structure for ordinary quantum field theory.\footnote{Note that it does not imply that the $A_{\infty }$ products can be written in terms of the $L_{\infty }$ products for ordinary field theory: 
the $A_{\infty }$ or $L_{\infty }$ products we consider are not elementary objects of a given theory but special combinations of them. 
Even for closed string field theory, 
the $L_{\infty }$ products or string vertices are constructed by specifying conformal mappings or differential forms on the (puncture-symmetrized) moduli space of Riemann surfaces. }

\vspace{2mm}

As we see later, 
physical gradings, 
such as the space-time ghost number or Grassmann parity of fields, 
do not give the $A_{\infty }$ degree directly. 
In addition, 
by using appropriate (de-)suspension maps, 
the change of the grading of $A_{\infty }$ algebras does not change the physics. 
Hence, 
it is useful to set all $A_{\infty }$ products to have degree $1$, 
which we call a natural $A_{\infty }$ degree. 


\subsubsection*{Quantum $A_{\infty }$ structure}

An $A_{\infty }$ structure $\boldsymbol{\mu } = \boldsymbol{\mu }_{1} + \boldsymbol{\mu }_{2} + \cdots$ is a (co-)derivation acting on $\mathcal{T}(\mathcal{\widehat{H}} )$ such that $( \boldsymbol{\mu } )^{2}= 0$. 
For a given $\psi _{1} \otimes \cdot \cdot \cdot \otimes \psi _{n} \in \mathcal{T}(\mathcal{\widehat{H}} ) $ with fixed $n \geq 1$, 
the $A_{\infty }$ relations $( \boldsymbol{\mu } )^{2} =0$ can be represented as  
\begin{align} 
\label{def of A-relations} 
\sum _{k+l=n} \sum_{m=0}^{k} (-)^{\epsilon (\psi ) }
\boldsymbol{\mu }_{k+1} 
\big{(} \, \underbrace{\psi _{1} , ... , \psi _{m} }_{m} , \, 
\boldsymbol{\mu }_{l} (\psi _{m+1} , ... , \psi _{m+l} ) , \, \underbrace{\psi _{m+l+1} , ... , \psi _{n}}_{k-m} \, \big{)} = 0 \, , 
\end{align}
where $\epsilon (\psi )$ denotes the sign factor arising from $\boldsymbol{\mu }_{l}$ passing $\psi _{1} \otimes \cdot \cdot \cdot \otimes \psi _{m}$\,. 
Let $\omega $ be a graded symplectic structure of degree $-1$ and $\{ \hat{e}_{-s} , \hat{e}_{1+s} \} _{s \geq 0}$ be a set of complete basis such that $\omega ( \hat{e}_{-s} , \hat{e}_{1+s'} ) = (-)^{s} \delta _{s,s'}$. 
A cyclic $A_{\infty }$ structure is an $A_{\infty }$ structure $\boldsymbol{\mu }$ satisfying $\omega (\boldsymbol{\mu } \otimes 1 + 1 \otimes \boldsymbol{\mu } )=0$, 
which is the classical limit of a quantum $A_{\infty }$ structure. 

\vspace{2mm} 

A quantum $A_{\infty }$ structure $\boldsymbol{\mu } + \hbar \, \mathfrak{L}$ is a linear map acting on $\mathcal{T} (\mathcal{\widehat{H}} )$ such that $(\boldsymbol{\mu } + \hbar \, \mathfrak{L} )^{2} = 0$ 
where $\boldsymbol{\mu } = \sum_{n} \boldsymbol{\mu }_{n\, [0] } + \sum _{n,g } \hbar ^{g} \, \boldsymbol{\mu }_{n, [g]}$ is a (co-)derivation and $\mathfrak{L}$ is a second order (co-)derivation. 
For fixed $n\geq 1$ and $g \geq 0$, 
the quantum $A_{\infty }$ relations $(\boldsymbol{\mu } + \hbar \, \mathfrak{L} )^{2} = 0$ can be represented as 
\begin{align}
\label{def of quantum A-relations} 
& \hspace{7mm}  
\sum _{\substack{k+l=n \\ g_{1}+g_{2} =g } } \sum_{m=0}^{k} (-)^{\epsilon (\psi ) }
\boldsymbol{\mu }_{k+1 ,\, [g_{1}] } \big{(} \underbrace{\psi _{1} , ... , \psi _{m} }_{m} , \, 
\boldsymbol{\mu }_{l ,\, [g_{2} ] } (\psi _{m+1} , ... , \psi _{m+l} ) , \, \underbrace{\psi _{m+l+1} , ... , \psi _{n}}_{k-m} \big{)} 
\no & +
\sum _{s \in \mathbb{Z} } \sum_{i = 0}^{n} \sum_{j=0}^{n-i} (-)^{\epsilon (s,i,j)}
\boldsymbol{\mu }_{n+2 , [g-1] } \big{(} \underbrace{\psi _{1} , ... , \psi _{i} }_{i} , \hat{e}_{-s} , \underbrace{\psi _{i+1} , ... , \psi _{i+j} }_{j} , \hat{e}_{1+s} , \psi _{i+j+1} , ... , \psi _{n} \, \big{)} 
= 0 \, , 
\end{align}
where the sign factor $\epsilon (s,i,j)$ arises from $\hat{e}_{1+s}$ passing $\psi _{1} \otimes \! \cdot \! \cdot \! \cdot \! \otimes \psi _{i+j}$ and $\hat{e}_{-s}$ passing $\psi _{1} \otimes \! \cdot \! \cdot \! \cdot \! \otimes \psi _{i}$\,. 


\subsubsection*{Quantum $L_{\infty }$ structure}

An $L_{\infty }$ structure $\boldsymbol{\mu }^{\mathsf{sym }}= \boldsymbol{\mu }^{\mathsf{sym} }_{1} + \boldsymbol{\mu }^{\mathsf{sym} }_{2} + \cdots$ is a (co-)derivation acting on $\mathcal{S}(\mathcal{\widehat{H}} )$ such that $( \boldsymbol{\mu }^{\mathsf{sym} } )^{2}= 0$. 
For fixed $n \geq 1$, 
the $L_{\infty }$ relations $( \boldsymbol{\mu }^{\mathsf{sym }} )^{2} =0$ can be represented as follows 
\begin{align} 
\label{def of L-relations}
\sum _{k+l=n} \sum_{\sigma \in \mathsf{S}_{l,k} } (-)^{\sigma (\psi ) }
\boldsymbol{\mu }^{\mathsf{sym}}_{k+1} \big{(} \, \boldsymbol{\mu }^{\mathsf{sym}}_{l} (\psi _{\sigma (1) } , ... , \psi _{\sigma (l) } ) , \psi _{\sigma (l+1)} , ... , \psi _{\sigma (n) } \, \big{)} = 0 \, , 
\end{align}
where $\sigma (\psi )$ denotes the sign factor arising from the $(l,k)$-unshuffle of $\psi _{\sigma (1)} \wedge \cdot \cdot \cdot \wedge \psi _{\sigma (n)} \in \mathcal{S}(\mathcal{\widehat{H}} )$. 
A cyclic $L_{\infty }$ structure is an $L_{\infty }$ structure $\boldsymbol{\mu }^{\mathsf{sym} }$ satisfying $\omega (\boldsymbol{\mu }^{\mathsf{sym} } \otimes 1 + 1 \otimes \boldsymbol{\mu }^{\mathsf{sym} } ) =0$, 
which is the classical limit of a quantum $L_{\infty }$ structure.\footnote{When we consider other gradings, 
such as $2-n$ for $\mu _{n}$, 
the same relations hold as (\ref{def of A-relations}), 
(\ref{def of quantum A-relations}), 
(\ref{def of L-relations}) or (\ref{def of quantum L-relations}) except for the sign factors: 
(de-)suspension maps relate them. 
See also \cite{Jurco:2018sby, Markl:1997bj, Herbst:2006kt, Munster:2011ij}. }   

\vspace{2mm} 

A quantum $L_{\infty }$ structure $\boldsymbol{\mu }^{\mathsf{sym}} + \hbar \, \mathfrak{L}$ is a linear map acting on $\mathcal{S} (\mathcal{\widehat{H}})$ such that $(\boldsymbol{\mu }^{\mathsf{sym}} + \hbar \, \mathfrak{L})^{2} = 0$ 
where $\boldsymbol{\mu }^{\mathsf{sym} } = \sum_{n} \boldsymbol{\mu }_{n\, [0] }^{\mathsf{sym} } + \sum _{n,g } \hbar ^{g} \, \boldsymbol{\mu }_{n, [g]}^{\mathsf{sym} }$ is a (co-)derivation and $\mathfrak{L}$ is a second order (co-)derivation. 
For fixed $n > 0$ and $g \geq 0$, 
the quantum $L_{\infty }$ relations $(\boldsymbol{\mu }^{\mathsf{sym}} + \hbar \, \mathfrak{L})^{2} = 0$ can be represented as  
\begin{align}
\label{def of quantum L-relations} 
&
\sum_{\substack{k+l = n \\ g_{1} + g_{2} = g }} \sum_{\sigma \in \mathsf{S}_{l,k} } (-)^{\sigma (\psi ) } 
\boldsymbol{\mu }^{\mathsf{sym}}_{k+1, \,[g_{1} ] } \big{(} \, \boldsymbol{\mu }^{\mathsf{sym}}_{l , \, [g_{2} ] } (\psi _{\sigma (1) } , ... , \psi _{\sigma (l) } ) , \psi _{\sigma (l+1)} , ... , \psi _{\sigma (n) } \, \big{)} 
 \no & \hspace{25mm} 
 + \frac{1}{2} \sum_{s \in \mathbb{Z} }  \boldsymbol{\mu }^{\mathsf{sym}}_{n+2, \,[g-1] }  \big{(} \, \hat{e}_{-s} , \hat{e}_{1+s} , \psi _{1} , ... , \psi _{n} \, \big{)} = 0 \, .  
\end{align}


As we see in section 3, 
a quantum $A_{\infty }$ structure can be assigned to \textit{every} quantum field theory that solves the BV master equation. 
Most of our results will be presented in terms of $A_{\infty }$ since the results based on $A_{\infty }$ can be always switched to those based on $L_{\infty }$ for ordinary quantum field theory. 
As long as we consider an $A_{\infty }$ structure $\mu $ that can be represented by the form of (\ref{multilinear mu}), 
the (quantum) $A_{\infty }$ structure $\mu $ of \textit{commutative} quantum field theory reduces the (quantum) $L_{\infty }$ structure $\mu ^{\mathsf{sym }}$ automatically just as (\ref{A to L}). 
We end this subsection by giving two examples. 

\subsubsection*{$4$ point amplitude} 

The amplitudes of Lagrangian field theory have a quantum $A_{\infty }$ structure, 
which we will explain in more detail in section 5. 
Let us consider the cubic action, 
which is (\ref{classical physical action}) with $\mu _{n >2} = 0$. 
It can be a non-commutative field theory.
We write $\mu _{1}^{-1}$ for a propagator of this theory. 
The $4$ point amplitude $\mathcal{A}_{4}$ is given by  
\begin{align}
\label{ST rep}
\mathcal{A}_{4} \sim \la \psi _{0} , \,  
\mu _{2} ( \mu_{1}^{-1} \mu _{2} ( \psi _{1} , \psi _{2} ) , \psi _{3} ) \ra 
+ 
\la \psi _{0} , \, \mu _{2} ( \psi _{1} , \mu_{1}^{-1} \mu _{2} ( \psi _{2} , \psi _{3} ) ) \ra \, . 
\end{align}
It consists of the $S$-channel and $T$-channel. 
When multiplications of space-time fields are commutative, 
$\mu $ reduces to $\mu ^{\mathsf{sym}}$ as (\ref{A to L}). 
Then, 
the expression (\ref{ST rep}) reduces to 
\begin{align}
\mathcal{A}_{4} & \sim  
\la \psi _{0} , \, \mu ^{\mathsf{sym} }_{2} ( \mu_{1}^{-1} \mu ^{\mathsf{sym}}_{2} ( \psi _{1} , \psi _{2} ) , \psi _{3} ) \ra 
+\la \psi _{0} , \mu ^{\mathsf{sym} }_{2} ( \mu_{1} ^{-1} \mu ^{\mathsf{sym}}_{2} ( \psi _{2} , \psi _{3} ) , \psi _{1} ) \ra 
\no & \hspace{25mm} 
+ \la \psi _{0} , \, \mu ^{\mathsf{sym} }_{2} ( \mu_{1}^{-1} \mu ^{\mathsf{sym}}_{2} ( \psi _{3} , \psi _{1} ) , \psi _{2} ) \ra \, .
\end{align}
It consists of the $S$-channel, 
the $T$-channel and the $U$-channel. 
As is known, 
this is a $4$ point amplitude of commutative Lagrangian field theory. 

\subsubsection*{Yang-Mills theory} 

Let us consider the $A_{\infty }$ structure of the ordinary Yang-Mills action $S [A] = - \frac{1}{2} \int \, \langle F , \star \, F \rangle $\,, 
which is a commutative Lagrangian field theory. 
Yang-Mills fields $A$ are Lie-algebra-value $1$-forms. 
The first $A_{\infty }$ structure is given by the kinetic operator  
\begin{align}
\mu _{1} ( A_{1} ) & = d \, \star \, d \, A_{1} \, , 
\end{align} 
where $d$ denotes the exterior differential and $\star $ denotes the Hodge dual operation.  
By casting the Yang-Mills action as the form of (\ref{classical physical action}), 
vertices provides higher $A_{\infty }$ products 
\begin{align}  
\mu _{2} ( A_{1} , A_{2} ) & = d \, \star \big{(} A_{1} \wedge A_{2} \big{)} 
-  \big{(} \star \, d \, A_{1} \big{)} \wedge A_{2} 
+  A_{1} \wedge \big{(} \star \, d \, A_{2} \big{)} \, ,
\\ 
\mu _{3} (A_{1} , A_{2} , A_{3} ) & = A_{1} \wedge \big{(} \star ( A_{2} \wedge A_{3} ) \big{)} 
- \big{(} \star ( A_{1} \wedge A_{2} ) \big{)} \wedge A_{3} \, ,
\end{align}
where $\wedge $ denotes the exterior product of forms. 
Note that this $\wedge $ is different from the symmetrized tensor product of (\ref{symmetrized tensor product}). 
As a symmetrization of exterior products, 
we can consider the graded commutator of exterior products, 
$[A_{1} , A_{2} ]_{\wedge } \equiv A_{1} \wedge A_{2} - (-)^{A_{1} A_{2} } A_{2} \wedge A_{1}$. 
We find  
\begin{align} 
\mu ^{\mathsf{sym} }_{2} ( A_{1} , A_{2} ) & = d \, \star \big{[} \, A_{1} \, , \, A_{2} \, \big{]}_{\wedge } 
-  \big{[} \star d \, A_{1} \, , \,  A_{2} \, \big{]}_{\wedge } 
+  \big{[} \, A_{1} \, , \star \, d \, A_{2} \, \big{]}_{\wedge } \, ,
\\ 
\mu ^{\mathsf{sym} }_{3} (A_{1} , A_{2} , A_{3} ) & 
= \big{[} \, A_{1} \, , \star [ A_{2} , A_{3} ]_{\wedge } \, \big{]}_{\wedge } 
+ \big{[} \, A_{2} \, , \star [ A_{3} , A_{1} ]_{\wedge } \, \big{]}_{\wedge } 
+ \big{[} \, A_{3} \, , \star [ A_{1} , A_{2} ]_{\wedge } \, \big{]}_{\wedge } \, . 
\end{align}
These are the $L_{\infty }$ structure of the Yang-Mills theory. 
These $A_{\infty }$ and $L_{\infty }$ structures are related to each other by (\ref{A to L}). 
As is known, 
the above \textit{incomplete} $A_{\infty }$ (or $L_{\infty }$) structure is just a piece of the complete $A_{\infty }$ (or $L_{\infty }$) structure of the BV master action for the Yang-Mills theory. 
When we consider non-commutative quantum field theory, 
it has only an $A_{\infty }$ structure obtained by replacing the exterior product with the non-commutative product.

\subsection{Deformation retract} 

Let us consider a sequence of vector spaces $\mathcal{H} = \{ \mathcal{H}_{g} \}_{g \in \mathbb{Z}}$ equipped with morphisms $Q = \{ Q |_{g} : \mathcal{H}_{g} \rightarrow \mathcal{H}_{g+1} \}_{g \in \mathbb{Z}}$ of degree $1$. 
When $Q |_{g+1} \circ Q|_{g} = 0$ for any $g \in \mathbb{Z}$ is satisfied, 
we write $Q^{2} = 0$ and call $Q$ a differential on $\mathcal{H}$. 
A sequence of vector spaces equipped with a differential $Q$ is called a $Q$-complex, 
which we abbreviate to $( \mathcal{H} , Q )$ as follows   
\begin{align} 
\label{sequence: DR}
( \, \mathcal{H} \, , Q \, ) 
\,\, \overset{\mathrm{def}}{ \Longleftrightarrow } \,\, 
\cdots \,\, 
\overset{Q|_{g-1}}{\longrightarrow } \,\, 
\mathcal{H}_{g} \,\, 
\overset{Q|_{g}}{\longrightarrow } 
\mathcal{H}_{g+1} \,\, 
\overset{Q|_{g+1}}{\longrightarrow } \,\, 
\cdots  
\hspace{4mm} 
\mathrm{with} \hspace{4mm} 
Q |_{g+1} \circ Q|_{g} = 0 \, . 
\end{align} 
For example, 
the state space $\mathcal{H}$ of covariant strings and the string BRST operator $Q$ can be regarded as a BRST complex $(\mathcal{H} , Q)$ because of $Q^{2} = 0$, 
where each $\mathcal{H}_{g}$ denotes the ghost number $g$ sector and each $Q|_{g}$ is given by the restriction of the BRST operator onto $\mathcal{H}_{g}$. 

\vspace{2mm} 

We write $\mathcal{H}_{\mathrm{phys}} |_{g} \equiv  \mathrm{Ker} [Q|_{g}] / \mathrm{Im} [Q|_{g-1}]$ for the $Q$-cohomology on $\mathcal{H}_{g}$ and consider a sequence of $\mathcal{H}_{\mathrm{phys}} = \{ \mathcal{H}_{\mathrm{phys} }|_{g} \}_{g \in \mathbb{Z} }$ as well. 
With a trivial differential $0$, 
we regard the sequence $(\mathcal{H}_{\mathrm{phys} } , 0)$ as a complex. 
We write $\pi : \mathcal{H} \rightarrow \mathcal{H}_{\mathrm{phys } }$ for a projection onto the cohomology and $\iota : \mathcal{H}_{\mathrm{phys} } \rightarrow \mathcal{H}$ for a inclusion from the cohomology, 
so that the projection $(1-\iota \pi ): \mathcal{H} \rightarrow \mathcal{H}$ makes $Q$ invertible.  
We write $\kappa ^{-1} = \{ \kappa ^{-1}|_{g} : \mathcal{H}_{g} \rightarrow \mathcal{H}_{g-1} \}_{g \in \mathbb{Z}}$ for an inverse for $Q$, 
which is of degree $-1$ and called a contracting homotopy. 
Now we can add arrows of $\kappa $, $\pi$ and $\iota $ to the complex as follows 
\begin{align} 
\label{sequences: DR}
\mathcal{H} \,\,  
\overset{\pi }{ \underset{\iota }{ \substack{
\relbar \joinrel \! \relbar \joinrel \! \longrightarrow 
\\[-1mm]
  \longleftarrow \! \joinrel \relbar \! \joinrel \relbar 
} } }  \,\, 
 \mathcal{H}_{\mathrm{phys}} 
\hspace{4mm}  \mathrm{and} \hspace{4mm} 
\cdots \,\, 
\overset{Q|_{g-1}}{ \underset{\kappa ^{-1}|_{g} }{ \substack{
\relbar \joinrel \! \relbar \joinrel \! \longrightarrow 
\\[-1mm]
\longleftarrow \! \joinrel \relbar \! \joinrel \relbar } } } \,\,  
\mathcal{H}_{g} \,\, 
\overset{Q| _{g}}{ \underset{\kappa ^{-1}|_{g+1}}{ \substack{
\relbar \joinrel \! \relbar \joinrel \! \longrightarrow 
\\[-1mm]
  \longleftarrow \! \joinrel \relbar \! \joinrel \relbar } } } \,\, 
\mathcal{H}_{g+1}  \,\, 
\overset{Q|_{g+1}}{ \underset{\kappa ^{-1}|_{g+2}}{ \substack{
\relbar \joinrel \! \relbar \joinrel \! \longrightarrow 
\\[-1mm]
\longleftarrow \! \joinrel \relbar \! \joinrel \relbar } } } \,\, 
\cdots   
\, .
\end{align}
These arrows clarify where $Q$ is invertible and the $Q$-cohomology condenses.  
We require $\pi \, \iota = \mathrm{Id}$ on $\mathcal{H}_{\mathrm{phys} }$ and impose $\pi \kappa ^{-1} =\kappa ^{-1} \iota = (\kappa ^{-1})^{2} = 0$ on $\mathcal{H}$ as well. 
Each arrow satisfies a Hodge decomposition and we find the identity 
\begin{align} 
\label{Hodge: DR}
Q \, \kappa ^{-1} + \kappa ^{-1} \, Q = 1 - \iota \, \pi 
\, .
\end{align}
For example, 
the Siegel gauge propagator $b_{0} \, \int_{0}^{\infty } dt \, e^{- t \,L_{0} } = \frac{b_{0}}{L_{0}} (1 - ip )$ gives such an inverse for the string BRST operator $Q$, 
where $L_{0}$ denotes the Virasoro zero mode and $\iota \pi$ is the projection onto $\mathrm{Ker}[L_{0}]$ -- the space of level zero states. 
The decomposition (\ref{Hodge: DR}) is the key stone of the string no-ghost theorem \cite{Kato:1982im, Aisaka:2004ga, Matsunaga:2019fnc}. 

\vspace{2mm} 

A deformation retract is the relation between two sequences $(\mathcal{H} , Q)$ and $( \mathcal{H}_{\mathrm{phys}} , 0 )$ described by the arrows in sequences (\ref{sequences: DR}) and the decomposition (\ref{Hodge: DR}).    
The diagram 
\begin{align} 
\label{abbreviate}
\kappa ^{-1} \,\, \rotatebox{90}{$\hspace{-1mm} \curvearrowright $} \, 
( \, \mathcal{H} \, , Q \, ) \,\, \, 
\overset{\pi }{ \underset{\iota }{ \substack{
\relbar \joinrel \! \relbar \joinrel \! \longrightarrow 
\\[-1mm]
  \longleftarrow \! \joinrel \relbar \! \joinrel \relbar } } } 
\,\, \, 
( \, \mathcal{H}_{\mathrm{phys}} \, , \, q \, ) 
\end{align} 
is an abbreviate expression for the deformation retract. 
Note that we can consider a non-zero differential $q $ on $\mathcal{H}_{\mathrm{phys}}$ in general, 
although $q = 0$ above. 
As we see in the next section, 
the perturbative path-integral is described by the projection onto the BV-BRST cohomology.

\subsection{Homological perturbation}

A homological perturbation is a map from a given differential $Q$ to a new differential $Q + \delta $, 
which tells us a new complex $( \mathcal{H} , Q + \delta )$. 
In order to perform such a perturbation, 
we have to find out $\delta $ such that $(Q + \delta )^{2} = 0$ in the sense of (\ref{sequence: DR}) for a given $(\mathcal{H} , Q)$. 
The idea of homological perturbation can be extended to deformation retracts. 
From the original deformation retract (\ref{abbreviate}) and a given perturbation $\delta $, 
we can obtain a new deformation retract  
\begin{align} 
\label{new abbreviate}
K ^{-1} \,\, \rotatebox{90}{$\hspace{-1mm} \curvearrowright $} \, 
( \, \mathcal{H} \, , Q + \delta  \, ) \,\, \, 
\overset{P}{ \underset{I}{ \substack{
\relbar \joinrel \! \relbar \joinrel \! \longrightarrow 
\\[-1mm]
  \longleftarrow \! \joinrel \relbar \! \joinrel \relbar } } } 
\,\, \, 
( \, \mathcal{H}_{\mathrm{phys}} \, , \, q_{\mathrm{eff} } \, ) 
\end{align} 
by solving the recursive relations 
\begin{align}
P = \pi + \kappa ^{-1} \delta \, P \, , 
\hspace{5mm} 
I = \iota + I \, \delta \, \kappa ^{-1} \, , 
\hspace{5mm} 
K^{-1} = \kappa ^{-1} + K^{-1} \, \delta \, \kappa ^{-1} \, . 
\end{align}
The (perturbed) \textit{differential} $q_{\mathrm{eff}}$ on $\mathcal{H}_{\mathrm{phys} }$ that satisfies $(q_{\mathrm{eff} })^{2} = 0$ is given by 
\begin{align}
q_{\mathrm{eff} } \equiv q + P \, ( Q + \delta ) \, \iota = q + \pi \, ( Q + \delta ) \, I \, . 
\end{align}
The results are proved by straightforward calculations. 
The recursive construction of new data (\ref{new abbreviate}) from given data is known as the homological perturbation lemma. 
See \cite{Crainic} for details.

\section{Projection onto the BV-BRST cohomology} 

In this section, 
we show that a homological perturbation performs the perturbative path-integral in terms of the BV-BRST cohomology, 
and discuss several properties that effective theories have as a consequence of it.\footnote{
We will give corresponding results in section 4 in terms of Lagrangian's homotopy algebra. } 
We first explain that solving the BV master equation is equivalent to extracting a quantum $A_{\infty }$ structure intrinsic to each Lagrangian field theory. 
Next, 
after giving a brief review of basic facts of the BV formalism which are related to properties of the path-integral, 
%
we explain our idea and requirement that bring us to main results. 
Then, 
we show that the path-integral is nothing but a projection map onto the BV-BRST cohomology by constructing such a map explicitly.\footnote{The equivalence of performing the Feynman graph expansion and calculating the BRST cohomology is natural for physicists. 
Except for details,  
this kind of equivalence (or hypothesis) would be more or less known 
(or expected) as a mathematical aspect of the BV-BRST formalism.
}  
This section ends with presenting several properties of the effective $A_{\infty }$ structure. 
See appendix A before reading subsection 3.1. 

\vspace{2mm} 

Note that 
quantum field theories \textit{without} gauge degrees can be also treated within the BV formalism. 
Although it trivially solves the BV master equation, 
it provides non-trivial results.

\subsection{$A_{\infty }$ structure of the BV master equation}

Suppose that for a given Lagrangian field theory, 
its BV master action $S[\psi ]$ was obtained by solving the BV master equation. 
When the theory consists of physical degrees only, 
the BV master action is the classical action itself. 
We start with a given $S [\psi ]$\,: 
see appendix A for notation. 

\vspace{2mm} 

We first consider the simplest case. 
Suppose that a solution $S$ of the classical master equation $(S,S)=0$ also solves the quantum master equation $\hbar \, \Delta S + \frac{1}{2} (S , S) = 0$ \textit{without any modification}. 
The cyclic $A_{\infty }$ structure $\mu $ can be read from the derivative $(S , \psi  )$ as follows 
\begin{align}
\label{(S,psi) in sec2}
\big{(} \,  S \, , \, \psi  \, \big{)} 
= \sum_{g} (-)^{g} \bigg{[} \, \frac{\partial S}{\partial \psi _{g}} 
+ \frac{\partial S}{\partial \psi ^{\ast }_{g}}  \, \bigg{]} 
=  \sum_{n} (-)^{|\mu _{n}(\psi) |} \mu _{n} ( \psi , ... , \psi ) \, , 
\end{align}
where $\psi $ is the sum of all fields and antifields $\psi = \sum \psi _{g} + \sum \psi ^{\ast }_{g}$ and $|\mu _{n} (\psi )|$ denotes the total ghost number of the inputs of $\mu _{n}$. 
%
Let $\mathcal{H}$ be the state space of fields and antifields in the BV formalism: 
the tensor algebra $\mathcal{T}(\mathcal{H})$ on which $\mu $ acts consists of only one kind of \textit{field} $\psi $.
The BV master action $S[\psi ]$ has neutral ghost number and the BV derivation $(S , \,\,\, )$ has ghost number one, 
although $\psi = \sum \psi _{g} + \sum \psi ^{\ast }_{g}$ includes fields having different ghost numbers. 
We write $\mu _{n} (\psi , .. , \psi ) |_{-g}$ for the restriction onto the ghost number $-g$ sector. 
The $A_{\infty }$ relations can be read from 
\begin{align}
\label{A in sec2}
0 = 
\big{(} \, S \, , (  S , \psi ) \big{)}  
= \sum_{n , g} \sum_{k+l=n} \sum _{m=0}^{k}
(-)^{\epsilon (\psi ) } \mu_{k+1} ( \underbrace{\psi , ... , \psi }_{m}, \mu _{l} (\psi , ... , \psi ) , \underbrace{\psi , ... ,\psi }_{\epsilon (\psi )}) \Big{|}_{g} \, , 
\end{align}
where $\epsilon (\psi )$ denotes the sum of $\psi $'s ghost numbers. 
See appendix for adjusting the sign factors. 
In terms of these $A_{\infty }$ products $\{ \mu _{n} \} _{n}$, 
the BV master action $S[\psi ]$ can be always cast into the following form of homotopy Maurer-Cartan action,\footnote{
To extract the (quantum) $L_{\infty }$ structure, 
instead of (\ref{cl bv}) and (\ref{(S,psi) in sec2}), 
one should start with the $L_{\infty }$ homotopy Maurer-Cartan form $S [\psi ] = \sum_{n=1}^{\infty } \frac{1}{(n+1)!} \langle  \psi , \, \mu ^{\mathsf{sym} }_{n} ( \psi , \cdots , \psi ) \rangle $ and $(-)^{\psi }(S , \psi ) = - \sum_{n} \frac{1}{n!} \mu ^{\mathsf{sym}}_{n} (\psi , ... , \psi )$.} 
\begin{align}
\label{cl bv}
S [\psi ] = \frac{1}{2}  \la  \, \psi  \, , \, \mu _{1} \, \psi  \, \ra 
+ \sum_{n=2}^{\infty } \frac{1}{n+1} \la  \, \psi \, , \, \mu _{n} ( \psi , \cdots , \psi ) \, \ra \, , 
\end{align}
where $\langle \,\,\, , \,\,\, \rangle$ denotes a symplectic form in the BV formalism, 
which is explained in the appendix. 
Note that the space-time ghost number is not a natural grading of the $A_{\infty }$ structure $\mu$\,. 
Since $\mu $ consists of kinetic operators and interacting vertices, 
$\mu $ has neutral ghost number.

\vspace{2mm} 

Next, 
we consider a generic case.  
Suppose that a solution $S_{[0]}$ of the classical master equation $(S_{[0]} ,S_{[0]} ) = 0$ does not solve the quantum master equation, 
such as $\hbar \Delta S_{[0]} \not=0$\,. 
Then, 
we need to construct correcting terms $\hbar S_{[1]} + \hbar ^{2} S_{[2]} + \cdots$ such that $S \equiv S_{[0]} + \hbar S_{[1]} + \hbar ^{2} S_{[2]} + \cdots$ satisfies the quantum master equation $\hbar \Delta S + \frac{1}{2} (S , S) = 0$. 
In this case, 
the quantum BV master action $S$ induces the quantum $A_{\infty }$ structure $\mu _{n, [l]}$ as follows 
\begin{align}
(-)^{\psi +1} \big{(} \, S \, , \, \psi \, \big{)} 
= \sum_{g} \bigg{[} 
\frac{\partial S_{[0]}}{\partial \psi _{g}} 
+ \sum_{l >0} \hbar ^{l} \frac{\partial S_{[l]} }{\partial \psi _{g} } 
\bigg{]} 
= \sum_{n,g} \bigg{[} 
\mu_{n, [0]} (\psi , ... , \psi ) +
\sum_{l} \hbar ^{l} \mu _{n,[l]} (\psi , ... , \psi ) 
\bigg{]}_{-g} 
\, . 
\end{align}
The quantum BV master action $S$ provides a natural nilpotent operation $\Delta _{S}$ defined by 
\begin{align}
\label{BV diff}
\hbar \, \Delta _{S}  \equiv \hbar \, \Delta + ( \, S \, , \hspace{3mm} ) \, . 
\end{align}
The quantum $A_{\infty }$ relation is encoded in (\ref{BV diff}) as follows 
\begin{align}
\label{quantum A}
(\hbar \, \Delta _{S} )^{2} \psi ^{\ast }_{g} =
& \sum_{n,l} \bigg{[} \, 
\hbar \, \sum_{s \in \mathbb{Z} } \sum_{i=0}^{n} \sum_{j=0}^{n-i} (-)^{\epsilon (s,i,j) }
 \mu _{n+2 , [l-1]} \big{(} \underbrace{\psi , ... , \psi }_{i} , e_{-s} , \underbrace{\psi , ... , \psi }_{j} , e_{1+s} , \psi , ... , \psi  \big{)}
\no & \hspace{3mm} 
+ 
\sum_{\substack{n_{1}+n_{2}=n \\ l_{1}+l_{2} = l}} 
\sum_{m=0}^{n_{1} } (-)^{\epsilon (\psi ) } 
\mu_{n_{1}+1, [l_{1}]} \big{(} \underbrace{ \psi , ... , \psi }_{m } , \, 
\mu _{n_{2} , [l_{2}] } ( \psi , ... , \psi  ) , \,
\underbrace{\psi , ... , \psi }_{n_{1} - m} \big{)}    
\bigg{]}_{1-g} \, , 
\end{align} 
where the sign factor $\epsilon (s,i,j)$ arises from $e_{1+s}$ passing $\psi ^{\otimes (i+j)}$ and $e_{-s}$ passing $\psi ^{\otimes i}$. 
These complicated sign factors can be simplified when we use a degree $-1$ symplectic form $\omega $ and assign a basis carrying unphysical grading to each field or antifield: 
see appendix A. 
For $s \geq 0$, 
these $e_{-s}$ and $e_{1+s}$ are defined by $e_{-s} \equiv \frac{\partial }{\partial \psi _{s}} \psi$ and $e_{1+s} \equiv (-)^{s}  \frac{\partial }{\partial \psi ^{\ast }_{s}} \psi $ respectively. 
They enable us to get the following useful representation 
\begin{align} 
\Delta \, \mu _{n,[l]}(...) = \sum_{s \in \mathbb{Z}} (-)^{\epsilon (s)} \mu _{n , [l]} ( ... , e_{-s} , ... , e_{1+s} , ... ) \,. 
\end{align} 
Note that the condition $(\hbar \, \Delta _{S} )^{2} = 0$ is equivalent to the BV master equation (\ref{quantum BV master equation}). 
Hence, 
a solution of the BV master equation assigns a quantum $A_{\infty }$ structure to each Lagrangian field theory \cite{Barannikov:2017, Barannikov:2010np}: 
see \cite{Kajiura:2003ax, Kajiura:2001ng} for the case of cyclic $A_{\infty}$. 
In terms of these quantum $A_{\infty }$ products, 
the quantum BV master action $S[\psi ]$ can be cast into the form of homotopy Maurer-Cartan action 
\begin{align}
\label{master action}
S [\psi ] = S_{[0]} [\psi ] 
+ \sum_{n,l} \frac{\hbar ^{l}}{n+1} \la \psi , \, \mu _{n,[l]} ( \psi , \cdots , \psi ) \ra \, , 
\end{align}
where the classical master action $S_{[0]} [\psi ]$ takes the same form as (\ref{cl bv}) with $\mu _{n,[0]} \equiv \mu _{n}$\,.

\subsection{On the BV-BRST cohomology} 

We consider the role of the nilpotent operator $\Delta _{S}$ given by (\ref{BV diff}) from physicist's point of view. 
In the classical theory, 
a solution of the equations of motion determines physical states up to gauge degrees. 
These information are encoded into the classical BV differential 
\begin{align}
\label{classical BV diff}
Q_{S} \equiv ( \, S \, , \hspace{3mm} ) \,   
\end{align}
acting on the state space $\mathcal{H}$ of given fields and antifields. 
The BV differential $( \mathcal{H},  Q_{S})$ can be naturally lift to an operator acting on the space of functionals 
of fields $( \mathcal{F} (\mathcal{H} ) , Q_{S} )$ 
%
or the tensor algebra $( \mathcal{T} (\mathcal{H} ) , Q_{S} )$. 
For a given master action (\ref{master action}), 
the equation of motion for the field $\psi _{g}$ can be represented by using the BV differential and its antifield $\psi ^{\ast }_{g}$ as follows 
\begin{align}
\label{stationary pt}
0 = (-)^{g} \frac{\partial S }{\partial \psi _{g}}  =\sum_{n} (-)^{g} \mu _{n} (\psi , \dots , \psi ) \Big{|}_{-g} 
=  Q_{S} \, \psi ^{\ast }_{g}  \, . 
\end{align}
It implies that the on-shell states are $Q_{S}$-closed subspace of $\mathcal{H}$.\footnote{More precisely, 
the on-shell states are specified by a critical point of $Q_{S}$. 
When a field theory including all non-perturbative effects is solved, 
we can give a manifold $\mathcal{M}$ whose points are dynamical fields, 
in which this $Q_{S}$ is nothing but a homological vector field. 
In this paper, 
we assume that only free theories are solved around the perturbative vacuum, 
and we know just one patch around it, 
in which we consider the standard perturbation based on the Feynman graph expansion. 
The state space $\mathcal{H}$ is a tangent space of $\mathcal{M}$ that corresponds to fluctuations around the perturbative vacuum and then the critical point of $Q_{S}$ is Gaussian.} 
Likewise, 
the gauge transformation of the master action implies that its gauge degrees are $Q_{S}$-exact subspace of $\mathcal{H}$. 
The space of the physical states $\mathcal{H}_{\mathrm{phys}}$ are described by the cohomology of $(\mathcal{H} , Q_{S} )$. 
In this sense, 
solving the classical theory is equivalent to finding 
a deformation retract 
\begin{align}
\label{cl BV cohomology}
\big{(} \, \mathcal{F} ( \mathcal{H}  ),  \, Q_{S} \,  \big{)} 
\hspace{3mm} 
\overset{\pi }{\underset{\iota }{
\scalebox{2}[1]{$\rightleftarrows $}
}}
\hspace{3mm} 
\big{(} \, \mathcal{F} ( \mathcal{H}_{\mathrm{phys} } ) , \, 0 \, \big{)} 
 \, , 
\end{align}
where $\pi $ denotes a restriction to on-shell and $\iota $ denotes an embedding to off-shell. 
Since (\ref{classical BV diff}) is invertible except for $\mathcal{H}_{\mathrm{phys}}$, 
we can find Hodge decomposition with respect to $Q_{S}$ for a solved theory. 
The classical observables $F$ are given by functionals of physical states, 
$F \in \mathcal{F} (\mathcal{H}_{\mathrm{phys} } )$. 
See text books including \cite{Henneaux} for details of the (classical) BV-BRST cohomology. 

\vspace{2mm} 

In the quantum theory, 
the stationary point of (\ref{stationary pt}) does not completely determine the physical states. 
In addition to solve (\ref{stationary pt}), 
we need to replace functionals $F$ of physical states by their expectation values $\langle F \rangle$, 
which is given by the path-integral 
\begin{align}
\label{expectation value}
F \,\, \overset{P
}{\longrightarrow } \,\, 
\la F \ra \equiv \int \mathcal{D}[ \psi ] \, F \, e^{S [\psi ]} \, . 
\end{align} 
Recall that solving the BV master equation was necessary to define a regular Hessian for $S[\psi ]$, 
such that (\ref{expectation value}) can be available for the perturbation theory.
As the case of $F = 1$, 
the integrand $Fe^{S}$ must be $\Delta$-closed in order to obtain the gauge independent path-integral. 
Hence, 
for a given theory $S[\psi ]$, 
its observables $F = F [\psi ]$ satisfy 
\begin{align}
\label{Delta_S F}
\hbar \, \Delta _{S} \, F [\psi ] = 0 \, . 
\end{align}
As $Q_{S}$ is important in the classical theory, 
the nilpotent operator $\Delta _{S} = \Delta + \frac{1}{\hbar } Q_{S}$ given by (\ref{BV diff}) play a central role in the quantum theory. 
The equation of motions can be cast as $\hbar \, \Delta _{S} \psi = 0$ as (\ref{Delta_S F}). 
Note however that the $\Delta _{S}$-exact transformation, 
such as $\delta \psi = \hbar \, \Delta _{S} \, \epsilon $, 
is not the invariance of the action. 
It is the invariance of the partition function or expectation values defined by the above path-integral: 
for example, 
the $\Delta _{S}$-exact deformation $F \mapsto F + \Delta _{S} \Lambda $ does not change $\langle F \rangle $ because of $\int \mathcal{D}[\psi ] \Delta (...)=0$\,. 
In this sense, 
the physics of BV-quantizable field theory is described by a deformation retract with respect to the nilpotent operator $\Delta _{S}$, 
such as 
\begin{align}
\label{BV cohomology}
\big{(} \, \mathcal{F} ( \mathcal{H} ) , \, \hbar \, \Delta + Q_{S} \,  \big{)} 
\hspace{3mm} 
\overset{P}{\underset{I}{
\scalebox{2}[1]{$\rightleftarrows $}
}}
\hspace{3mm} 
\big{(} \, \mathcal{F} ( \mathcal{H}_{\mathrm{q\mathchar`-phys} } ) , \, 0 \, \big{)} 
\, , 
\end{align}
where $\mathcal{H}_{\mathrm{q\mathchar`-phys}}$ denotes the space of physical state in the quantum theory. 

\subsection{Key ingredient}

We assume that $\mathcal{H}_{\mathrm{q\mathchar`-phys}}$ is isomorphic to $\mathcal{H}_{\mathrm{phys}}$ and write 
$\mathrm{ev} : \mathcal{F} ( \mathcal{H}_{\mathrm{phys}} ) \rightarrow \mathcal{F} ( \mathcal{H}_{\mathrm{q\mathchar`-phys}} )$ for the corresponding isomorphism on cohomology. 
We claim that the path-integral can be identified with 
\begin{align}
P =\mathrm{ev} \circ \pi 
\end{align} 
and be divided into two operations: 
the classical projection $\pi $ and some map $\mathrm{ev}$. 
Since $\mathrm{ev}$ is a map detecting quantum corrections in functionals of classically-realizable field configurations, 
it would be worth studying how to construct such a map and elucidating its explicit form.  
%
In this section, 
we show that our claim $P = \mathrm{ev} \circ \pi $ is valid for the perturbative case by constructing such a projection map explicitly. 

\vspace{2mm} 

When we consider free theories, 
we notice that for the unit $1 \in \mathcal{F} (\mathcal{H}^{\otimes 0})$, 
the value $P (1)$ must be the same as performing the Gaussian integral (\ref{Gaussian}). 
We therefore claim $P(1) = 1$. 
We also claim that 
for any $F [\psi ] \in \mathcal{F} ( \mathcal{H}_{\mathrm{phys}} )$,  
the map $\mathrm{ev} : \mathcal{F} (\mathcal{H}_{\mathrm{phys} } ) \rightarrow \mathcal{F} ( \mathcal{H}_{\mathrm{q\mathchar`- phys}} )$ satisfies 
\begin{align} 
\label{imposing free BV}
\Delta _{S_{\mathrm{free}} } ( \mathrm{ev} \circ F [\psi ] ) = 
\Delta ( \mathrm{ev} \circ F [\psi ] ) = 0  \,  . 
\end{align}


\subsubsection*{Our set up and requirement}

In this paper, 
we consider the perturbative path-integral (\ref{Feynman rep}) around the perturbative vacuum, 
which is written in terms of the free theory (\ref{Gaussian}). 
We would like to emphasize the importance of the boundary condition on field configurations: 
we must impose \textit{the boundary condition such that the field equations have no solution except for the zero}. 
In other words, 
we must require that the on-shell projection $\pi : \mathcal{H} \rightarrow \mathcal{H}_{\mathrm{phys }}$ is nothing but \textit{the zero map} 
\begin{align} 
\label{zero map}
\pi ( \psi ) = 0 
\end{align} 
for any classical fields $\psi \in \mathcal{H}$ \textit{due to the boundary condition that we imposed}.\footnote{In general cases, 
field theories may have interactions and non-trivial classical solutions with some boundary condition. 
Then, for any $\psi = \psi _{\mathrm{phys }} + \psi _{\mathrm{gauge} } + \psi _{\mathrm{unphys}} \in \mathcal{H}$, 
an on-shell projection just picks $\pi ( \psi ) = \psi _{\mathrm{phys}}$ up by eliminating all gauge-and-unphysical degrees. 
We consider the case of $\pi (\psi ) = p_{\mathrm{phys} } = 0$. }  
Physically, 
this situation can be realized by taking the boundary condition of the Feynman propagator. 
Under the requirement (\ref{zero map}), 
we find that for each $n \geq 0$, 
the map $\mathrm{ev} : \mathcal{F} ( \mathcal{H}_{\mathrm{phys} } ) \rightarrow \mathcal{F} (\mathcal{H}_{\mathrm{q\mathchar`-phys }} )$ assigns $0 \in  \mathcal{F} ( \mathcal{H}_{\mathrm{phys} } ^{\otimes n} ) $ to appropriate constant\footnote{It is a function of space-time points, which are now frozen.} $c_{n} \in \mathcal{F} (\mathcal{H}_{\mathrm{q\mathchar`-phys }} ^{\otimes n} )$ so that the $n$-point multiplication of fields $\psi _{1} \cdots \psi _{n} \in  \mathcal{F} ( \mathcal{H} )$ is mapped to its vacuum expectation value evaluated by the Wick theorem 
\begin{align}
P (\psi _{1} \cdots \psi _{n}) = c_{n} 
\equiv \langle \psi _{1} \cdots \psi _{n} \rangle _{\mathrm{free} } \, . 
\end{align}
The zeroth value $c_{0} = 1$ is fixed by hand. 
%
%
%
The map $P = \mathrm{ev} \circ \pi $ must reproduce the expectation value $\langle \psi  \rangle _{\mathrm{free}} =0$ of a free field $\psi $, 
which fixes the first value 
\begin{align}
\label{ev}
P (\psi ) 
= c_{1} = 0 \, .  
\end{align} 
The property (\ref{ev}) follows from the requirement (\ref{zero map}), 
and $P ( \psi ) = c_{1}$ can take nonzero value iff we do not impose (\ref{zero map}): 
This is why we need (\ref{zero map}) and how (\ref{zero map}) works in the path-integral. 
%
%
%
We show that the homological perturbation for the BV differential (or the $A_{\infty }$ structure) of the free theory determines the other values of $\{ c_{n} \} _{n\geq 0}$ appropriately. 

\vspace{2mm} 

We do not need to require (\ref{imposing free BV}) and (\ref{zero map}) explicitly when we consider the path-integral to obtain $S$-matrix or to remove gauge and unphysical degrees. 
This is because (\ref{zero map}) is automatic for purely off-shell fields. 
The condition (\ref{zero map}) should be explicitly imposed when we consider the path-integral of fields that can contribute to the cohomology: for example, 
integrating massive modes or higher-momentum modes (beyond some cut-off scale) out.

\subsection{Path-integral preserves the BV master equation}

For a given BV master action $S[\psi ]$, 
we split fields $\psi $ into two components $\psi '$ and $\psi ''$, 
\begin{align}
\psi = \psi ' + \psi '' \, . 
\end{align}
By performing the path-integral of the fields $\psi ''$, 
we obtain the BV effective action $A[\psi ' ]$ from the original BV action $S[\psi '+ \psi '' ]$.\footnote{
Notice that it does not have to keep given invariance in $S[\psi ]$ or $\mathcal{D}[\psi ]$ manifestly, 
such as an exact renormalization group flow of gauge theory, 
so that generic $A[\psi ']$ needs the \textit{quantum} BV master equation.} 
The effective action can be written as follows 
\begin{align} 
\label{effective BV master action}
A [\psi ' ] \equiv \ln \int \mathcal{D} [\psi '' ] \, e^{S[\psi ' + \psi '' ] } \, . 
\end{align} 
Although it is independent of $\psi ''$ on the mass shell, 
the form of $A[\psi ']$ depends on the Lagrangian submanifold of $\psi ''$: 
the path-integral of $\psi ''$ imposes nonlinear constraints arising from $\frac{\partial S }{\partial \psi ''} = 0$ on the remaining fields $\psi '$ when two fields $\psi '$ and $\psi ''$ interact. 
It is well-known that the BV effective action $A[\psi ']$ also solves the BV master equation 
\begin{align}
\label{effective BV master equation}
\hbar \, \Delta ' A [\psi ' ]
+ \frac{1}{2} \big{(} \, A [\psi ' ] \, , \, A [\psi '] \, \big{)}' = 0 \, . 
\end{align}
The effective BV Laplacians $\Delta '$ and $\Delta ''$ of $\Delta = \Delta ' + \Delta ''$ are defined by 
\begin{align}
\Delta'  \equiv \sum (-)^{g} \frac{\partial }{\partial \psi '_{g} }  \frac{\partial }{\partial \psi _{g}^{\prime \, \ast }} \,  ,  
\hspace{5mm} 
\Delta '' \equiv \sum (-)^{g} \frac{\partial }{\partial \psi ''_{g} }  \frac{\partial }{\partial \psi _{g}^{\prime \prime \, \ast }} \,  . 
\end{align}
As $\Delta $ provides the BV bracket (\ref{antibracket}), 
the effective BV Laplacian $\Delta '$ also provides the effective BV bracket $(-)^{A}(A ,B)' \equiv \Delta ' (AB) - (\Delta ' A)B - (-)^{A} A (\Delta ' B)$. 
Because of the effective BV master equation (\ref{effective BV master equation}), 
the operator $\hbar \, \Delta '_{A} \equiv \hbar \, \Delta ' + ( A , \,\, )'$ satisfies $(\hbar \, \Delta _{A}' )^{2} = 0$. 
Hence, 
the effective action also has a quantum $A_{\infty }$ structure $\mu '$ and takes the homotopy Maurer-Cartan form 
\begin{align}
\label{effective BV action}
A [\psi ' ] = 
\sum_{n} \frac{1}{n+1} \la \psi ' , \, \mu '_{n,[0]} ( \psi ' , ... , \psi ' ) \ra '
+ \sum_{n,l} \frac{\hbar ^{l}}{n+1} \la \psi ' , \, \mu '_{n,[l]} ( \psi ' , ... , \psi ' ) \ra ' \, , 
\end{align}
where $\langle \,\, , \,\, \rangle '$ denotes the BV symplectic form defined for unintegrated fields. 
As we see in the rest of this section, 
$\mu '$ and $\langle \,\, ,\,\, \rangle '$ can be read from the construction of an effective Lagrangian. 
We will explain details of (\ref{effective BV action}) again in the next section. 

\vspace{2mm} 

This fact implies that the path-integral of fields $\psi ''$ gives a morphism $P$ preserving the BV master equation such that 
\begin{align}
P \, \Delta _{S}  =  \Delta '_{A } \, P \, .
\end{align} 
As long as the original action $S[\psi ] $ satisfies the BV master equation, 
these operators $\Delta _{S}$ and $\Delta '_{A}$ are nilpotent and the morphism $P$ preserves the cohomology. 
Because of $\mu(\psi , ..., \psi ) |_{-g} = \Delta _{S} ( \psi ^{\ast }_{g})$ and $\mu ' (\psi ' , ... , \psi ' ) |_{-g} = \Delta '_{A} ( \psi ^{\prime \, \ast }_{g})$, 
as we will explain in section 4, 
the path-integral $P$ induces a morphism $\textsf{p}$ between these $A_{\infty }$ structures $\mu $ and $\mu '$ such that   
\begin{align}
\textsf{p} \,  ( \mu _{1} + \mu _{2} + \cdots ) 
= (\mu '_{1} + \mu '_{2} + \cdots )  \, \textsf{p} \, . 
\end{align}
In the rest of this section, 
we show that the path-integral can be understood as a morphism $P$ preserving the cohomology of the BV differentials, 
\begin{align}
\big{(} \, \mathcal{F} ( \mathcal{H}' \oplus \mathcal{H}'' ) , \, \Delta _{S} \,  \big{)} 
\hspace{3mm} 
\overset{P}{\underset{I}{
\scalebox{2}[1]{$\rightleftarrows $}
}}
\hspace{3mm} 
\big{(} \, \mathcal{F} ( \mathcal{H}' \oplus \mathcal{H}''_{\mathrm{phys} } ) , \, \Delta '_{A} \, \big{)} 
 \, , 
\end{align}
where $\mathcal{H}'$ and $\mathcal{H}''$ denotes the state spaces of $\psi ' $ and $\psi ''$ respectively, 
$\mathcal{H}''_{\mathrm{phys} } $ denotes the physical space of $\psi ''$. 
On the basis of the homological perturbation, 
we can construct this morphism $P$ explicitly and show that $P$ gives 
\begin{align}
\label{morphism P}
P ( F [\psi ] ) = \la \, F [ \psi ''+ \psi '' ] \, \ra ''   
\equiv Z_{\psi '}^{-1} \int \mathcal{D} [\psi '' ] \, F[\psi ] \, e^{S[\psi ' + \psi '' ] } \, , 
\end{align} 
where $F [\psi ]$ is any functional of fields $\psi$ and $Z_{\psi '}$ is defined by 
\begin{align}
Z_{\psi '} \equiv \int \mathcal{D} [\psi '' ] \, e^{S[\psi ' + \psi '' ] } \, . 
\end{align} 
The related arguments are presented in \cite{Kajiura:2003ax, Kajiura:2001ng} for the tree graph expansion, 
in \cite{Albert, Doubek:2017naz} for how to integrate out gauge and unphysical degrees, 
and in \cite{Gwilliam:2012jg, JohnsonFreyd:2012ww} for homological and graphical argument within finite dimensional models. 
The argument presented in \cite{Doubek:2017naz} is most explicit but not applicable for the case of physical degrees that contribute to the cohomology, 
such as massive or high energy modes, 
as it is. 
%
%
In the rest of this section, 
we improve the argument of \cite{Doubek:2017naz} by presenting that the condition (\ref{ev}) and the $i \epsilon $-trick solve this problem, 
and give an explicit and direct derivation of the standard formula (\ref{Feynman rep}) that can be applied to physical degrees as well. 

\subsection{Homological perturbation performs the path-integral I} 

We first construct a morphism $\widehat{P}$ performing the path-integral without normalization. 
The normalized path-integral $P$ is constructed by using this $\widehat{P}$ in the next subsection. 
We split the action $S  = S_{\mathrm{free}}+ S_{\mathrm{int}} $ into the kinetic part $S_{\mathrm{free}} $ and interacting pert $S_{\mathrm{int} }$. 
Since the perturbative path-integral is based on the free theory, 
we construct a map $\widehat{P}$ such that 
\begin{align}
\label{pre-statement}
\widehat{P} ( e^{S_{\mathrm{int} }[\psi '' ] } ) 
= \la \, e^{S_{\mathrm{int}} [ \psi '' ] } \, \ra _{\mathrm{free} } '' 
\equiv \int \mathcal{D} [\psi ''] \, e^{S[\psi '' ] } \, , 
\end{align}
where we set $\psi ' = 0$ for simplicity. 
Clearly, 
such $\widehat{P}$ satisfies $\widehat{P} (1) = 1$ as (\ref{Gaussian}) and describes the perturbative path-integral based on the free field theory. 

\vspace{2mm} 

We assume that the kinetic terms of $\psi '$ and $\psi ''$ have no cross term $S_{\mathrm{free} } [\psi ' + \psi ''] = S_{\mathrm{free} } [\psi '] + S_{\mathrm{free}} [\psi '']$, 
and that the free theory of $\psi ''$ is solved for a given boundary condition. 
We consider 
\begin{align}
S_{\mathrm{free}} [ \psi '' ] 
= \frac{1}{2} \la \, \psi '' , \, \mu _{1}'' \, \psi '' \, \ra 
= \frac{1}{2} \la \, \psi ''_{0} , \, K_{0} \, \psi ''_{0} \, \ra 
+ \sum_{g} \la \, \psi ^{\prime \prime \, \ast }_{g-1} , \,  K_{g} \, \psi ''_{g} \, \ra 
\, , 
\end{align} 
where $\psi ''_{g}$ is the $g$-th ghost field of $\psi '' = \sum_{g} \psi ''_{g} + \sum _{g} \psi ^{\prime \prime \, \ast }_{g}$ and $K_{g}$ is its kinetic operator. 
We write $K^{-1}_{g}$ for a propagator of the kinetic operator $K_{g}$, 
whose form is specified by the boundary condition. 
In order to derive the propagators, 
we need to add an appropriate gauge-fixing fermion into the action with trivial pairs. 
Note that we are considering the path-integral over a corresponding Lagrangian submanifold: $\psi ^{\prime \prime \, \ast }$ should be understood as functionals of fields and trivial pairs determined by the gauge-fixing fermion.\footnote{Recall that propagators are singular in the gauge invariant basis. 
If we want propagators not to be singular, 
we need to switch to appropriate gauge-fixed basis via canonical transformations of fields-antifields.}

\vspace{2mm} 

Let us consider a projection $\pi  : \mathcal{H}'' \rightarrow \mathcal{H}''_{\mathrm{phys}}$ onto the physical space of the $\psi ''$ fields, 
in which the free equations of motion $\mu ''_{1} \, \psi = 0$ holds. 
We may represent $\iota  \, \pi  = e^{-\infty |\mu ''_{1} |}$ by using a natural embedding $\iota : \mathcal{H}''_{\mathrm{phys}} \rightarrow \mathcal{H}''$ satisfying $\mu ''_{1} \, \iota  = 0$\,. 
We write $K^{-1} = \sum_{g} K^{-1}_{g}$ and $\mu _{1}'' =\sum_{g} K_{g}$ for brevity. 
Once $K^{-1}$ is given, 
we get the abstract Hodge decomposition 
\begin{align}
\label{dec in qft} 
\mu ''_{1} \, K^{-1} + K^{-1} \, \mu ''_{1} = 1 - \iota  \, \pi  \, . 
\end{align}
Note that we have $\iota  \, \pi ( \psi '' ) = 0$ when the boundary condition of the Feynman propagator is imposed. 
The decomposition (\ref{dec in qft}) induces a homotopy contracting operator $\mathsf{k}^{-1}_{\psi ''}$ for $Q_{S_{\mathrm{free}} [\psi '' ] } = ( S_{\mathrm{free}} [\psi '' ]  , \hspace{2mm} )$ and it provides an alternative expression of (\ref{dec in qft}) 
\begin{align}
\label{Hodge relation}
Q_{S_{\mathrm{free}} [\psi '' ] } \, \mathsf{k}^{-1}_{\psi ''} + \mathsf{k}^{-1}_{\psi ''} \, Q_{S_{\mathrm{free}} [\psi '' ] } = 1 - \iota  \, \pi  \, . 
\end{align}
Note that $\mathsf{k}^{-1}_{\psi ''}$ decreases space-time ghost number $1$ since $Q_{S_{\mathrm{free}} [\psi '' ] } $ increases $1$. 
We often impose the conditions $\pi \, \mathsf{k}^{-1}_{\psi ''} = 0$, 
$\mathsf{k}^{-1}_{\psi ''}\, \iota = 0$ and $( \mathsf{k}^{-1}_{\psi ''} )^{2}=0$, 
which is always mathematically possible without additional assumptions \cite{Crainic}. 
As we see later, 
this situation are physically realized by a ramification of the $i \epsilon$-trick in the Feynman propagator. 
%
In terms of the kinetic operators $K_{g}$ and their propagators $K_{g}^{-1}$, 
these BV operations can be represented as  
\begin{align} 
Q_{S_{\mathrm{free}} [\psi '' ]} & = - K_{0} \, \psi ''_{0} \frac{\partial }{\partial \psi _{0}^{\prime \prime \, \ast } }  
- \sum_{g>0} 
K_{g} \, \Big{[} 
\psi _{g-1}^{\prime \prime \, \ast } \frac{\partial }{\partial \psi _{g}^{\prime \prime \, \ast }} 
+ \psi ''_{g} \frac{\partial }{\partial \psi ''_{g-1}} \Big{]} \, , 
\\ 
\mathsf{k}^{-1}_{\psi ''} & = - \frac{K^{-1}_{0} }{n_{0} } \, \psi _{0}^{\prime \prime \, \ast } \, \frac{\partial }{\partial \psi ''_{0} } 
-  \sum_{g>0} 
\frac{K^{-1}_{g} }{n_{g} } \, \Big{[} 
\psi _{g}^{\prime \prime \, \ast } \, \frac{\partial }{\partial \psi _{g-1}^{\prime \prime \, \ast } }
+ \psi ''_{g-1} \, \frac{\partial }{\partial \psi ''_{g} } \Big{]} \, , 
\end{align} 
where $n_{0}$ and $n_{g}$ are determined by the relation (\ref{Hodge relation}). 
%
%
In the above normalization, 
the operator $n_{g}$ counts the $\psi ''_{g}$-$\psi ''_{g-1}$ polynomial degree as $n_{g} ( \psi ''_{g} )^{\otimes m} (\psi ''_{g-1})^{\otimes n} = (m+n) (\psi ''_{g} )^{\otimes m} (\psi ''_{g-1})^{\otimes n}$\,. 
Likewise, 
by identifying $\psi ''_{-g-1} = \psi ^{\prime \prime \, \ast }_{g}$ for $g\geq 0$, 
we find $n_{0} ( \psi ''_{0} )^{\otimes m} (\psi ^{\prime \prime \, \ast }_{0})^{\otimes n} = (m+n) (\psi ''_{0} )^{\otimes m} (\psi ^{\prime \prime \, \ast }_{0})^{\otimes n}$ and $n_{g} ( \psi _{g-1} ^{\prime \prime \, \ast })^{\otimes m} (\psi ^{\prime \prime \, \ast }_{g})^{\otimes n} = (m+n) (\psi _{g-1} ^{\prime \prime \, \ast } )^{\otimes m} (\psi ^{\prime \prime \, \ast }_{g})^{\otimes n}$\,. 
We thus obtain $n_{g} (\psi '' )^{\otimes n} = n \, (\psi '')^{\otimes n}$. 

\vspace{2mm}

Now, 
we have the following homological data of the classical theory of free fields 
\begin{align}
\label{old dr}
\mathsf{k}^{-1}_{\psi ''} \, 
\rotatebox{-70}{$\circlearrowright $} \, 
\big{(} \, \mathcal{F} ( \mathcal{H}' \oplus \mathcal{H}'' ) , \, Q_{S_{\mathrm{free} } [\psi ']}
+ Q_{S_{\mathrm{free} } [ \psi '' ] } \, \big{)} 
\hspace{3mm} 
\overset{\pi }{\underset{\iota }{
\scalebox{2}[1]{$\rightleftarrows $}
}}
\hspace{3mm} 
\big{(} \, \mathcal{F} (\mathcal{H}' \oplus \mathcal{H}''_{\mathrm{phys}} ), \, Q_{S_{\mathrm{free} } [\psi '] } \,  \big{)} 
 \, , 
\end{align} 
which is called a deformation retract. 
Note that we must solve the equations of motion of $\psi ''$ to specify $\pi $ or $\iota $. 
In order to define a propagator $\mathsf{k}^{-1}_{\psi ''}$, 
we have to specify the off-shell and carry out its gauge-fixing if $\psi ''$ has any gauge or unphysical degree. 
Therefore, 
we must know how to solve the theory to obtain this homological data. 

\vspace{2mm} 

We expect that the perturbative path-integral (\ref{pre-statement}) can be found by transferring the relation (\ref{old dr}) into its quantum version (\ref{BV cohomology}) \textit{without} interactions since (\ref{pre-statement}) is an expectation value of the free theory. 
The homological perturbation lemma enables us to perform such a transfer of homological data. 
Clearly, 
we can take a perturbation $\hbar \, \Delta $ since $\hbar \, \Delta _{S_{\mathrm{free} } [\psi '+ \psi '' ] } = \hbar \, \Delta + Q_{S_{\mathrm{free} } [\psi '+ \psi '' ] } $ is nilpotent. 
%
%
Aa a result of the homological perturbation, 
we obtain a new deformation retract 
\begin{align}
\label{new dr}
\widehat{\mathsf{K}}^{-1} \, 
\rotatebox{-70}{$\circlearrowright $} \, 
\big{(} \, \mathcal{F} ( \mathcal{H}' \oplus \mathcal{H}'' ) , \, \hbar \, \Delta _{S_{\mathrm{free} } [\psi '+ \psi '' ] }  \, \big{)} 
\hspace{3mm} 
\overset{\widehat{P} }{\underset{\widehat{I} }{
\scalebox{2}[1]{$\rightleftarrows $}
}}
\hspace{3mm} 
\big{(} \, \mathcal{F} ( \mathcal{H}' \oplus \mathcal{H}''_{\mathrm{phys }} ) , \, \hbar \, \Delta '_{S_{\mathrm{free} } [\psi '] } \,  \big{)} 
 \, , 
\end{align}
where morphisms $\iota $, $\pi $ and a contracting homotopy $\mathsf{k}^{-1}_{\psi ''}$ of the initial data (\ref{old dr}) are replaced by perturbed ones 
\begin{align}
\widehat{I} 
= \big{(} 1 + \mathsf{k}^{-1}_{\psi ''} \hbar \,\Delta \big{)}^{-1} \, \iota  
\, ,
\hspace{5mm} 
\widehat{P} 
= \pi  \, \big{(} 1+ \hbar \, \Delta \, \mathsf{k}^{-1}_{\psi ''} \big{)}^{-1}  \, , 
\hspace{5mm} 
\widehat{\mathsf{K}}^{-1} & = \mathsf{k}^{-1}_{\psi '' } \, \big{(} 1 + \hbar \, \Delta \, \mathsf{k}^{-1}_{\psi ''} \big{)}^{-1} \, .
\end{align} 
As (\ref{Hodge relation}), 
these operators satisfy the abstract Hodge decomposition with $\hbar \, \Delta _{S_{\mathrm{free} [\psi '+ \psi '' ] } }$ on the left side of (\ref{new dr}).  
Note that $\widehat{I} = \iota $ follows from $\mathsf{k}^{-1}_{\psi ''}  \Delta ' + \Delta ' \mathsf{k}^{-1}_{\psi '' } = 0$, 
$(\Delta )^{2} = (\Delta ')^{2} = 0$ and $\mathsf{k}^{-1}_{\psi ''} ( \Delta '') \, \iota  = 0$. 
On the right side of (\ref{new dr}), 
a new differential operator is given by 
\begin{align}
\hbar \, \Delta '_{S_{\mathrm{free} }[\psi ']} 
& \equiv Q_{S_{\mathrm{free} } [\psi ' ] } 
+ \pi  \,  \hbar \, \Delta \, \widehat{I} 
= Q_{S_{\mathrm{free} } [\psi ' ] } 
+ \hbar \, \pi  \, \Delta  \, \iota   \, .
\end{align} 
Note that the differential $\pi \, \Delta '' \iota $ must vanish on $\mathcal{F} (\mathcal{H}''_{\mathrm{phys }} )$ to obtain $\Delta ' = \pi \, \Delta \, \iota $, 
which is automatic when we consider the path-integral of off-shell states or gauge-and-unphysical degrees.\footnote{For the Hodge decomposition $\psi = \psi _{p} + \psi _{g} + \psi _{u}$, 
the BV Laplacian $\Delta $ takes the form $\frac{\partial }{\partial \psi } \frac{\partial }{\partial \psi ^{\ast } } = \frac{\partial }{\partial \psi _{p} } \frac{\partial }{\partial \psi ^{\ast }_{p} } + \frac{\partial }{\partial \psi _{g} } \frac{\partial }{\partial \psi ^{\ast }_{u} } + \frac{\partial }{\partial \psi _{u} } \frac{\partial }{\partial \psi ^{\ast }_{g} }$. 
We find $\pi \, \frac{\partial }{\partial \psi } \frac{\partial }{\partial \psi ^{\ast } }  \, \iota = \pi \, \frac{\partial }{\partial \psi _{p} } \frac{\partial }{\partial \psi ^{\ast }_{p} }  \, \iota $. 
} 
In order to consider the other cases, 
at this point, 
we must assume (\ref{imposing free BV}) or (\ref{zero map}) explicitly.  

\vspace{2mm} 

We show that the above $\widehat{P}$ obtained as a result of the homological perturbation indeed realizes the perturbative path-integral (\ref{pre-statement}). 
Note that when we impose $\pi \, \mathsf{k}_{\psi ''}^{-1} = \mathsf{k}_{\psi ''}^{-1} \, \iota = ( \mathsf{k}_{\psi ''}^{-1} )^{2} = 0$ in (\ref{dec in qft}), 
the operator $\mathsf{k}^{-1}_{\psi ''}$ commutes with $\Delta '$ and vanishes on $\mathcal{H}''_{\mathrm{phys} }$. 
We first consider off-shell fields $\pi \, \psi '' = 0$, 
where $\psi '' = \sum _{g \in \mathbb{Z}} \psi ''_{g}$ with $\psi ''_{-g} \equiv \psi ^{\prime \prime \, \ast }_{g-1}$ having ghost number $-g$. 
We find 
\begin{align}
\widehat{P} (\psi ''{}^{\otimes 2n}) 
& = \pi  \, ( \hbar \, \Delta \, \mathsf{k}^{-1}_{\psi ''} )^{n} (\psi '' {}^{\otimes 2n})
 =  \pi \, \frac{1}{n!} \Big{(} \frac{\hbar }{2} \, \sum_{g} K^{-1}_{g} \frac{\partial }{\partial \psi ''_{g } } \, \frac{\partial }{\partial \psi ''_{-g}  } \Big{)}^{n} (\psi '' {}^{\otimes 2n})
\end{align}  
because of $\pi  \, ( \hbar \, \Delta \, \mathsf{k}^{-1}_{\psi ''} )^{m} (\psi '' {}^{\otimes 2n}) = 0$ for $m \not=n$ and obtain 
\begin{align}
\pi \, ( \hbar \, \Delta \, \mathsf{k}^{-1}_{\psi ''} )^{n} (\psi '' {}^{\otimes 2n})
& = \pi \,  ( \hbar \, \Delta \, \mathsf{k}^{-1}_{\psi ''} )^{n-1} 
\underbrace{ \Big{(} \frac{\hbar }{2n} \, \sum_{g} K^{-1}_{g} \frac{\partial }{\partial \psi ''_{g } } \, \frac{\partial }{\partial \psi ''_{-g}  } \Big{)} (\psi '' {}^{\otimes 2n}) }_{\psi '' {}^{\otimes 2(n-1)} }
\no 
& = \pi \,  ( \hbar \, \Delta \, \mathsf{k}^{-1}_{\psi ''} )^{n-2} 
\frac{1}{n (n-1)} \underbrace{ \Big{(} \frac{\hbar }{2} \, \sum_{g} K^{-1}_{g} \frac{\partial }{\partial \psi ''_{g } } \, \frac{\partial }{\partial \psi ''_{-g}  } \Big{)}^{2} (\psi ''{}^{\otimes 2n}) }_{\psi '' {}^{\otimes 2(n-2)} } \, . 
\end{align} 
It leads the Feynman graph expansion (\ref{Feynman rep}) iff all $\psi ''$ are purely off-shell fields. 
Hence, 
the requirement (\ref{zero map}) makes the projection $\widehat{P}$ a map performing the path-integral. 
Then, we find  
\begin{align}
\label{integrated psi''}
\widehat{P} \big{(} e^{S_{\mathrm{int} } [\psi '+ \psi '' ] } \big{)}  
& = \pi  \, \sum_{n=0}^{\infty }( \hbar \, \Delta \, \mathsf{k}^{-1}_{\psi '' } )^{n} \big{(} e^{S_{\mathrm{int} } [\psi ' + \psi '' ] } \big{)}
\no 
& = \pi \exp{ \bigg{[} \frac{\hbar }{2} \, \sum_{g} K^{-1}_{g} \frac{\partial }{\partial \psi ''_{g } } \, \frac{\partial }{\partial \psi ''_{-g} }  \bigg{]} } \big{(} e^{S_{\mathrm{int} } [\psi ' + \psi '' ] } \big{)} \, ,  
\end{align} 
which gives a functional of  $\psi ' $ due to $\pi ( \psi ' + \psi '') = \psi ''$ under (\ref{zero map}).  
We notice that by introducing a source term $e^{J \psi ''}$ as (\ref{Feynman rep}), 
the formula (\ref{integrated psi''}) can be represented as follows  
\begin{align}
\label{realizing widehat P}
\widehat{P} \big{(} e^{S_{\mathrm{int} } [\psi ' + \psi '' ]  + J \psi '' } \big{)} 
= 
\exp{ \bigg{[} \frac{\hbar }{2} \, \sum_{g} K^{-1}_{g} \frac{\partial }{\partial \psi ''_{\mathrm{ev}\, g } } \, \frac{\partial }{\partial \psi ''_{\mathrm{ev} \, -g} }  \bigg{]} } \big{(} e^{S_{\mathrm{int} } [\psi ' + \psi ''_{\mathrm{ev}} ] + J \psi ''_{\mathrm{ev}} } \big{)}  \bigg{|}_{\psi ''_{\mathrm{ev}} 
=0 } 
\, . 
\end{align}

\vspace{3mm} 

The boundary condition of the Feynman propagator realizes the situation of $\pi ( \psi '' ) = 0$ for any fields, so that the requirement (\ref{zero map}) would not be restrictive. 
We notice that the conditions $\pi \, \mathsf{k}_{\psi ''}^{-1} = \mathsf{k}_{\psi ''}^{-1} \, \iota = 0$ can be regarded as a ramification of the $i \epsilon $-trick in the Feynman propagator: 
each operators of the Hodge decomposition (\ref{Hodge relation}) are $i \epsilon$-modified when we apply the $i \epsilon $-trick to the propagator in order for choosing a contour avoiding the on-shell poles.\footnote{
Besides modifying the mass shell itself to avoid poles, 
there is another option: 
we can replace the propagator $\mathsf{k}_{\psi ''}^{-1}$ by $\mathsf{k} _{\psi ''}^{-1} ( 1 - \iota \pi )$, 
which is a propagator without propagation of on-shell states. 
What is special about the $i \epsilon $-trick in the Feynman propagator is that it automatically sets the conditions $\iota \pi (\psi '' ) = 0$ and $\mathsf{k}_{\psi ''}^{-1} \iota = \pi \mathsf{k}_{\psi ''}^{-1} = 0$. }   

%

%

\vspace{2mm} 

Once the normalization can be fixed as $\widehat{P} (1) = 1$, 
we succeed to construct the projection $P = \mathrm{ev} \circ p \equiv \frac{1}{\widehat{P} (1)} \widehat{P} $. 
The map $\mathrm{ev} $ is given by $\mathrm{ev} \circ \pi ( F ) = \pi \circ \frac{1}{1 + \hbar \Delta \mathsf{k}_{\psi ''}^{-1} } ( F)= \pi \circ e^{\frac{\hbar }{2} K^{-1} (\partial _{\psi '' }  )^{2}} F $ for any $F \in \mathcal{F} ( \mathcal{H})$, 
which is a pullback of $\pi $ by $e^{\frac{\hbar }{2} K^{-1} (\partial _{\psi '' }  )^{2}} : \mathcal{F} ( \mathcal{H}) \rightarrow \mathcal{F} ( \mathcal{H}) [\![ \hbar ]\!]$.

\vspace{2mm} 

The above (\ref{integrated psi''}) and (\ref{realizing widehat P}) are nothing but the Feynman graph expansion (\ref{Feynman rep}) in the perturbative quantum field theory. 
Note that the term $\frac{\hbar }{2} K^{-1}_{0} (\partial _{\psi _{\mathrm{cl} } })^{2} $ consists of classical fields and their propagators 
and thus we obtain non-zero value after removing all antifields (and also ghosts) from (\ref{realizing widehat P}). 
Hence, 
quantum field theory \textit{without} gauge degrees can be treated within this framework and does not provide trivial results after the homological perturbation, 
although its BV master action is the same as the classical action and the BV master equation looks trivial.


\subsection{Homological perturbation performs the path-integral II} 

We construct a morphism $P$ performing the perturbative path-integral such that 
\begin{align}
\label{statement} 
P ( ... ) \equiv  Z_{\psi '}^{-1} \, \widehat{P} \big{(} (...)e^{S [\psi ' + \psi '' ] - S_{\mathrm{free} } [\psi ''] }  \big{)}  
\overset{(\ref{pre-statement})}{=} Z_{\psi '}^{-1} \int \mathcal{D} [\psi ''] \, (...) \, e^{S[\psi ' + \psi '' ] } \, .
\end{align} 
We expect that it can be found by transferring the homological data of (\ref{old dr}) into its fully quantum version including interactions. 
Again, 
the homological perturbation enables us to perform such a transfer. 
We can take $\hbar \, \Delta _{S_{\mathrm{int}}[ \psi ] } = \hbar \, \Delta + Q_{S_{\mathrm{int} [\psi ] }}$ as a perturbation since $\hbar \, \Delta _{S [\psi ] } = \hbar \, \Delta + Q_{S [\psi ] } $ is nilpotent. 
After the perturbation, 
we obtain a new homological data as follows 
\begin{align}
\label{quantum dr}
\mathsf{K}^{-1} \, 
\rotatebox{-70}{$\circlearrowright $} \, 
\big{(} \, \mathcal{F} ( \mathcal{H}' \oplus \mathcal{H}'' ) , \, \hbar \, \Delta _{S[\psi ' + \psi '' ] } \, 
\big{)} 
\hspace{3mm} 
\overset{P}{\underset{I}{
\scalebox{2}[1]{$\rightleftarrows $}
}}
\hspace{3mm} 
\big{(} \, \mathcal{F} ( \mathcal{H}' \oplus \mathcal{H}''_{\mathrm{phys }} ) , \, \hbar \, \Delta '_{A[\psi ' ]}  \, 
\big{)} 
 \, . 
\end{align}
The perturbation lemma tells us how to construct morphisms $I$ and $P$ explicitly, 
\begin{align}
I = \big{(} 1 + \mathsf{k}^{-1}_{\psi ''} \hbar \,\Delta _{S_{\mathrm{int}}[ \psi ] } \big{)}^{-1} \, \iota  \, ,
\hspace{5mm}  
P = \pi  \, \big{(} 1+ \hbar \, \Delta _{S_{\mathrm{int}}[ \psi ] } \, \mathsf{k}^{-1}_{\psi ''} \big{)}^{-1}  \, . 
\end{align} 
Likewise, 
a contracting homotopy for $\hbar \, \Delta _{S[\psi ] }$ and the induced differential $\hbar \, \Delta '_{A[\psi ' ] }$ are given by  
\begin{align}
\mathsf{K}^{-1}  = \mathsf{k}^{-1}_{\psi ''} \, \big{(} 1 + \hbar \, \Delta _{S_{\mathrm{int}}[ \psi ] } \, \mathsf{k}^{-1}_{\psi ''} \big{)}^{-1} 
\, \hspace{5mm}  
\hbar \, \Delta '_{A[\psi ']}  =
P \, \hbar \, \Delta _{S[\psi ] } \, \iota  
= \pi  \, \hbar \, \Delta _{S [\psi ] } \, I  \, . 
\end{align} 
The statement (\ref{statement}) can be proved by tedious but direct calculations as section 2.4. 
We however takes another pedagogical approach given by \cite{Doubek:2017naz}. 
See also \cite{Albert} and \cite{Plumann}. 

\vspace{2mm} 

As is known, 
the homological perturbation transfers a given deformation retract to a new deformation retract. 
It therefore enables us to obtain the new Hodge decomposition 
\begin{align}
\label{Hodge dec of F}
\big{(} \, 1 -  I \, P \, \big{)} \, F  = \Big{[} ( \hbar \, \Delta _{S[\psi ] } ) \, \mathsf{K}^{-1}  +  \mathsf{K}^{-1} \, ( \hbar \, \Delta _{S[\psi ] } ) \Big{]} \, F   
\end{align}
for any $F [ \psi ] \in \mathcal{H}' \oplus \mathcal{H}''$\,. 
Note that since $[ \, \mathsf{k}^{-1}_{\psi ''} , \Delta \, ] = \sum_{g} n_{g}^{-1} K^{-1}_{g} \partial _{\psi ''_{-g}} \partial _{\psi ''_{g}} $ acts on the off-shell states satisfying $K \, \psi '' \not= 0$, 
it does not act on $\iota  (...)$\,. 
Because of $\widehat{P} \, I \, (...) = \widehat{P} \, \iota (...)$ with $\pi  \, \mathsf{k}^{-1}_{\psi ''} =0$\,, 
we find the following property of $I$ and $\widehat{P}$, 
\begin{align}
\widehat{P} \, \Big{(} \, I \, (...) \, e^{S_{\mathrm{int} } [\psi ' + \psi '' ] } \Big{)} 
& = 
\pi  \, \Big{[} \, 
\big{(} ( 1 + \hbar \, \Delta '' \, \mathsf{k}^{-1}_{\psi ''} )^{-1} \, e^{S_{\mathrm{int}} [\psi ' + \psi '' ] } \big{)} \, 
\iota \, (...)  \Big{]}
\no & 
= 
\widehat{P} \, \Big{(} e^{S_{\mathrm{int}} [\psi ' + \psi '' ] } \Big{)} \, 
\pi \, \iota  \, (...) \, . 
\end{align} 
In other words, 
since $\pi  \, \iota  =1$ on $\mathcal{H}''_{\mathrm{phys} }$, 
we proved that $I(...)$ passes the $\psi ''$ integral as follows  
\begin{align}
 Z_{\psi '}^{-1} \int \mathcal{D} [\psi ''] \, I \, ( P F ) \, e^{S [\psi ' + \psi '' ]  } 
=  Z_{\psi '}^{-1} \int \mathcal{D} [\psi ''] \, \iota  \, ( P F ) \, e^{S [\psi ' + \psi '' ]  } 
 = 
 P F \, . 
\end{align} 

\vspace{2mm} 

The abstract Hodge decomposition (\ref{Hodge dec of F}) elucidates that our morphism $P$, 
a result of homological perturbation, 
indeed performs the perturbative path-integral as follows 
\begin{align}
P (F) 
= Z_{\psi '}^{-1} \int \mathrm{D} [\psi ''] \, F \, e^{S [\psi ' + \psi '' ]  } 
- Z_{\psi '}^{-1}  (\mathrm{extra}) \, . 
\end{align}
We show that the extra term vanishes 
\begin{align} 
(\mathrm{extra}) \equiv  
\hbar \, \widehat{P} \, \Big{(} 
\big{(} \, \Delta _{S[\psi ] } \, \mathsf{K}^{-1} F  +  \mathsf{K}^{-1} \, \Delta _{S[\psi ] } \, F \, \big{)}  \, e^{S [\psi ' + \psi '' ] - S_{\mathrm{free} } [\psi '' ] }  \Big{)} = 0 \,  . 
\end{align}
The second term is trivially zero when we use $\mathsf{k}^{-1}_{\psi ''}$ satisfying the subsidiary condition $(\mathsf{k}^{-1}_{\psi ''})^{2} =0$ and $\pi \, \mathsf{k}^{-1}_{\psi ''} = 0$, which can be always imposed by dressing the old $\mathsf{k}^{-1}_{\psi ''}$ without any additional condition \cite{Crainic}. 
Note that $(\mathsf{k}^{-1}_{\psi ''})^{2} =0$ gives $\widehat{P} \, \mathsf{K}^{-1} = \pi \, \mathsf{K}^{-1}$. 
Thus, 
$\pi \, \mathsf{k}^{-1}_{\psi ''} = 0$ provides  
\begin{align}
\widehat{P} \, \Big{(} \, \mathsf{K}^{-1}(...) \, e^{S_{\mathrm{int} } [\psi ' + \psi '' ] } \, \Big{)} 
= \pi  \, \Big{[} \, 
\mathsf{k}^{-1}_{\psi ''} \, (1 + \hbar \, \Delta _{S_{\mathrm{int} } [\psi ] } \mathsf{k}^{-1}_{\psi '' } )^{-1} \, (...)
e^{S_{\mathrm{int}} [\psi ' + \psi '' ] } \, 
 \Big{]} 
 = 0 \, . 
\end{align}
This fact implies that after the path-integral, 
as expected, 
the $\mathsf{K}^{-1}$-exact quantities vanish 
\begin{align}
\int \mathcal{D} [\psi '' ] \, \mathsf{K}^{-1} (...) \, e^{S[\psi ' + \psi '' ]} = 0 \, .
\end{align}
Actually, 
the first term vanishes for similar reasons. 
The morphism $\widehat{P}$ satisfies $\widehat{P} \, \Delta _{S_{\mathrm{free} }[\psi '']} = \Delta ' \, \widehat{P} $ because of its defining properties $\widehat{P} \, \Delta _{S_{\mathrm{free} } [\psi ] } = \Delta '_{S_{\mathrm{free} } [ \psi ' ] } \, \widehat{P}$ and $\widehat{P} \, e^{S_{\mathrm{free} } [\psi ' ] } =  e^{S_{\mathrm{free} } [\psi ' ]  } \, \widehat{P}$\,. 
We find 
\begin{align}
\widehat{P} \, \Big{(} \big{[} \Delta _{S [\psi ] } (...) \big{]} \, e^{S [\psi ] - S_{\mathrm{free} }[\psi '' ] } \, \Big{)} 
= \widehat{P} \, \Big{(} \Delta _{S_{\mathrm{free} } [\psi '' ] } \big{[} (...) e^{S [\psi ] - S_{\mathrm{free} }[\psi '' ] } \big{]} \, \Big{)} 
=   \Delta ' \widehat{P} \, \Big{(} \big{[} (...) e^{S [\psi ] - S_{\mathrm{free} } [\psi '' ] } \big{]} \, \Big{)} 
\, .
\end{align} 
It implies that the $\psi ''$ integral maps the $\Delta _{S[\psi ]}$-exacts into $\Delta '$-exact quantities, 
\begin{align}
\label{PD=D'P}
\int \mathcal{D} [\psi '' ] \, \Delta _{S [\psi ] } (...) \, e^{S[\psi ' + \psi '' ]} 
= \Delta '  \, \bigg{[} \int \mathcal{D} [\psi '' ] \, (...) \, e^{S[\psi ' + \psi '' ]} \bigg{]} \, .
\end{align}
After applying this property, 
the integrand of the first term becomes $\mathsf{K}^{-1}$-exact and gives zero. 
Hence, 
the statement (\ref{statement}) is proved. 
Note also that 
because of $Z_{\psi '} \, P(1) = \widehat{P} (1) = e^{A[\psi '] }$, 
the relation (\ref{PD=D'P}) is nothing but the condition of morphism 
\begin{align}
P \, \Delta _{S[\psi ]} = Z_{\psi '}^{-1} \, \Delta ' \, Z_{\psi '} \, P = \Delta '_{A[\psi ']} \, P \, . 
\end{align}

\subsection{$A_{\infty }$ structure of the effective theory}

In the rest of this section, 
we explain several properties that effective theories have as a result of the homological perturbation. 
As far as we know, 
explicit calculations and consistency checks given in subsections 3.7 and 3.8 would be new, 
which are natural results for physicists. 
We consider the (quantum) $A_{\infty }$ structure of the effective theory, 
\begin{align}
\mu ' ( \psi ', ... , \psi ' ) = \mu '_{1} (\psi ' ) + \mu '_{\mathrm{int}} ( \psi ' , ... , \psi ' ) \, , 
\end{align}
which is given by $\mu ' (\psi ' , ... , \psi ') \equiv \hbar \, \Delta '_{A[\psi ' ]} \psi ' $ for $\psi ' = \sum_{g} [ \psi '_{g} + \psi ' {}^{\ast }_{g} ]$. 
The $A_{\infty }$ structure of the effective theory can be obtained by calculating the perturbed BV differential $\hbar \, \Delta '_{A[\psi ' ] } $. 
Since $\mathsf{k}^{-1}_{\psi ''}$ commutes with $\Delta '$, 
we find that it takes  
\begin{align}
\hbar \, \Delta '_{A[\psi ' ] } 
= Q_{S_{\mathrm{free} } [\psi ' ] } 
+ \pi \, \sum_{n} ( \hbar \, \Delta ''_{S_{\mathrm{int}} [\psi ' +\psi ''] } \, \mathsf{k}^{-1}_{\psi ''} )^{n} \, 
\hbar \, \Delta ' _{ S_{\mathrm{int } } [ \psi ' + \psi '' ] }
\, \iota \, .   
\end{align}
Note that the commutator of the full perturbation $\hbar \, \Delta _{S_{\mathrm{int}} [\psi ] }$ and the propagator $\mathsf{k}^{-1}_{\psi ''}$, 
\begin{align}
\big{[} \, \hbar \,\Delta _{S_{\mathrm{int}}[ \psi ] } , \, \mathsf{k}^{-1}_{\psi ''} \, \big{]}  
= \hbar \, \sum_{g} \frac{K^{-1}_{g}}{n_{g} } \frac{\partial }{\partial \psi ''_{-g} } \frac{\partial }{\partial \psi ''_{g} } 
+ \sum_{g} \frac{K^{-1}_{g}}{n_{g} } \, \mu _{\mathrm{int} } ( \psi , ... \psi ) \big{|}_{g}  
\frac{\partial }{\partial \psi ''_{g} } \, , 
\end{align} 
naturally includes the loop operator $L_{\check{\psi }'' \check{\psi }'' }$ and the tree grafting operator $T_{\check{\psi }''}$ defined by  
\begin{align} 
\hbar \, L_{\check{\psi }'' \check{\psi }'' } \equiv \hbar \, \sum_{g} \frac{K^{-1}_{g}}{n_{g} } \frac{\partial }{\partial \psi ''_{-g} } \frac{\partial }{\partial \psi ''_{g} } \, , 
\hspace{5mm} 
T_{\check{\psi }'' } \equiv \sum_{g} \frac{K^{-1}_{g}}{n_{g} } \, \mu _{\mathrm{int} } ( \psi '+ \psi '' , ... , \psi ' + \psi '') \big{|}_{g} \frac{\partial }{\partial \psi ''_{g} } \, . 
\end{align}
These provide basic manipulations of the $\psi ''$ Feynman graphs as follows 
\begin{align}
L_{\check{\psi }'' \check{\psi }'' } \, \mu _{n+2} ( 
\psi , ... , \psi 
)  \big{|}_{\psi '' = 0} 
& = \frac{1}{2} \sum_{s \in \mathbb{Z} } \sum_{i,j} \mu _{n+2} \big{(} \underbrace{ 
\psi ' , ... , \psi '}_{i} , \, 
K^{-1}_{s} e_{s} , \, \underbrace{\psi ' , ... , \psi ' }_{n-i-j} , \, 
e_{-s} , \, \underbrace{\psi ' , ... ,\psi ' 
}_{j} \big{)} \, ,  
\\ 
T_{\check{\psi }'' } \, \mu _{n+1} ( 
\psi, ... , \psi 
)  \big{|}_{\psi '' = 0} 
& = \sum_{k} \mu _{n+1} \big{(} 
\underbrace{\psi ' , 
... , \psi ' }_{k} , \, 
\mu_{\mathrm{int} } ( 
\psi ' , ... , \psi ' ) , \, 
\underbrace{\psi ' , ... , \psi ' 
}_{n-k} 
\big{)} \, . 
\end{align}
Note that 
since $n_{g} \, \psi ''{}^{\otimes m} = m \, \psi ''{}^{\otimes m}$, 
each graph has appropriate coefficient, 
such as 
\begin{align}
\label{mention}
T_{\check{\psi } '' } \, T_{\check{\psi }'' } \, \, 
\mu _{\mathrm{int}} \big{|}_{\psi ''=0} 
= & \sum \mu _{\mathrm{int} } 
\big{(} ... , \, K^{-1} \mu_{\mathrm{int} } ( ... , \, K^{-1} \mu_{\mathrm{int} } \, , \, ... ) , \, ... \big{)} 
\no &\hspace{5mm} 
+ 2 \sum  \, \frac{1}{2} \, 
\mu_{\mathrm{int} } ( ... , K^{-1} \mu_{\mathrm{int} } \, , ... , K^{-1} \mu_{\mathrm{int} } \, , ... ) \, . 
\end{align}
We write $\pi (\psi ' + \psi '' ) = \pi ( \psi ') =  \varphi $ and $\iota (\varphi ) = \psi ' $ for clarity. 
From $\hbar \, \Delta '_{A[\varphi ]}$ acting on $\varphi $, 
we obtain the quantum $A_{\infty }$ structure of the effective theory as follows 
\begin{align}
\label{effective A} 
\mu ' ( \varphi , ... , \varphi ) 
= \mu _{1} (\varphi ) +
\sum_{n=0}^{\infty } \Big{[} \, 
\hbar \, L_{\check{\psi }''_{\mathrm{ev} } \check{\psi }''_{\mathrm{ev}} } + T_{\check{\psi }''_{\mathrm{ev} } } \, 
\Big{]}^{n} 
\mu _{\mathrm{int} } ( \varphi + \psi ''_{\mathrm{ev} } , ... , \varphi + \psi ''_{\mathrm{ev} } ) \bigg{|}_{\psi ''_{\mathrm{ev} }= 0} \, . 
\end{align} 
Note that the effective vertices $\mu '_{\mathrm{int}} = \mu '_{2} + \mu '_{3} + \cdots $ have the $\hbar $ dependent parts, 
\begin{align} 
\label{expansion of mu'}
\mu '_{n} (\varphi , ... \varphi ) = 
\mu '_{n, [0]} (\varphi , ... \varphi ) 
+ \hbar \, \underbrace{ \mu ' _{n,[1]} (\varphi , ... \varphi ) 
+ \hbar ^{2} \, \mu '_{n,[2] } (\varphi , ... \varphi ) 
+ \cdots }_{\mathrm{made \,\, from}\,\, L_{\check{\psi }''_{\mathrm{ev} } \check{\psi }''_{\mathrm{ev} } } } \,. 
\end{align}  
We consider $\varphi (t)$ such that $\varphi (0) = 0$ and $\varphi (1) = \varphi$ for $t \in \mathbb{R}$. 
As a functional of $\varphi $, 
by using $\varphi (t)$, 
the effective action (\ref{effective BV action}) can be cast as 
\begin{align}
\label{effective BV action alt}
A [\varphi ] = \int _{0}^{1} \, dt \, \la \, \partial _{t} \, \varphi (t) , \, \mu ' \big{(} \, \varphi (t) , \dots , \varphi (t) \, \big{)} \, \ra \, . 
\end{align}

\subsection{The classical limit and cyclic $A_{\infty }$} 

The classical part of the effective theory has a cyclic $A_{\infty }$ structure.  
The effective $A_{\infty }$ structure (\ref{effective A}) has the non-trivial classical limit $\mu '_{\mathrm{tree} } \equiv  \lim _{\hbar \rightarrow 0} \mu ' $, 
which is obtained by setting $\hbar \rightarrow 0$ in (\ref{expansion of mu'}) as follows,  
\begin{align}
\label{tree A}
\mu '_{\mathrm{tree}} ( \varphi  , ... , \varphi ) 
= 
\sum_{n=0}^{\infty } \bigg{[} 
\sum_{g} \frac{K^{-1}_{g}}{n_{g} } \, \mu _{\mathrm{int} } ( \varphi +\psi ''_{\mathrm{ev} } , ... ,  \varphi +\psi ''_{\mathrm{ev} } ) \Big{|}_{g}  
\frac{\partial }{\partial \, \psi ''_{\mathrm{ev} \, g} } \bigg{]}^{n} \mu ( \varphi + \psi ''_{\mathrm{ev} } ) \bigg{|}_{\psi ''_{\mathrm{ev} } = 0} \, . 
\end{align} 
We write $A_{\mathrm{tree}} [\varphi ]$ for the classical part of the effective action $A[\varphi ]$, 
which consists of tree graphs only. 
By construction of (\ref{effective A}) and $\mu '_{\mathrm{tree} } (\varphi ) = Q_{A_\mathrm{tree} [\varphi ]} \, \varphi$, 
we find  
\begin{align} 
Q_{A_{\mathrm{tree} } [\varphi ] } = \pi  \, Q_{S [\psi ] } \, I_{\mathrm{tree}} 
= P_{\mathrm{tree}} \, Q_{S [ \psi ] } \, \iota  \, , 
\end{align}
where $I_{\mathrm{tree}}$ and $P_{\mathrm{tree}}$ are the classical limits of $I$ and $P$ respectively, 
\begin{align}
I_{\mathrm{tree} } = \big{(} 1 + 
\mathsf{k}^{-1}_{\psi ''} \, Q_{S_{\mathrm{int}}[ \psi ] } \,  \big{)}^{-1} \,  \iota \, ,
\hspace{5mm} 
P_{\mathrm{tree} } = \pi  \, \big{(} 1 + 
Q_{S_{\mathrm{int}}[ \psi ] } \, \mathsf{k}^{-1}_{\psi ''} \, \big{)}^{-1} \, . 
\end{align} 
We can obtain these classical limits as a result of the perturbation $Q_{S_{\mathrm{int}} [ \psi ]}$ to (\ref{old dr}), 
\begin{align} 
\label{classical dr} 
\mathsf{K}_{\mathrm{tree} }^{-1} \, 
\rotatebox{-70}{$\circlearrowright $} \, 
\big{(} \, \mathcal{F} ( \mathcal{H}' \oplus \mathcal{H}'' ) , \, Q_{S[ \psi ] } \, 
\big{)} 
\hspace{3mm} 
\overset{P_{\mathrm{tree} }}{\underset{I_{\mathrm{tree} }}{
\scalebox{2}[1]{$\rightleftarrows $}
}}
\hspace{3mm} 
\big{(} \, \mathcal{F} ( \mathcal{H}' \oplus \mathcal{H}''_{\mathrm{phys }} ) , \, Q_{A[\varphi ]}  \, 
\big{)} 
 \, . 
\end{align}
This fact implies that a morphism $P_{\mathrm{tree} }$ performs the classical part of the perturbative path-integral, 
or the Feynman graph expansion grafting only trees, 
as follows 
\begin{align}
\label{tree partition function}
P_{\mathrm{tree} } ( ... ) =  (Z^{\mathrm{tree} } _{\varphi } )^{-1} 
\lim _{\hbar \rightarrow 0 } 
\int \mathcal{D} [\psi '' ] \, (...) \, e^{S[\psi ' + \psi ''] } 
\, . 
\end{align} 
The same result was obtained in \cite{Matsunaga:2019fnc} in terms of cyclic $A_{\infty }$ algebras. 
In (\ref{tree partition function}), 
we assumed that the perturbative partition function $Z_{\varphi }$ splits into the tree and loop parts, 
\begin{align}
Z_{\varphi } = Z_{\varphi }^{\mathrm{tree}} \, \cdot \,  Z_{\varphi }^{\mathrm{loop}}  \, , 
\hspace{8mm} 
Z_{\varphi }^{\mathrm{tree} } \equiv e^{ A_{\mathrm{tree} } [\varphi ]  } 
\, . 
\end{align} 
Thus, 
if we interested in the tree part only, 
the classical perturbation (\ref{classical dr}) is enough. 
Actually, 
by using these $I_{\mathrm{tree} }$, $P_{\mathrm{tree} }$, 
a first few terms of (\ref{tree A}) are also calculated as follows 
\begin{align*}
& \hspace{15mm} \mu '_{\mathrm{tree} ,\, 1} (\varphi )  = \mu _{1} ( \varphi ) \, , 
\hspace{5mm} 
\mu '_{\mathrm{tree} ,\, 2} (\varphi , \varphi ) = \mu _{2} ( \varphi , \varphi ) \, , 
\\ & \hspace{-5mm} 
\mu '_{\mathrm{tree} ,\, 3} (\varphi , \varphi , \varphi )  = \mu _{3} (\varphi , \varphi , \varphi ) 
+ \mu _{2} ( K^{-1} \mu _{2} ( \varphi , \varphi ) , \varphi ) 
+ \mu _{2} ( \varphi , K^{-1} \mu _{2} (\varphi , \varphi ) ) \, , 
\\ 
\mu '_{\mathrm{tree} ,\, 4} (\varphi , ... , \varphi ) & = \mu _{4} (\varphi , ... , \varphi ) 
+ \sum \mu _{3} ( \varphi , \varphi , K^{-1} \mu _{2}  (\varphi , \varphi ) ) 
+ \mu_{2} ( K^{-1} \mu _{2}  (\varphi , \varphi ) ,  K^{-1} \mu _{2}  (\varphi , \varphi ) )
\no & \hspace{5mm} 
+ \sum  \mu _{2} ( \varphi , K^{-1} \mu _{3}  (\varphi , \varphi , \varphi ) ) 
+ \sum \mu_{2} ( \varphi , K^{-1} \sum \mu _{2} ( \varphi , K^{-1} \mu _{2}  (\varphi , \varphi ) ) ) \, , 
\end{align*}
where $\Sigma $ denotes the cyclic sum. 
Note that as we mentioned in (\ref{mention}), 
the propagator $\mathsf{k}^{-1}_{\psi ''}$ adjusts the coefficients and the restriction $\psi ''_{\mathrm{ev} } = 0$ picks up the appropriate contributions.


\section{Path-integral as a morphism of $A_{\infty }$}


All results obtained in section 3 can be written in terms of the (quantum) $A_{\infty }$ algebra and its morphism directly, 
so that we reach the statement that the path-integral can be performed in terms of Lagrangian's homotopy algebra as discussed in \cite{Kajiura:2003ax, Kajiura:2001ng, Nakatsu:2001da, Konopka:2015tta,  Doubek:2017naz, Braun:2017ikg, Jurco:2018sby, Matsunaga:2019fnc, Macrelli:2019afx, Arvanitakis:2019ald,  Jurco:2019yfd} and references therein. 
We would like to emphasize that this statement is somewhat unclear without both of the explicit derivation of the standard Feynman graph expansion formula that we gave in section 3 and an identity that connects the BV-BRST operator and Lagrangian's homotopy algebraic structure directly, 
which is our motivation in this paper. 

\vspace{2mm} 

In this section, 
we explicitly construct an effective $A_{\infty }$ structure $\mu '$ and a morphism $\textsf{p}$ between two $A_{\infty }$ structures $\mu $ and $\mu '$ such that   
\begin{align}
\textsf{p} \,  ( \mu _{1} + \mu _{2} + \cdots ) 
= (\mu '_{1} + \mu '_{2} + \cdots )  \, \textsf{p} \, , 
\end{align}
where $\mu = \mu _{1} + \mu _{2} + \cdots$ and $\mu ' = \mu '_{1} + \mu '_{2} + \cdots $ are (higher order) differentials acting on tensor algebras. 
The perturbative path-integral map $P$ induces such $\mathsf{p}$, 
which we explain. 

\vspace{2mm} 

We would like to emphasize the importance of the boundary condition on field configurations again: 
the on-shell projection $\pi $ is required to satisfy $\pi ( \psi ) = 0$ for any classical fields $\psi \in \mathcal{H}$ \textit{due to the boundary condition that we imposed}. 
Note that all deformation retracts and Hodge decompositions that we consider in this paper are obtained by imposing the boundary condition such that the field equations have no nontrivial solution. 

\subsection{Tensor trick} 

In order to extract $A_{\infty }$ products from (\ref{old dr}), 
we first consider the state space $\mathcal{H}$ instead of $\mathcal{F}(\mathcal{H} )$, 
on which $Q_{S_{\mathrm{free}} [\psi ] } \psi = \mu _{1} (\psi )$ and $\mathsf{k}^{-1}_{\psi ''} \psi = K^{-1} ( \psi )$ hold. 
For brevity, 
we write (\ref{old dr}) as 
\begin{align}
\label{dr to co-dr}
\kappa ^{-1}_{\psi ''} \, 
\rotatebox{-70}{$\circlearrowright $} \, 
\big{(} \, \mathcal{H} , \, \mu _{1} \, \big{)} 
\hspace{3mm} 
\overset{\pi  }{\underset{\iota  }{
\scalebox{2}[1]{$\rightleftarrows $}
}}
\hspace{3mm} 
\big{(} \, \mathcal{H}' , \, \mu _{1}' \,  \big{)}  
\, . 
\end{align} 
By applying the tensor trick to each component of (\ref{dr to co-dr}), 
we can obtain corresponding deformation retract of tensor algebras. 
The identity map $\mathbb{I}$ of $\mathcal{H}$ and morphisms $\pi $ and $\iota $ can be extended to the identity $1 = 1_{\mathcal{T}(\mathcal{H} )}$ of $\mathcal{T}(\mathcal{H})$ and morphisms $\pi $ and $\iota $ of tensor algebras by defining 
\begin{align}
1 |_{\mathcal{H}^{\otimes n } } = (\mathbb{I} )^{\otimes n}  \, , 
\hspace{5mm} 
\pi |_{\mathcal{H}^{\otimes n} } \equiv  (\pi  )^{\otimes n} \, , 
\hspace{5mm} 
\iota |_{\mathcal{H ' }^{\otimes n} } \equiv  (\iota  )^{\otimes n} \, . 
\end{align}
These are morphisms of tensor algebra preserving the cohomology 
\begin{align}
\label{morph prop}
\pi \, (1 \otimes 1 ) = \pi \otimes \pi \, , 
\hspace{5mm} 
\iota \, (1 \otimes 1 ) = \iota \otimes \iota \, ,  
\end{align}
where $\otimes $ is the product $\otimes : \mathcal{T}(\mathcal{H} ) \otimes \mathcal{T}(\mathcal{H} ) \rightarrow \mathcal{T}(\mathcal{H} )$ of the tensor algebra. 
The tensor algebra $\mathcal{T}(\mathcal{H} )$ can be regarded as a coalgebra. 
Note that these $\pi $ and $\iota $ are also coalgebra morphisms 
\begin{align}
\label{comor prop}
\widehat{\Delta } \, \pi =  ( \pi \otimes \pi ) \, \widehat{\Delta } \, , 
\hspace{5mm} 
\widehat{\Delta }\, \iota = ( \iota \otimes \iota ) \, \widehat{\Delta } \, , 
\end{align}
where $\widehat{\Delta }$ denotes the coproduct $\widehat{\Delta } : \mathcal{T}(\mathcal{H} ) \rightarrow \mathcal{T}(\mathcal{H} ) \otimes \mathcal{T}(\mathcal{H} )$ of coalgebra. 
The $k$-linear map $\mu _{k} : \mathcal{H}^{\otimes k} \rightarrow \mathcal{H}$ can be extended to a linear map $\mu _{k}$ acting on the tensor algebra, 
which becomes a (co-)derivation on $\mathcal{T} (\mathcal{H} )$, 
and the contracting homotopy $\mathsf{k}^{-1}_{\psi ''}$ between $\mathbb{I}$ and $\iota \, \pi $ becomes a homotopy $\kappa ^{-1}$ between two morphisms $1$ and $\iota \, \pi $ of $\mathcal{T}(\mathcal{H} )$ by defining 
\begin{align}
\label{deriv}
\mu _{k} |_{\mathcal{H}^{\otimes n} } \equiv \sum_{l} \mathbb{I}^{\otimes n-l } \otimes \mu _{k} \otimes \mathbb{I}^{\otimes l-k } \, , 
\hspace{5mm} 
\kappa ^{-1} |_{\mathcal{H}^{\otimes n}} \equiv \sum_{l} \mathbb{I}^{\otimes n-l-1} \otimes \kappa ^{-1}_{\psi ''} \otimes ( \iota \, \pi )^{\otimes l} \, . 
\end{align}
Namely, 
$\mu _{k}$ is a derivation on the tensor products and $\kappa ^{-1}$ is a $(1, \iota \pi )$-derivation 
\begin{align}
\label{deriv prop}
\mu _{k} \, (1 \otimes 1 ) = \mu _{k} \otimes 1 + 1 \otimes  \mu _{k} \, ,
\hspace{5mm} 
\kappa ^{-1} \, ( 1 \otimes 1 ) = \kappa ^{-1} \otimes \iota \pi  + 1 \otimes \kappa ^{-1} \, . 
\end{align}
Note that $\mu _{k}$ is a coderivation acting on $\mathcal{T} (\mathcal{H})$ and $\kappa ^{-1}$ is also a $(1, \iota \pi )$-coderivation 
\begin{align}
\label{coder prop}
\widehat{\Delta } \, \mu _{k} \, (1 \otimes 1 ) = ( \mu _{k} \otimes 1 + 1 \otimes \mu _{k} ) \,\widehat{\Delta } \, , 
\hspace{5mm} 
\widehat{\Delta } \, \kappa ^{-1}  = ( \kappa ^{-1} \otimes \iota \pi  + 1 \otimes \kappa ^{-1} ) \, \widehat{\Delta } \, , 
\end{align}
and thus $\kappa ^{-1}$ satisfies the characteristic property with the coproduct $\widehat{\Delta }$ as follows 
\begin{align}
(\kappa ^{-1} \otimes 1 - 1 \otimes \kappa ^{-1} ) \, \widehat{\Delta } \, \kappa ^{-1} = ( \kappa ^{-1} \otimes \kappa ^{-1} ) \, \widehat{\Delta } \, . 
\end{align}

We obtain the abstract Hodge decomposition on $\mathcal{T}(\mathcal{H} )$ 
\begin{align}
1 - \iota \, \pi = \mu _{1} \, \kappa ^{-1} + \kappa ^{-1} \, \mu _{1} \, , 
\end{align}  
and thus we can consider a deformation retract of tensor algebras, 
induced from (\ref{dr to co-dr}),  
\begin{align}
\label{free co-dr}
\kappa ^{-1} \, 
\rotatebox{-70}{$\circlearrowright $} \, 
\big{(} \, \mathcal{T} (\mathcal{H} ) , \, \mu _{1} \, \big{)} 
\hspace{3mm} 
\overset{ \pi }{\underset{\iota }{
\scalebox{2}[1]{$\rightleftarrows $}
}}
\hspace{3mm} 
\big{(} \, \mathcal{T}(\mathcal{H}' ) , \, \mu '_{1} \, \big{)} 
 \, , 
\end{align}
which can be also regarded as a deformation retract of coalgebras. 
The similar construction can be applied to (\ref{new dr}) or (\ref{classical dr}). 
Note that $\pi $ and $\iota $ are $A_{\infty }$ morphisms such that 
\begin{align} 
\label{pi-iota}
\pi \, \mu _{1} = \mu '_{1} \, \pi \, , 
\hspace{5mm} 
\mu _{1} \, \iota = \iota \, \mu '_{1} \, .  
\end{align}

\vspace{2mm} 

In the rest of this paper, 
for simplicity, 
we use the degree $0$ fields $\Psi \in \mathcal{\widehat{H}}$, 
the degree $1$ linear maps $\{ \boldsymbol{\mu }_{n} \}_{n}$ acting on $\mathcal{T} ( \hat{\mathcal{H}} )$ and the degree $-1$ symplectic $\omega $ by using a set of bases $\{ \hat{e}_{a} \}_{a} $ carrying unphysical gradings such that $\Psi = \sum _{g} e_{-g} \otimes \psi _{g} + \sum _{g} e_{1+g} \otimes \psi ^{\ast }_{g}$, 
which is explained in the appendix. 
For a given derivation $\boldsymbol{\mu }_{1}$ acting on $\mathcal{T} ( \hat{\mathcal{H}} )$, 
which can be defined as (\ref{adjusted A}) for a given $\mu _{1}$, 
there is a contracting homotopy $\boldsymbol{\kappa }^{-1}$ satisfying 
\begin{align}
\boldsymbol{\kappa }^{-1} \, (\Psi _{-g} )  \equiv - \mathsf{k}^{-1}_{\psi } \Psi _{g} 
= (-)^{g+1} \hat{e}_{1+g} \otimes K^{-1} \, (\psi _{-g}) \, . 
\end{align}
Likewise, 
morphisms $\pi $ and $\iota $ of (\ref{free co-dr}) are extended in a natural way 
\begin{align}
\boldsymbol{\pi } (\Psi _{g} ) \equiv \hat{e}_{-g} \otimes \pi (\psi _{g} ) \, , 
\hspace{5mm} 
\boldsymbol{\iota } (\Psi _{g} ) \equiv \hat{e}_{-g} \otimes \iota (\psi _{g} ) \, . 
\end{align} 
Notice that these degree-adjusted operators can be defined for given $\mu _{1}$, 
$\kappa ^{-1}$, 
$\pi$ and $\iota$ uniquely once a set of symplectic bases is fixed. 
They satisfy the abstract Hodge decomposition 
\begin{align}
\label{Hodge ad}
1 - \boldsymbol{\iota } \, \boldsymbol{\pi } = \boldsymbol{\mu }_{1} \, \boldsymbol{\kappa }^{-1} + \boldsymbol{\kappa }^{-1} \, \boldsymbol{\mu }_{1} \, ,  
\end{align} 
where $1$ denotes the unit of the tensor algebra $\mathcal{T} (\hat{\mathcal{H} } )$. 
Therefore, 
instead of (\ref{free co-dr}), 
we can consider a deformation retract of the degree-adjusted $A_{\infty }$ algebra  
\begin{align}
\label{dr free}
\boldsymbol{\kappa }^{-1} \, 
\rotatebox{-70}{$\circlearrowright $} \, 
\big{(} \, \mathcal{T} (\hat{\mathcal{H}} ) , \, \boldsymbol{\mu }_{1} \, \big{)} 
\hspace{3mm} 
\overset{ \boldsymbol{\pi } }{\underset{ \boldsymbol{\iota } }{
\scalebox{2}[1]{$\rightleftarrows $}
}}
\hspace{3mm} 
\big{(} \, \mathcal{T}( \hat{\mathcal{H}}' ) , \, \boldsymbol{\mu }'_{1} \, \big{)} 
 \, , 
\end{align}
which has the same algebraic or coalgebraic properties as (\ref{free co-dr}).

\subsection{The identities connecting BV and homotopy algebras}

We consider the degree-adjusted $A_{\infty }$ structure (\ref{adjusted A}) and the perturbation of (\ref{dr free}). 
We would like to extract information of the perturbation of $A_{\infty }$ structure from (\ref{classical dr}) or (\ref{quantum dr}). 
Let us consider a derivation $Q_{S_{\mathrm{int} } [\Psi ]}$ acting on the tensor algebra $\mathcal{T} (\hat{\mathcal{H} } )$, 
via the tensor trick. 
We find 
\begin{align}
Q_{S_{\mathrm{int} } [\Psi ] } \, \Psi ^{\otimes n} = 
\sum_{k=1}^{n} 
\Psi ^{\otimes k-1} \otimes \sum_{m \geq 2} \boldsymbol{\mu }_{m} ( \Psi ^{\otimes m} ) \otimes \Psi ^{\otimes n - k} 
= \sum_{m \geq 2} \boldsymbol{\mu }_{m} \, \Psi ^{\otimes n+m-1} \, . 
\end{align}
These derivations give the same results on the tensor algebra $\mathcal{T} (\hat{\mathcal{H}} ) $, 
which becomes explicit if we take the sum of $\Psi ^{\otimes n}$. 
We thus consider the group-like element of the tensor algebra $\mathcal{T} ( \hat{\mathcal{H} } )$, 
\begin{align}
\frac{1}{1- \Psi } \equiv \mathbb{I} + \Psi + \Psi \otimes \Psi + \Psi \otimes \Psi \otimes \Psi + \cdots  + \Psi ^{\otimes n}  + \cdots \, . 
\end{align}
By using this element $(1-\Psi )^{-1} \in \mathcal{T} ( \hat{\mathcal{H}} )$, 
we find the equality of (co-)derivations 
\begin{align}
\label{Q=mu}
\Big{[} \, 
Q_{S_{\mathrm{int} } [\Psi ] } - \boldsymbol{\mu }_{\mathrm{int} } \, \Big{]} \, \frac{1}{1- \Psi } = 0 \, . 
\end{align} 
Hence, 
as long as we consider the operators acting on 
the vector space $\mathcal{T} ( \hat{\mathcal{H} } )$, 
we obtain the same results as the previous section even if we replace $Q_{S_{\mathrm{int}} }$ by $\boldsymbol{\mu }_{\mathrm{int} }$ in the homological perturbation. 

\vspace{2mm} 

We can extend the BV Laplacian $\Delta $ to a linear map acting on the tensor algebra $\mathcal{T} (\hat{\mathcal{H} } )$, 
a second order derivation of the tensor algebra $\mathcal{T} ( \hat{\mathcal{H} } )$,  
which provides 
\begin{align}
\hbar \, \Delta \, \Psi ^{\otimes n } 
= \hbar \, \sum_{k,l}  \sum_{s \in \mathbb{Z}} 
\Psi ^{\otimes k-1} \otimes \hat{e}_{-s} \otimes \Psi ^{\otimes n -k - l} \otimes \hat{e}_{1+s} \otimes \Psi ^{l-1} 
= \hbar \, 
\mathfrak{L} \, \Psi ^{\otimes n-2} \, . 
\end{align} 
The higher order coderivation $\mathfrak{L}$ is defined as follows. 
For a given base $\hat{e}_{-s} (= \mathbb{I} \otimes \hat{e}_{-s}) \in \hat{\mathcal{H}}$\,, 
we consider a derivation $\hat{e}_{-s}$ acting on $\mathcal{T}( \hat{\mathcal{H}})$ by defining $\hat{e}_{-s} |_{\hat{\mathcal{H}}^{\otimes n} }  : \hat{\mathcal{H}}^{n} \rightarrow \hat{\mathcal{H}}^{n+1}$ as follows  
\begin{align}
\hat{e}_{-s} |_{\hat{\mathcal{H}}^{\otimes n} } =  \sum_{l} 
\mathbb{I}^{\otimes l} \otimes \hat{e}_{-s} \otimes \mathbb{I}^{\otimes n-l} 
\, ,
\end{align}
which is also a coderivation $\widehat{\Delta } \, \hat{e}_{-s} = ( \hat{e}_{-s} \otimes 1 + 1 \otimes \hat{e}_{-s} ) \, \widehat{\Delta }$\,. 
Then, 
a higher order coderivation $\mathfrak{L}$ is defined by 
\begin{align}
\mathfrak{L} |_{\mathcal{H}^{\otimes n} } 
= \sum_{l,m} \sum_{s \in \mathbb{Z} } 
\mathbb{I}^{\otimes l} \otimes \hat{e}_{-s} \otimes \mathbb{I}^{\otimes m} \otimes \hat{e}_{1+s} \otimes \mathbb{I}^{\otimes n-l-m} \, . 
\end{align}
This $\mathfrak{L}$ does not satisfy (\ref{coder prop}) as $\widehat{\Delta } \, \mathfrak{L} = ( \mathfrak{L} \otimes 1 + \sum_{s} (-)^{s} \hat{e}_{-s} \otimes \hat{e}_{1+s} + 1 \otimes \mathfrak{L} ) \, \widehat{\Delta }$\,. 
Instead, 
it satisfies the relation of order 2 coderivation as follows 
\begin{align}
( \widehat{\Delta }  \otimes 1 ) \widehat{\Delta }  \, \mathfrak{L} 
- \sigma \circ (\widehat{\Delta } \otimes 1 ) \, (  \mathfrak{L} \otimes 1  ) \, \widehat{\Delta } 
+ \sigma \circ ( \mathfrak{L} \otimes 1\otimes 1 ) \, 
( \widehat{\Delta }  \otimes 1) \widehat{\Delta }  = 0 \, , 
\end{align}
where $\sigma$ denotes the sum of cyclic permutation of operators: $\sigma \circ ( f_{1} \otimes f_{2} \otimes f_{3} ) = f_{1} \otimes f_{2} \otimes f_{3} + (-)^{|f_{1}| (|f_{2}| + |f_{3}| )} f_{2} \otimes f_{3} \otimes f_{1} + (-)^{|f_{3}| (|f_{1}| + |f_{2}| )} f_{3} \otimes f_{1} \otimes f_{3}$. 
The equivalence of $\hbar \, \Delta$ and $\hbar \, \mathfrak{L}$ becomes manifest if we take the sum of $\Psi ^{\otimes n}$,  
\begin{align}
\label{Delta=L}
\Big{[} \, 
\hbar \, \Delta -  \hbar \, \mathfrak{L} \, 
\Big{]} \, \frac{1}{1- \Psi } = 0 \, . 
\end{align} 
Hence, 
we can obtain the perturbed (quantum) $A_{\infty }$ structure directly by replacing $\hbar \, \Delta _{S_{\mathrm{int}} }$ with $\hbar \, \mathfrak{L} + \boldsymbol{\mu }_{\mathrm{int} }$ in the homological perturbation. 
The equations (\ref{Q=mu}) and (\ref{Delta=L}) tell us how to switch from the BV-BRST operator to Lagrangian's homotopy algebra \cite{Jurco:2018sby, Jurco:2019yfd}.

\subsection{Morphism of the cyclic $A_{\infty }$ structure}

Recall that the perturbed BV differential $h \Delta '_{A}$ provides the perturbed $A_{\infty }$ structure of (\ref{effective A}), 
which is a result of the homological perturbation (\ref{new dr}). 
We can derive the perturbed (quantum) $A_{\infty }$ structure $\boldsymbol{\mu }'$ directly by applying homological perturbation to this coalgebraic homological data (\ref{dr free}). 
We first consider the classical part with the equality (\ref{Q=mu}). 
We can take 
\begin{align}
\boldsymbol{\mu }_{\mathrm{int} } = \boldsymbol{\mu }_{2} + \boldsymbol{\mu }_{3} + \boldsymbol{\mu }_{4} + \cdots  
\end{align}
acting on $\mathcal{T} ( \mathcal{H} )$ as a perturbation for (\ref{dr free}) because of the $A_{\infty }$ relations of $\boldsymbol{\mu }= \boldsymbol{\mu }_{1} + \boldsymbol{\mu }_{\mathrm{int}}$\,. 
Note that this $\boldsymbol{\mu }_{\mathrm{int}}$ is a coderivation 
$\widehat{\Delta } \, \boldsymbol{\mu }_{\mathrm{int}} = ( \boldsymbol{\mu }_{\mathrm{int}} \otimes 1 + 1 \otimes \boldsymbol{\mu }_{\mathrm{int}} ) \, \widehat{\Delta }$ and the coderivation $ \boldsymbol{\mu } = \boldsymbol{\mu }_{1} + \boldsymbol{\mu }_{\mathrm{int} } $ is nilpotent $(\boldsymbol{\mu })^{2} = 0$\,. 
We obtain the deformation retract of tensor algebras or coalgebras 
\begin{align}
\label{transfer}
\mathsf{k}^{-1} \, 
\rotatebox{-70}{$\circlearrowright $} \, 
\big{(} \, \mathcal{T} (\hat{\mathcal{H}} ) , \, \boldsymbol{\mu }_{1} + \boldsymbol{\mu }_{\mathrm{int}}  \, \big{)} 
\hspace{3mm} 
\overset{\mathsf{p} }{\underset{\mathsf{i}}{
\scalebox{2}[1]{$\rightleftarrows $}
}}
\hspace{3mm} 
\big{(} \, \mathcal{T}(\hat{\mathcal{H}}' ) , \, \boldsymbol{\mu }'_{1} + \boldsymbol{\mu }'_{\mathrm{int}} \, \big{)} 
 \, , 
\end{align}
where $\mathsf{p}$ and $\mathsf{i}$ are morphisms preserving its cohomology and $\mathsf{k}^{-1}$ is a contracting homotopy. 
The perturbed data also satisfy the abstract Hodge decomposition  
\begin{align}
1 - \mathsf{i} \mathsf{p} = \boldsymbol{\mu } \, \mathsf{k}^{-1} + \mathsf{k}^{-1} \, \boldsymbol{\mu } \, . 
\end{align} 
The morphisms of tensor algebra $\mathsf{p}$ and $\mathsf{i}$ and the contracting homotopy $\mathsf{k}^{-1}$ of tensor algebra are obtained by solving recursive relations 
\begin{align}
\label{defining properties}
\mathsf{p} = \boldsymbol{\pi } - \mathsf{p} \, \boldsymbol{\mu }_{\mathrm{int} } \, \boldsymbol{\kappa }^{-1}  \, , 
\hspace{5mm} 
\mathsf{i} = \boldsymbol{\iota } - \boldsymbol{\kappa }^{-1} \, \boldsymbol{\mu }_{\mathrm{int} } \, \mathsf{i} \, , 
\hspace{5mm} 
\mathsf{k}^{-1} = \boldsymbol{\kappa }^{-1} - \boldsymbol{\kappa }^{-1} \, \boldsymbol{\mu }_{\mathrm{int}} \, \mathsf{k}^{-1} \, , 
\end{align}
where $\boldsymbol{\pi }$ and $\boldsymbol{\iota }$ are morphisms satisfying (\ref{morph prop}) and (\ref{comor prop}), 
$\boldsymbol{\mu }_{\mathrm{int}}$ is a (co-)derivation satisfying (\ref{deriv prop}) and (\ref{coder prop}), 
and $\boldsymbol{\kappa }$ is a contracting homotopy of the tensor algebra satisfying (\ref{deriv prop}) and (\ref{coder prop}). 
By using this morphism $\mathsf{p}$ or $\mathsf{i}$\,, 
the effective $A_{\infty }$ structure $\boldsymbol{\mu }' = \boldsymbol{\mu }'_{1} + \boldsymbol{\mu }'_{\mathrm{int}}$ is given by 
\begin{align} 
\label{effective co tree A}
\boldsymbol{\mu }'_{\mathrm{int}} 
\equiv \mathsf{p} \, \boldsymbol{\mu }_{\mathrm{int} } \, \boldsymbol{\iota }
= \boldsymbol{\pi } \, \boldsymbol{\mu }_{\mathrm{int} } \, \mathsf{i}  \, , 
\end{align} 
where the second equality follows form (\ref{defining properties}) quickly 
\begin{align}
\mathsf{p} \,  \boldsymbol{\mu }_{\mathrm{int}} \, ( \mathsf{i} +  \boldsymbol{\kappa }^{-1} \,  \boldsymbol{\mu }_{\mathrm{int}} \, \mathsf{i} )
= ( \mathsf{p} + \mathsf{p} \, \boldsymbol{\mu }_{\mathrm{int}} \, \boldsymbol{\kappa }^{-1} ) \,  \boldsymbol{\mu }_{\mathrm{int}} \, \mathsf{i} 
\, . 
\end{align}

The above results are well-known in the context of string field theory (cf. \cite{Kajiura:2003ax, Kajiura:2001ng, Konopka:2015tta, Matsunaga:2019fnc}). 
A proof is as follows. 
As a result of the perturbation, 
the $A_{\infty }$ relations are \textit{automatic}
\begin{align} 
\big{(} \, \boldsymbol{\mu }'_{1} + \boldsymbol{\mu }'_{\mathrm{int} } \, \big{)} ^{2} = 0 \, , 
\end{align}
which come from $( \boldsymbol{\mu }_{1} + \boldsymbol{\mu }_{\mathrm{int}} )^{2} = 0$ and the defining properties (\ref{defining properties}),  
\begin{align}
( \boldsymbol{\mu }'_{\mathrm{int}} )^{2} & = \mathsf{p} \, \boldsymbol{\mu }_{\mathrm{int} } \, ( \boldsymbol{\iota } \, \boldsymbol{\pi  } ) \, \boldsymbol{\mu }_{\mathrm{int} } \, \mathsf{i}
=  \mathsf{p} \, ( \boldsymbol{\mu }_{\mathrm{int}} )^{2} \, \mathsf{i}
+ \underbrace{( - \mathsf{p} \, \boldsymbol{\mu }_{\mathrm{int} } \, \boldsymbol{\kappa } )}_{\mathsf{p} - \boldsymbol{\pi } }  \, \boldsymbol{\mu }_{1} \, \boldsymbol{\mu }_{\mathrm{int} } \, \mathsf{i}
+ \mathsf{p} \, \boldsymbol{\mu }_{\mathrm{int} } \, \boldsymbol{\mu }_{1} \, 
\underbrace{( - \boldsymbol{\kappa } \, \boldsymbol{\mu }_{\mathrm{int} } \, \mathsf{i} ) }_{\mathsf{i} - \boldsymbol{\iota } }
\no 
& =  \mathsf{p} \, \Big{[}  ( \boldsymbol{\mu }_{\mathrm{int}} )^{2} + \boldsymbol{\mu }_{\mathrm{int}} \, \boldsymbol{\mu }_{1} + \boldsymbol{\mu }_{1} \, \boldsymbol{\mu }_{\mathrm{int}} \Big{]} \, \mathsf{i}
- \boldsymbol{\mu }'_{1} \, \underbrace{( \boldsymbol{\pi } \,  \boldsymbol{\mu }_{\mathrm{int}} \, \mathsf{i} )}_{ \boldsymbol{\mu }'_{\mathrm{int}} }
- \underbrace{( \mathsf{p} \, \boldsymbol{\mu }_{\mathrm{int}} \, \boldsymbol{\iota } )}_{ \boldsymbol{\mu }'_{\mathrm{int}} } \, \boldsymbol{\mu }'_{1} \, . 
\end{align}
This phenomenon, or the perturbed data (\ref{transfer}), 
is often called homotopy transfer in the context of mathematics. 
Likewise, 
the morphisms $\mathsf{p}$ and $\mathsf{i}$ become $A_{\infty }$ morphisms such that  
\begin{align}
\label{p-i}
\mathsf{p} \, \boldsymbol{\mu } = \boldsymbol{\mu }' \, \mathsf{p} \, , 
\hspace{5mm} 
\mathsf{i} \, \boldsymbol{\mu }' = \boldsymbol{\mu } \, \mathsf{i} \, , 
\end{align}
as long as the assumptions of the perturbation $\boldsymbol{\mu }_{\mathrm{int}} \, \boldsymbol{\kappa }^{-1} \not= -1$ and $\boldsymbol{\kappa }^{-1} \boldsymbol{\mu }_{\mathrm{int}} \not= -1$ are provided. 
Apparently, 
when $\sum_{n}(- \boldsymbol{\mu }_{\mathrm{int}} \, \boldsymbol{\kappa }^{-1})^{n}$ converges, 
$1 +  \boldsymbol{\mu }_{\mathrm{int}} \, \boldsymbol{\kappa }^{-1}$ is invertible and (\ref{p-i}) follows from  
\begin{align}
\label{p mu = mu' p}
\mathsf{p} \, \boldsymbol{\mu } -  \boldsymbol{\mu }' \, \mathsf{p}   
& = ( \boldsymbol{\pi } - \mathsf{p} \, \boldsymbol{\mu }_{\mathrm{int} } \, \boldsymbol{\kappa }^{-1} ) \, \boldsymbol{\mu }_{1}
+ \mathsf{p} \, \boldsymbol{\mu }_{\mathrm{int}} \, (
\boldsymbol{\kappa }^{-1} \, \boldsymbol{\mu }_{1} 
+ \boldsymbol{\iota } \, \boldsymbol{\pi } 
+ \boldsymbol{\mu }_{1}  \, \boldsymbol{\kappa }^{-1} )
- \boldsymbol{\mu }' \, \mathsf{p}   
\no
& = ( \boldsymbol{\mu }'_{1} + \mathsf{p} \, \boldsymbol{\mu }_{\mathrm{int}} \boldsymbol{\iota } ) \,  \boldsymbol{\pi }  
+ \mathsf{p} \, ( \boldsymbol{\mu }_{\mathrm{int}} \, \boldsymbol{\mu }_{1} ) \, \boldsymbol{\kappa }^{-1} 
- \boldsymbol{\mu }' \, \mathsf{p}  
\no 
& = ( \boldsymbol{\mu }'_{1} + \boldsymbol{\mu }'_{\mathrm{int}} ) \, ( \boldsymbol{\pi } - \mathsf{p} )   
- \mathsf{p} \,( \boldsymbol{\mu }_{1} + \boldsymbol{\mu }_{\mathrm{int} } ) \, \boldsymbol{\mu }_{\mathrm{int}} \, \boldsymbol{\kappa }^{-1} 
\no 
& = \big{(} \boldsymbol{ \mu }' \, \mathsf{p} 
- \mathsf{p} \, \boldsymbol{\mu } \, \big{)} \,  \boldsymbol{\mu }_{\mathrm{int}} \, \boldsymbol{\kappa }^{-1} \, . 
\end{align}
Likewise, 
we find $\boldsymbol{\mu } \, \mathsf{i} - \mathsf{i} \, \boldsymbol{\mu }' = - \boldsymbol{\kappa }^{-1} \, \boldsymbol{\mu }_{\mathrm{int}} \, \big{(} \boldsymbol{\mu } \, \mathsf{i} - \mathsf{i} \, \boldsymbol{\mu }' \big{)}$ and (\ref{p-i}) follows from it. 

\vspace{2mm} 

By solving (\ref{defining properties}), 
we obtain the following expression of morphisms $\mathsf{p}$ and $\mathsf{i}$\,, 
\begin{align}
\label{geo} 
\mathsf{p} =  \boldsymbol{\pi } \, \frac{1}{1+  \boldsymbol{\kappa }^{-1} \,  \boldsymbol{\mu }_{\mathrm{int} } } \, , 
\hspace{5mm} 
\mathsf{i} = \frac{1}{1+  \boldsymbol{\kappa }^{-1} \,  \boldsymbol{\mu }_{\mathrm{int} } } \,  \boldsymbol{\iota } \, . 
\hspace{5mm} 
\end{align}
Note that these expressions enable us to obtain (\ref{p-i}) directly from the second line of (\ref{p mu = mu' p}), 
$\mathsf{p} \, (  \boldsymbol{\mu }_{1} +  \boldsymbol{\mu }_{\mathrm{int}} ) = (  \boldsymbol{\mu }'_{1} +  \boldsymbol{\mu }'_{\mathrm{int}} ) \,  \boldsymbol{\pi }    
- \mathsf{p} \,(  \boldsymbol{\mu }_{1} +  \boldsymbol{\mu }_{\mathrm{int} } ) \,  \boldsymbol{\mu }_{\mathrm{int}} \,  \boldsymbol{\kappa }^{-1}$. 
By substituting (\ref{geo}) into (\ref{effective co tree A}), 
the effective cyclic $A_{\infty }$ structure can be cast as follows 
\begin{align} 
 \boldsymbol{\mu }' =  \boldsymbol{\mu }'_{1} 
+  \boldsymbol{\pi } \,  \boldsymbol{\mu }_{\mathrm{int} } \, \frac{1}{1 +  \boldsymbol{\kappa }^{-1} \,  \boldsymbol{\mu }_{\mathrm{int} }} \,  \boldsymbol{\iota } \, . 
\end{align}

In this tree-level case, 
because of the coalgebraic properties (\ref{coder prop}) of $ \boldsymbol{\mu }_{\mathrm{int} }$\,,
tensor algebra morphisms $\mathsf{p}$ and $\mathsf{i}$ are also coalgebra morphisms and the derivation $ \boldsymbol{\mu }'$ is also a coderivation  
\begin{align}
\label{co p-i}
\widehat{\Delta } \, \mathsf{p}  = (\mathsf{p} \otimes \mathsf{p} )\, \widehat{\Delta }  \, , 
\hspace{5mm} 
\widehat{\Delta } \, \mathsf{i} = ( \mathsf{i} \otimes \mathsf{i} ) \, \widehat{\Delta } \, , 
\hspace{5mm} 
\widehat{\Delta } \,  \boldsymbol{\mu }' = (  \boldsymbol{\mu }' \otimes 1 + 1 \otimes  \boldsymbol{\mu }'  ) \, \widehat{\Delta } \, . 
\end{align}
The third property quickly follows from the first or second property. 
As long as $ \boldsymbol{\mu }_{\mathrm{int} }$ is a well-defined perturbation such that $ \boldsymbol{\mu }_{\mathrm{int}} \,  \boldsymbol{\kappa }^{-1} \not= -1$ or $ \boldsymbol{\kappa }^{-1}  \boldsymbol{\mu }_{\mathrm{int}} \not=-1$,    
the first property follows from 
\begin{align}
(  \mathsf{p} \otimes \mathsf{p} ) \, \widehat{\Delta } \, (1+  \boldsymbol{\mu }_{\mathrm{int} }  \boldsymbol{\kappa }^{-1}) =
(  \boldsymbol{\pi } \otimes  \boldsymbol{\pi } ) \, \widehat{\Delta } = \widehat{\Delta } \, \boldsymbol{ \pi } = \widehat{\Delta } \,   \mathsf{p}  \, 
(1+  \boldsymbol{\mu }_{\mathrm{int} } \,  \boldsymbol{\kappa }^{-1} ) \, , 
\end{align}
where the first equality follows from direct computation 
\begin{align}
( \mathsf{p} \otimes  \mathsf{p} ) \widehat{\Delta } (
 \boldsymbol{\mu }_{\mathrm{int}} \,  \boldsymbol{\kappa }^{-1} ) 
& = 
( \mathsf{p} \,  \boldsymbol{\mu }_{\mathrm{int}} \otimes  \mathsf{p} +  \mathsf{p} \otimes  \mathsf{p} \,  \boldsymbol{\mu }_{\mathrm{int}} ) \widehat{\Delta } ( \boldsymbol{\kappa }^{-1} ) 
\no
& = 
\Big{[} \, 
\mathsf{p} \,  \boldsymbol{\mu }_{\mathrm{int}} \otimes  \boldsymbol{\pi } +  \boldsymbol{\pi } \otimes  \mathsf{p} \,  \boldsymbol{\mu }_{\mathrm{int}}  
- \underbrace{ (  \mathsf{p} \,  \boldsymbol{\mu }_{\mathrm{int}} \otimes  \mathsf{p} \,  \boldsymbol{\mu }_{\mathrm{int}} \,  \boldsymbol{\kappa }^{-1} +  \mathsf{p} \,  \boldsymbol{\mu }_{\mathrm{int}} \,  \boldsymbol{\kappa }^{-1} \otimes  \mathsf{p} \,  \boldsymbol{\mu }_{\mathrm{int}} ) 
\Big{]}  
\widehat{\Delta } (  \boldsymbol{\kappa }^{-1} )   }_{ (  \boldsymbol{\kappa }^{-1} \otimes 1 - 1 \otimes  \boldsymbol{\kappa }^{-1} ) \widehat{\Delta } (  \boldsymbol{\kappa }^{-1} ) = (  \boldsymbol{\kappa }^{-1} \otimes  \boldsymbol{\kappa }^{-1} ) \widehat{\Delta } } 
\no
& = 
( \underbrace{ \mathsf{p} \,  \boldsymbol{\mu }_{\mathrm{int}}  \,  \boldsymbol{\kappa }^{-1} }_{ \boldsymbol{\pi } -  \mathsf{p}} \otimes  \boldsymbol{\pi }  
+  \boldsymbol{\pi } \otimes \underbrace{ \mathsf{p} \,  \boldsymbol{\mu }_{\mathrm{int}} \,  \boldsymbol{\kappa }^{-1} }_{  \boldsymbol{\pi } -  \mathsf{p} } ) \, \widehat{\Delta } 
- ( \underbrace{ \mathsf{p} \,  \boldsymbol{\mu }_{\mathrm{int}} \,  \boldsymbol{\kappa }^{-1} }_{  \boldsymbol{\pi } -  \mathsf{p}} \otimes \underbrace{ \mathsf{p} \,  \boldsymbol{\mu }_{\mathrm{int}} \,  \boldsymbol{\kappa }^{-1} }_{  \boldsymbol{\pi } -  \mathsf{p}} ) \, \widehat{\Delta } 
\, . 
\end{align} 
Likewise, 
the second property of (\ref{co p-i}) holds. 
Again, 
(\ref{geo}) can solve (\ref{co p-i}) easily.

\subsection{Cyclicity of the effective $A_{\infty }$ structure} 

Before adding the quantum part, 
we consider the cyclicity. 
Note that the cyclic property of the (perturbed) $A_{\infty }$ structure is manifest from the beginning because we consider the (effective) $A_{\infty }$ structure of the (effective) BV master action. 
However, 
we would like to show that the homological perturbation itself preserves the cyclic $A_{\infty }$ structure whenever a contracting homotopy satisfies the compatibility condition (\ref{bpz of h}). 
In quantum field theory, 
it is nothing but a Hermitian property of the propagators. 
In string field theory, 
it is the BPZ property. 

\vspace{2mm} 

We may write $\langle \omega | A \otimes B \equiv \omega ( A , B )$ for the symplectic structure on $\hat{\mathcal{H}}$ for simplicity. 
The cyclic property of the $A_{\infty }$ structure $\boldsymbol{\mu } = \boldsymbol{\mu }_{1} + \boldsymbol{\mu }_{\mathrm{int} }$ can be cast as follows 
\begin{align} 
\label{bpz of mu}
\la \, \omega \, \big{|} \, \boldsymbol{\mu }_{n} \otimes \mathbb{I} = 
- \la \, \omega \, \big{|} \, \mathbb{I} \otimes \boldsymbol{\mu }_{n} \, . 
\end{align}
The perturbed $A_{\infty }$ structure $\boldsymbol{\mu }'$ provides an effective homotopy Maurer-Cartan action 
\begin{align}
\label{hMC A}
\hat{A} 
[ \Psi ' ] = \sum_{n} \frac{1}{n+1} \omega ' \big{(} \,  \Psi ' , \, \boldsymbol{\mu }'_{n} ( \Psi ' , ... , \Psi ' ) \, \big{)} \, , 
\end{align} 
where $\Psi ' \in \hat{\mathcal{H} }' \equiv E \otimes \mathcal{H}'$ denotes effective fields. 
When $\boldsymbol{\mu }$ is given by (\ref{adjusted A}), 
it equals to (\ref{effective BV action}). 
The symplectic structure $\omega ' $ on $\hat{\mathcal{H}}'$ is defined by using the inner product on $\mathcal{H}'$, 
\begin{align}
\la \, A' , \, B' \, \ra ' 
= \la \, \iota \, A' , \, \iota \, B' \, \ra 
\hspace{10mm} 
 A' , B' \in \mathcal{H}' \, , 
\end{align}
and the symplectic form $\hat{\omega }$ on $E$, 
just as (\ref{symplec}). 
We write $\langle \omega ' | A' \otimes B' = \omega ' ( A' , B' ) $ for this symplectic structure on $\hat{\mathcal{H}}$ for simplicity, 
which provides 
\begin{align} 
\la \, \omega ' \, \big{|} = \la \, \omega \, \big{|} \, \boldsymbol{\iota } \otimes \boldsymbol{\iota } \, .  
\end{align} 
When we take Hermit propagators $K^{-1}$, 
we find $\omega ( \boldsymbol{\kappa }^{-1} A , B )  = 
(-)^{A} \omega ( A  , \boldsymbol{\kappa }^{-1} B )$ quickly. 
This compatibility of $\boldsymbol{\kappa }^{-1}$ and $\omega $ can be cast as follows   
\begin{align}
\label{bpz of h}  
\la \, \omega \, \big{|} \, \boldsymbol{\kappa }^{-1} \otimes \mathbb{I} 
=  \la \, \omega \, \big{|} \, \mathbb{I} \otimes  \boldsymbol{\kappa }^{-1}  \, .  
\end{align}
As we see, 
this property (\ref{bpz of h}) ensures the cyclicity of the perturbed $A_{\infty } $ structure. 

\vspace{2mm} 

When we have (\ref{bpz of mu}) and (\ref{bpz of h}), 
the abstract Hodge decomposition (\ref{Hodge ad}) implies 
\begin{align} 
\la \, \omega \, \big{|} \, \boldsymbol{\iota } \, \boldsymbol{\pi } \otimes \boldsymbol{\iota } 
= \la \, \omega \, \big{|} \, \mathbb{I} \otimes \boldsymbol{\iota } \, , 
\hspace{5mm} 
\la \, \omega \, \big{|} \, \boldsymbol{\iota } \otimes \boldsymbol{\iota } \, \boldsymbol{\pi } 
= \la \, \omega \, \big{|} \,\boldsymbol{\iota } \otimes  \mathbb{I} \, . 
\end{align}
Because of (\ref{pi-iota}), 
it provides the cyclic property of $\boldsymbol{\mu }'_{1} = \boldsymbol{\pi } \, \boldsymbol{\mu }_{1} \, \boldsymbol{\iota }$ quickly  
\begin{align}
\la \, \omega ' \, \big{|} \, \boldsymbol{\mu }'_{1} \otimes \mathbb{I}' = 
- \la \, \omega ' \, \big{|} \, \mathbb{I}' \otimes \boldsymbol{\mu }'_{1} \, , 
\end{align}
where $\langle \omega ' | = \langle \omega | ( \boldsymbol{\iota } \otimes \boldsymbol{\iota } )$ is the symplectic form on $\hat{\mathcal{H} }'$ and $\mathbb{I}' \equiv \boldsymbol{\pi } \, \boldsymbol{\iota }$ denotes the unit of $\hat{\mathcal{H}}'$\,. 
Likewise, 
(\ref{bpz of mu}) and (\ref{bpz of h}) guarantees the cyclic property of $\boldsymbol{\mu }'_{\mathrm{int}} = \boldsymbol{\pi } \, \boldsymbol{\mu }_{\mathrm{int} } \, \mathsf{i}$ as follows  
\begin{align}
\la \, \omega \, \big{|} \, \boldsymbol{\mu }_{\mathrm{int} } \mathsf{i}  \otimes \underbrace{(\mathsf{i} + \boldsymbol{\kappa }^{-1} \boldsymbol{\mu }_{\mathrm{int} } \mathsf{i} )}_{\boldsymbol{\iota }} 
= - \la \, \omega \, \big{|} \underbrace{(\mathsf{i} + \boldsymbol{\kappa }^{-1} \boldsymbol{\mu}_{\mathrm{int} } )}_{\boldsymbol{\iota }} \otimes   \boldsymbol{\mu }_{\mathrm{int} } \mathsf{i} 
\, .
\end{align} 
Hence, 
as long as we take $ \boldsymbol{\kappa }^{-1}$ satisfying (\ref{bpz of h}), 
the cyclic property of $\boldsymbol{\mu }'$ is manifest 
\begin{align} 
\la \, \omega ' \, \big{|}\, ( \boldsymbol{\mu }'_{1} + \boldsymbol{\mu }'_{\mathrm{int} } ) \otimes \mathbb{I}'  
= - \la \, \omega ' \, \big{|} \, \mathbb{I}' \otimes ( \boldsymbol{\mu }'_{1} + \boldsymbol{\mu }'_{\mathrm{int} } ) \, . 
\end{align} 
%

\vspace{2mm} 

Note also that in the context of the quantum $A_{\infty }$ algebra, 
the cyclic property is a part of the quantum $A_{\infty }$ relations and thus manifest by definition. 
If the path-integral or corresponding results of homological perturbation can be understood as a morphism of the quantum $A_{\infty }$ structure, 
the above result arises as a natural consequence of its classical limit. 

\subsection{Morphism of the quantum $A_{\infty }$ structure}

Finally, 
we include the quantum part with (\ref{Delta=L}). 
Suppose that a solution $S$ of the classical master equation also solves the quantum one $\Delta S = 0$.  
Then, 
the cyclic $A_{\infty }$ structure $\mu = \mu _{1} + \mu _{\mathrm{int} }$ induced from $S$ satisfies the quantum $A_{\infty }$ relation 
\begin{align}
\label{co quantum A}
\big{(} \,  \boldsymbol{\mu } + \hbar \, \mathfrak{L} \, \big{)}^{2} = 0 \, , 
\end{align}
which is the coalgebraic representation of (\ref{quantum A}). 
Since $\boldsymbol{\mu } + \hbar \, \mathfrak{L}$ is a nilpotent linear map acting on \textit{the vector space} $\mathcal{T} ( \hat{\mathcal{H}} )$, 
we can take the following perturbation for (\ref{dr free}), 
\begin{align}
\label{quantum mu}
 \boldsymbol{\mu }_{\mathrm{int}} + \hbar \, \mathfrak{L} \, . 
\end{align}
As a result of the homological perturbation, 
we obtain the deformation retract describing the perturbative path-integral of quantum field theory 
\begin{align}
\mathsf{K}^{-1} \, 
\rotatebox{-70}{$\circlearrowright $} \, 
\big{(} \, \mathcal{T} ( \hat{\mathcal{H}} ) , \,  \boldsymbol{\mu } 
+ \hbar \, \mathfrak{L}   \, \big{)} 
\hspace{3mm} 
\overset{\mathsf{P} }{\underset{\mathsf{I}}{
\scalebox{2}[1]{$\rightleftarrows $}
}}
\hspace{3mm} 
\big{(} \, \mathcal{T}( \hat{\mathcal{H}}' ) , \,  \boldsymbol{\mu }'_{1} +  \boldsymbol{\mu }'_{\mathrm{eff}} \, \big{)} 
 \, . 
\end{align}
Morphisms $\mathsf{P}$ and $\mathsf{I}$ and a contracting homotopy $\mathsf{K}^{-1}$ satisfy the abstract Hodge decomposition 
\begin{align}
1 - \mathsf{I} \mathsf{P} = (  \boldsymbol{\mu } + \hbar \, \mathfrak{L} ) \, \mathsf{K}^{-1} + \mathsf{K}^{-1} \, ( 
 \boldsymbol{\mu } + \hbar \, \mathfrak{L} )  \, . 
\end{align} 
These morphisms $\mathsf{P}$ and $\mathsf{I}$ are obtained by solving recursive relations 
\begin{align}
\label{def of P-I}
\mathsf{P} =  \boldsymbol{\pi } - \mathsf{P} \, ( \, \hbar \, \mathfrak{L} +  \boldsymbol{\mu }_{\mathrm{int} } \, ) \,  \boldsymbol{\kappa }^{-1}  \, , 
\hspace{5mm} 
\mathsf{I} =  \boldsymbol{\iota } -  \boldsymbol{\kappa }^{-1} \, ( \hbar \, \mathfrak{L} +  \boldsymbol{\mu }_{\mathrm{int} } \, ) \, \mathsf{I} \, . 
\end{align}
Note however that these $\mathsf{P}$ and $\mathsf{I}$ are not coalgebra morphisms and do not satisfy (\ref{comor prop}) because $\mathfrak{L}$ is a higher coderivative and does not satisfy (\ref{coder prop}). 
%
The morphism $\mathsf{P}$ or $\mathsf{I}$ enables us to obtain the effective quantum $A_{\infty }$ structure $ \boldsymbol{\mu }' =  \boldsymbol{\mu }'_{1} +  \boldsymbol{\mu }'_{\mathrm{eff}}$ with 
\begin{align} 
\label{effective co quantum A}
 \boldsymbol{\mu }'_{\mathrm{eff}} \equiv  \boldsymbol{\pi } \, ( \hbar \, \mathfrak{L} +  \boldsymbol{\mu }_{\mathrm{int} } ) \, \mathsf{I} = \mathsf{P} \, ( \hbar \, \mathfrak{L} +  \boldsymbol{\mu }_{\mathrm{int} } ) \,  \boldsymbol{\iota } \, . 
\end{align} 
Note also that $ \boldsymbol{\mu }'$ is not a coderivation and does not satisfy (\ref{coder prop}), 
which may be regarded as a higher order coderivation of $IBL_{\infty }$ (or $IBA_{\infty }$) algebra \cite{Munster:2011ij, Munster:2012gy}, 
since $\mathfrak{L}$ is a second order. 
These maps $\mathsf{P}$ and $\mathsf{I}$ are just morphisms of the vector space $\mathcal{T} ( \hat{\mathcal{H}})$ such that 
\begin{align}
\label{P-I}
\mathsf{P} \, (  \boldsymbol{\mu } + \hbar \, \mathfrak{L} ) = (  \boldsymbol{\mu }'_{1} + \boldsymbol{\mu }'_{\mathrm{int}} ) \, \mathsf{P} \, , 
\hspace{5mm} 
(  \boldsymbol{\mu } + \hbar \, \mathfrak{L} ) \, \mathsf{I} =   \mathsf{I} \, (  \boldsymbol{\mu }'_{1} + \boldsymbol{\mu }'_{\mathrm{int}} ) \, , 
\end{align}
which we call \textit{a morphism of the (quantum)} $A_{\infty }$ \textit{structure}. 
These relations (\ref{P-I}) are proven by the same way as (\ref{p-i}), 
which follows from the homological perturbation lemma as shown in \cite{Jurco:2019yfd}. 

\vspace{2mm} 

In general, 
when a given solution $S_{[0]}$ of the classical master equation $(S_{[0]} , S_{[0]} ) = 0$ does not solve the quantum master action, 
it necessitates quantum corrections $\hbar \, S_{[1]} + \hbar ^{2} \, S_{[2]} + \cdots $ such that $S \equiv S_{[0]} + \hbar \, S_{[1]} + \hbar ^{2} \, S_{[2]} + \cdots $ satisfies $\hbar \, \Delta S + \frac{1}{2} ( S , S) = 0$\,.  
Then, 
the cyclic $A_{\infty }$ structure $\mu _{[0]}$ induced from $S_{[0]}$ does not satisfy the quantum $A_{\infty }$ relation, 
$( \boldsymbol{\mu }_{[0]} + \hbar \, \mathfrak{L} )^{2} \not=0$. 
Each correction $S_{[l]}$ gives components of quantum $A_{\infty }$ maps $\mu _{n , \, [l] } : \mathcal{H}^{\otimes n} \rightarrow \mathcal{H}$\,, 
which is extended to coderivation of $\mathcal{T} (\hat{\mathcal{H}})$ by defining $ \boldsymbol{\mu }_{n , \, [l] } |_{\hat{\mathcal{H}}^{\otimes m} } : \hat{\mathcal{H}}^{m} \rightarrow \hat{\mathcal{H}}^{\otimes m-n+1}$ for $m \geq n$ otherwise zero as (\ref{deriv}). 
For a given $S_{[l]}$\,, 
we obtain the coderivation  
\begin{align}
 \boldsymbol{\mu }_{[l]} =  \boldsymbol{\mu }_{0 , [l]} +  \boldsymbol{\mu }_{1, [l] } +  \boldsymbol{\mu }_{2, [l] } +  \cdots \, , 
\end{align}
which are necessary for the quantum $A_{\infty }$ relations $( \boldsymbol{\mu }_{[0]} + \sum_{l} \hbar^{l} \,  \boldsymbol{\mu }_{[l]} + \hbar \, \mathfrak{L} )^{2} =0$\,. 
Hence, 
in this case, 
the above $ \boldsymbol{\mu }_{\mathrm{int}}$ of (\ref{quantum mu}) must be replaced by  
\begin{align}
 \boldsymbol{\mu }_{\mathrm{int}} =  \boldsymbol{\mu }_{\mathrm{int} , [0] } + \sum_{l\geq1} \hbar ^{l} \,  \boldsymbol{\mu }_{ [l] } \, . 
\end{align}
This replacement of (\ref{quantum mu}) enables us to obtain the appropriate perturbed data in the same way.  
We can express the solutions of the defining equations (\ref{def of P-I}) as follows 
\begin{align}
\mathsf{P} =  \boldsymbol{\pi } \, \frac{1}{ 1+ ( \, \hbar \, \mathfrak{L} +  \boldsymbol{\mu }_{\mathrm{int} } \, ) \, \boldsymbol{ \kappa }^{-1} }   \, , 
\hspace{5mm} 
\mathsf{I} = \frac{1}{1+  \boldsymbol{ \kappa }^{-1} \, ( \hbar \, \mathfrak{L} +  \boldsymbol{\mu }_{\mathrm{int} } \, ) } \,  \boldsymbol{\iota } \, . 
\end{align}
The perturbed quantum $A_{\infty }$ structure can be written as  
\begin{align}
\label{solution mu'}
 \boldsymbol{\mu }' 
=  \boldsymbol{\mu }'_{1} 
+  \boldsymbol{\pi } \, ( \hbar \, \mathfrak{L} +  \boldsymbol{\mu }_{\mathrm{int} } \, ) \, \frac{1}{1+   \boldsymbol{\kappa }^{-1} \, ( \hbar \, \mathfrak{L} + \boldsymbol{ \mu }_{\mathrm{int} } \, ) } \,  \boldsymbol{\iota } \, , 
\end{align}
which takes the same form as (\ref{effective A}). 
Its homotopy Maurer-Cartan action is given by (\ref{hMC A}) with replacing $\boldsymbol{\mu }'$ by (\ref{solution mu'}), 
which equals to (\ref{effective BV action}) or (\ref{effective BV action alt}) derived in section 3.

\subsection{Weak $A_{\infty }$ and source terms} 

We show that the BV master action including source terms or external fields gives an effective theory with a weak $A_{\infty }$ structure.\footnote{It is an $A_{\infty }$ structure including the zeroth product $\mu '_{0}$, which is also called a twisted $A_{\infty }$ or curved $A_{\infty }$. } 
It yields that when we consider field theories having two (or higher) point interactions with external fields, 
every effective theory always has a weak $A_{\infty }$ structure. 
For a given BV master action, 
we can couple $V$ as follows 
\begin{align}
S_{V} [ \Psi  ] \equiv S [\Psi ] - \omega \big{(} \,  \Psi , \, V \, \big{)} 
\end{align}
and suppose that this $S_{V} [\Psi ]$ and its free part also satisfy the BV master equation $\hbar \, \Delta \, e^{S_{V} [\Psi ] } = 0$\,. 
Then, 
source terms $V$ must satisfy the following properties with the $A_{\infty }$ structure of $S[\Psi ]$, 
\begin{align}
\mu _{1} (V) = 0 \, , 
\hspace{5mm} 
\sum _{n} \sum_{k=0}^{n} \mu _{n+1} (  \underbrace{\Psi , ... , \Psi }_{k} , V , \Psi , ... , \Psi ) = 0 \, , 
\end{align}
which we call gauge invariant source terms. 
Then, 
the source terms $\mu _{0} \equiv V$ become the zeroth product of a weak $A_{\infty }$ structure $\mu + V$. 
Note that $\mu $ itself is the $A_{\infty }$ structure and thus this $\mu +V$ is stronger than a generic weak $A_{\infty }$ structure. 
If we add such $V$ to our perturbation, 
we find that the effective $A_{\infty }$ structure is shifted by $\kappa ^{-1} V$ as follows 
\begin{align}
\mu _{V} ' = e^{- \kappa ^{-1} V} \, \mu ' \, e^{\kappa ^{-1} V} + V \, . 
\end{align} 
It becomes a weak (quantum) $A_{\infty }$ structure, 
whose zeroth element is  
\begin{align}
\mu _{V , 0}' = V + \sum_{k=0}^{\infty } 
\mu '_{k} \big{(} \, ( \kappa ^{-1} V)^{\otimes k } \, \big{)} \ . 
\end{align}
Note that $\mu '$ itself is the $A_{\infty }$ structure and 
the shifted $n$-product is given by 
\begin{align} 
\mu '_{V , n} ( \Psi ^{ \prime \, \otimes n}  ) = \sum_{k=0}^{\infty } \sum _{k_{0} + \cdots + k_{n} =k} 
\mu '_{n+k} \big{(} \, (\kappa ^{-1} V)^{\otimes k_{0} } , \Psi ' , (\kappa ^{-1} V)^{\otimes k_{1} } , 
... , 
\Psi ' , ( \kappa ^{-1} V)^{\otimes k_{n}} \, \big{)} \, . 
\end{align}

\subsection{The Wick theorem via quantum $A_{\infty }$}

We show that the Wick theorem follows from the projection again by using the above homotopy algebraic description.     
Let us consider free theories for simplicity: we have  
\begin{align}
\widehat{\mathsf{P} } = \boldsymbol{\pi } \frac{1}{1 + \hbar \, \mathfrak{L} \, \boldsymbol{\kappa }^{-1}  } 
\, . 
\end{align}
In the homotopy algebraic description, 
we must identify $\boldsymbol{\kappa }^{-1}$ defined by lifting $K^{-1} \phi ^{a} \frac{\delta }{\delta \phi ^{\ast }_{a}}$ with the operator replacing $\hat{e}_{1+a} \equiv (-)^{a} \hat{e}_{-a}^{\ast }$ by $K^{-1} \, \hat{e}_{a}$ when it acts on $\frac{1}{1-\Psi }$ as in the case of (\ref{Delta=L}): 
Then, 
we have to pay attention to the action of composition operators.\footnote{The ordering of composition operators can be ignored when we consider the transfer of the tree revel $A_{\infty }$ structure because of the symmetry of $\boldsymbol{\mu }_{int } \frac{1}{1 - \boldsymbol{\kappa }^{-1} \boldsymbol{\mu }_{int }}$. }  
Note that since $\boldsymbol{\mu }$, $\boldsymbol{\kappa }^{-1}$ and $\mathfrak{L}$ are made of the differential of antifields, differential of fields and Laplacian, 
compositions of these operators are defined like compositions of ``pullbacks''. 
For instance, 
the composition operator $\mathfrak{L} \, \boldsymbol{\kappa }^{-1} (...)$ is computed like a pullback of $\boldsymbol{\kappa }^{-1} (...)$ by $\mathfrak{L}$ as follows 
\begin{align}
 \mathfrak{L} \, \boldsymbol{\kappa }^{-1} (...)  \Big{[} \frac{1}{1-\Psi } \Big{]} 
 & = \boldsymbol{\kappa }^{-1} (...)  \Big{[} \sum_{s \in \mathbb{Z} }\frac{1}{1- \Psi } \otimes \hat{e}_{-s} \otimes \frac{1}{1-\Psi } \otimes  \hat{e}_{1+s} \otimes \frac{1}{1- \Psi } \Big{]} 
 \, . 
\end{align} 
%
By using $\sum_{s} \hat{e}_{-s} \otimes \boldsymbol{\kappa }^{-1} ( \hat{e}_{1+s} ) = \frac{1}{2} \sum_{s} K_{s}^{-1} \hat{e}_{-s} \wedge \hat{e}_{s}$ and $(\boldsymbol{\kappa }^{-1} )^{2} = 0$, 
we find  
\begin{align} 
( \mathfrak{L} \, \boldsymbol{\kappa }^{-1} )^{n}  \big{(} \, 1 \, \big{)} 
& =  (  \mathfrak{L} \, \boldsymbol{\kappa }^{-1} )^{n-1} \frac{1}{2} 
\sum_{s \in \mathbb{Z} } K_{s}^{-1}  \hat{e}_{-s}  \wedge  \hat{e}_{s} 
\no 
& = (  \mathfrak{L} \, \boldsymbol{\kappa }^{-1} )^{n-2} \frac{1}{4 \cdot 2}  
 \sum_{s , s'  \in \mathbb{Z} } K_{s}^{-1} 
 \hat{e}_{-s} \wedge  \hat{e}_{s} \,  \wedge K_{s'}^{-1}  
 \hat{e}_{-s'}   \wedge \hat{e}_{s'} 
 \no 
 & = \frac{1}{(2n)!!} \sum_{s_{1}, ... s_{n} } ( K^{-1}_{s_{1} } \, \hat{e}_{s_{1} } \wedge \hat{e}_{-s_{1} } ) \wedge  \dots \wedge ( K^{-1}_{s_{n}} \, \hat{e}_{s_{n} } \wedge \hat{e}_{-s_{n}} ) \, . 
\end{align}
We write $\boldsymbol{\hat{e}}_{-g}$ for the coderivation (of 0-linear map) just inserting $\hat{e}_{-g}$ for brevity, 
which (graded) commutes each other. 
Because of $\boldsymbol{\pi } (\Psi ) = 0$, 
we obtain the following expression  
\begin{align}
\widehat{\mathsf{P}} \, \frac{1}{1- \Psi } & = \boldsymbol{\pi } \, \sum_{n=1}^{\infty } \hbar^{n}  ( \mathfrak{L} \, \boldsymbol{\kappa }^{-1} )^{n-1} \, \Big{[}
 \frac{1}{1- \Psi } \Big{]} 
\no 
& = \boldsymbol{\pi } \, \sum_{n=0}^{\infty } \hbar^{n}  \, \frac{1}{(2n)!!} \, \bigg{[} \prod_{i=1}^{n} \Big{(} \sum_{s_{i}  \in \mathbb{Z} } K^{-1} \, \boldsymbol{\hat{e}}_{-s_{i}} \, \boldsymbol{\hat{e}}_{s_{i}} \Big{)} \, \bigg{]}  \frac{1}{1-\Psi } 
\no
& = \sum_{n=0}^{\infty } \frac{1}{(2n)!!}  ( \hbar \, K^{-1} \, \sum_{s_{1}} \hat{e}_{-s_{1} } \wedge \hat{e}_{s_{1} }  ) \wedge \dots \wedge (  \hbar \, K^{-1} \, \sum_{s_{n}}  \hat{e}_{-s_{n} } \wedge \hat{e}_{s_{n} } ) 
\end{align}
Recall that $\Psi = \sum_{s \in \mathbb{Z} } \hat{e}_{s} \otimes \psi _{s}$ is the sum of all possible BV fields and for each $n \in \mathbb{N}$, 
the relation $\Psi ^{\otimes n } = \frac{1}{n!} \Psi ^{\wedge n}$ holds. 
The summand can be cast as follows 
\begin{align}
\Big{(} 
\hbar \, K^{-1} \, \sum_{s}  \hat{e}_{-s} \wedge \hat{e}_{s} 
\Big{)}^{\wedge n}  
& = \bigg{(} 
\hbar \, K^{-1} \, \sum_{s} \frac{\delta }{\delta \psi _{s}} \Psi  \wedge \frac{\delta }{\delta \psi _{-s}} \Psi  
\bigg{)}^{\wedge n -1} \frac{1}{2} \, 
\Big{[} \, \hbar  \, K^{-1} \, \sum_{s} \frac{\delta }{\delta \psi _{s}}  \, \frac{\delta }{\delta \psi _{-s}} \Big{]} \, \Psi \wedge \Psi  
\no 
& =  \Big{[} \, \hbar  \, K^{-1} \, \sum_{s} \frac{\delta }{\delta \psi _{s}}  \, \frac{\delta }{\delta \psi _{-s}} \Big{]}^{n} \, 
\overbrace{\frac{1 }{(2n)!} \, \Psi ^{\wedge 2n } }^{\Psi ^{\otimes n}} 
\, . 
\end{align} 
Hence, the Wick theorem is derived from the (quantum) $A_{\infty }$ (quasi-iso)morphism $\widehat{\mathsf{P}}$ as follows 
\begin{align}
\widehat{\mathsf{P} } \, \frac{1}{1- \Psi } 
& = \sum_{n=0}^{\infty }
\frac{1}{n! } \Big{[} \, \frac{\hbar }{2} \sum_{s} K_{s}^{-1} \, \frac{\delta }{\delta \psi _{s}}  \, \frac{\delta }{\delta \psi _{-s}} \Big{]}^{n} \, \Psi ^{\otimes 2 n}   
\no 
& =  \exp  \Big{[} \, \frac{\hbar }{2} \sum_{s} K_{s}^{-1} \, \frac{\delta }{\delta \psi _{s}}  \, \frac{\delta }{\delta \psi _{-s}} \Big{]} \frac{1}{1-\Psi } \bigg{|}_{\Psi = 0} 
\, . 
\end{align}
In this sense, 
we conclude that the perturbative path-integral is a morphism $\mathsf{P}$ of Lagrangian's quantum $A_{\infty }$ structure and the $\mathsf{P}$ is performed via the homological perturbation. 

\section{Application to string field theory} 

In most string field theories, 
fortunately, 
we have classical or quantum BV master actions. 
Hence, 
we can apply the previous results and perform the perturbative path-integral of string fields on the basis of the homological perturbation. 
Let us consider the BV master action of string field theory 
\begin{align}
S [\Psi ] 
= \frac{1}{2} \omega  \big{(} \, \Psi  , \, \boldsymbol{\mu }_{1} (  \Psi  ) \, \big{)}
+ \sum_{n} \frac{1}{n+1} \omega  \big{(} \, \Psi  , \, \boldsymbol{\mu }_{n} ( \Psi , ... , \Psi  ) \, \big{)} \, ,  
\end{align}
where $\mu _{1} = Q$ is the BRST operator of strings, 
$\{ \boldsymbol{\mu }_{n} \}_{n}$ denotes the set of the string products and $\omega $ is the degree $-1$ symplectic form induced from the BPZ inner product. 
A state $\Psi = \sum_{g \in \mathbb{Z}} \hat{e}_{-g} \otimes \psi _{g}$ is a string field that includes all fields and antifields: 
classical fields $\psi _{0} (x) $, 
ghosts $\{ \psi _{g} (x) \}_{g>0}$ and antifields $\{ \psi _{g} (x) \}_{g<0}$ are associated with the (suspended) complete basis $\{ \hat{e}_{-g} \} _{g \in \mathbb{Z}}$ of conformal field theory, 
where $\Psi $ corresponds to (\ref{total state}) and we write $\mathcal{H}$ for the state space spanned by $\Psi$ in this section. 
The action and the sum of string products $\boldsymbol{\mu } (\Psi ) = \sum_{n} \boldsymbol{\mu }_{n} (\Psi ^{\otimes n})$ satisfy (\ref{def of A-relations}) or (\ref{def of quantum A-relations}) with $\boldsymbol{\mu } (\Psi ) = \hbar \, \Delta _{S} \Psi $. 
By solving the free theory, 
we obtain the Hodge decomposition 
\begin{align} 
\label{solving free theory}
1 - \iota \pi = \mu _{1}  \, \kappa ^{-1} + \kappa ^{-1} \, \mu_{1}  \, , 
\end{align}
which is the starting point of performing the perturbative path-integral.

\subsection{Effective theories with finite $\alpha '$} 

Each effective theory based on the perturbative path-integral, 
\begin{align}
\label{effective sft}
A [ \Psi ' ] 
= \frac{1}{2} \omega ' \big{(} \, \Psi ' , \, \boldsymbol{\mu }'_{1} (  \Psi ' ) \, \big{)}
+ \sum_{n} \frac{1}{n+1} \omega ' \big{(} \, \Psi ' , \, \boldsymbol{\mu }'_{n} ( \Psi ' , ... , \Psi ' ) \, \big{)} \, , 
\end{align}
always has the (quantum) $A_{\infty }$ or $L_{\infty }$ structure trivially, 
as a result of the homological perturbation, 
as long as the original action $S[\Psi ]$ solves the BV master equation. 
In addition, 
when the original action includes source terms $\omega ( \Psi ,V )$, 
the effective action (\ref{effective sft}) has a weak (quantum) $A_{\infty }$ structure $\mu '_{V} = e^{-\kappa ^{-1} V } \mu ' e^{\kappa ^{-1} V }$ as shown in the previous section.  

\vspace{2mm} 

We can integrate all massive space-time fields $\Psi _{\mathrm{massive}}$ out from the string field $\Psi  = \Psi _{\mathrm{massless}} + \Psi _{\mathrm{massive}}$ and get an effective field theory that consists of massless (plus auxiliary) fields $\Psi _{\mathrm{massless}}$ by using the Hodge decomposition  
\begin{align}
1 - (\iota \pi )_{\mathrm{massless} } = \mu _{1}  \, \kappa _{\mathrm{massive} }^{-1} + \kappa _{\mathrm{massive} }^{-1} \, \mu_{1} \, , 
\end{align}
where $\kappa _{\mathrm{massive}}^{-1}$ denotes propagators of massive fields and $(\iota \pi )_{M=0}$ denotes a projector onto the massless fields $\Psi _{\mathrm{massless}} = (\iota \pi )_{\mathrm{massless}} \Psi  $\,. 
We can construct these $\psi _{\mathrm{massless}}$, 
$\iota _{\mathrm{massless}} $ and $\kappa _{\mathrm{massive}}^{-1}$ explicitly by solving the free theory, 
which gives the effective action (\ref{effective sft}) with $\Psi ' = \pi _{\mathrm{massless}} ( \Psi ) = \pi _{\mathrm{massless}} ( \Psi _{\mathrm{massless}}) \in \mathcal{H}'$\,. 
Likewise, 
we can integrate space-time fields $\Psi _{\mathrm{UV}}$ having higher momentum $p > \Lambda $ out from the string field $\Psi = \Psi _{\mathrm{IR}} + \Psi _{\mathrm{UV}}$ and construct a Wilsonian effective action with the cut-off scale $\Lambda$ perturbatively. 
It is obtained by using the Hodge decomposition 
\begin{align}
1 - (\iota \pi )_{p \leq \Lambda } = \mu _{1}  \, \kappa _{\mathrm{UV} }^{-1} + \kappa _{\mathrm{UV}}^{-1} \, \mu_{1}  \, , 
\end{align}
where $(\iota \pi )_{p \leq \Lambda }$ denotes the restriction onto the lower momentum fields $(\iota \pi )_{p \leq \Lambda } \Psi  = \Psi _{\mathrm{IR} }$ and $\kappa _{\mathrm{UV}}^{-1}$ denotes propagators of the higher momentum fields.  
It provides (\ref{effective sft}) with $\Psi ' = \pi _{p \leq \Lambda } (\Psi ) \in \mathcal{H} '$\,. 
In the same manner, 
for any decomposition (\ref{solving free theory}), 
we can obtain corresponding effective action. 
 
\vspace{2mm} 

We would like to emphasize that the physically important information is in how to construct these projectors and (regular) propagators concretely: to give the Hodge decomposition is equivalent to solving the theory. 
We thus started from the free theory and considered perturbations.

\subsection{Light-cone reduction}

In string field theory, 
explicit Lorentz covariance is given in return for adding the gauge and unphysical degrees. 
Thus, 
while the light-cone theory consists of physical degrees, 
the covariant theory has the gauge invariance. 
We can remove theses extra degrees by using the path-integral and obtain a light-cone string field theory for each covariant string field theory \cite{Matsunaga:2019fnc, EM}. 

\vspace{2mm} 

We write $Q$ for the BRST operator of the world-sheet theory and $\omega $ for the BPZ inner product of its conformal field theory. 
We consider a covariant string field theory, 
\begin{align}
S [\Psi ] = \frac{1}{2} \omega \big{(} \Psi , \, Q \, \Psi \big{)} 
+ \frac{1}{3} \omega \big{(} \Psi , \, \boldsymbol{m}_{2} ( \Psi , \Psi ) \, \big{)} 
+ \cdots \, . 
\end{align}
It has an $A_{\infty }$ (or $L_{\infty }$) structure $\boldsymbol{m}$ with $\boldsymbol{m}_{1} = Q$ as long as it satisfies the BV master equation. 
There exists a similarity transformation $U$ connecting the BRST operator $Q$ and the kinetic operator $L_{0}^{lc}$ in the light-cone gauge plus the differential $d$ acting on the gauge and unphysical degrees \cite{Aisaka:2004ga}, 
which diagonalize physical and extra degrees as follows  
\begin{align}
Q = U^{-1} \, 
\big{(}  
c_{0} \, L_{0}^{lc} + d 
\, \big{)} \, 
U \, . 
\end{align} 
Note that $(c_{0} L^{lc}_{0} ) ^{2} = 0$ holds in addition to $(c_{0} L_{0}^{lc} + d \, )^{2} = 0$ and these are defined on the critical dimention.\footnote{ 
For bosonic open strings, 
by using $\lambda =1$ Virasoro generators $\tilde{L}_{n} = \tilde{L}_{n}^{\parallel }+ \tilde{L}_{n}^{\perp }$, 
these are given by 
\begin{align*}
U \equiv \exp \Big{[} - c_{0} \underbrace{\frac{1}{p^{+} } \sum_{n \not= 0} a^{+}_{-n} b_{n}  }_{h }
 \Big{]} 
 \exp \Big{[} \frac{1}{p^{+}} \sum _{n \not= 0} \frac{1}{n}  a^{+}_{-n} \tilde{L}_{n}  \Big{]} \, ,  
 \hspace{3mm} 
d \equiv - p^{+} \sum_{n \not= 0} a^{-}_{n} \, c_{-n} \, ,  
\hspace{3mm} 
\kappa ^{-1} \equiv h \int_{0}^{1} dt \, e^{- t \tilde{L}_{0}^{\parallel } } \, . 
\end{align*}
Note that that $\tilde{L}_{n}$ commutes with $a ^{+}_{n}$ and $L_{0}^{\parallel } = dh+hd$ counts excitations of $\{ a^{\pm }_{n}, b_{n} , c_{n} \}_{n\not= 0}$. 
See \cite{Aisaka:2004ga, Matsunaga:2019fnc}. }
The similarity transformation $U$ becomes an isomorphism $\boldsymbol{U} \,\boldsymbol{m}  = \boldsymbol{\mu } \, \, \boldsymbol{U}$ and gives the diagonalized $A_{\infty }$ structure $\boldsymbol{\mu }$ which is defined by $\boldsymbol{\mu }_{1} \equiv c_{0} \, L_{0}^{lc} + d$ and 
\begin{align}
\boldsymbol{\mu }_{n} \equiv U \, \boldsymbol{m}_{n} \,( U^{-1} \otimes \cdots \otimes U^{-1} ) \, . 
\end{align}
It gives a linear transformation of the conformal basis and thus provides a linear string-field redefinition $\Phi \equiv U \, \Psi$\,. 
We obtain the diagonalized action with the $A_{\infty }$ structure $\boldsymbol{\mu }$\,,  
\begin{align}
\label{cov} 
S [\Phi ] = \frac{1}{2} \omega \big{(} \Phi , \, ( \underbrace{ c_{0} \, L_{0}^{lc} + d }_{\mu _{1} } \, ) \Phi \big{)} 
+ \sum_{n=2}^{\infty } \frac{1}{n+1} \omega \big{(} \Phi , \, \boldsymbol{\mu }_{n} ( \underbrace{\Phi , \cdots , \Phi }_{n} ) \, \big{)} 
\, . 
\end{align}
The extra degrees become the BRST quartets and thus $d$ has no cohomology unless there is no quartet excitation. 
As is known, 
the integration of the BRST quartets is volume $1$ since bosonic and fermionic integrations exactly cancel each other. 
We can start with the BRST quartets,  
\begin{align}
\label{quartet}
\kappa ^{-1} \, 
\rotatebox{-70}{$\circlearrowright $} \, 
\big{(} \, \mathcal{H} , \,  d \, \big{)} 
\hspace{3mm} 
\overset{\pi }{\underset{\iota }{
\scalebox{2}[1]{$\rightleftarrows $}
}}
\hspace{3mm} 
\big{(} \, \mathcal{H}_{\mathrm{lc} } , \, 0\, \big{)} 
\, , 
\end{align} 
where $\kappa ^{-1} $ denotes the propagator for $d$ and $\mathcal{H}_{lc}$ is the state space of string fields in the light-cone gauge. 
We take $\pi : \mathcal{H} \rightarrow \mathcal{H}_{\mathrm{lc}}$ and $\iota : \mathcal{H}_{\mathrm{lc}} \rightarrow \mathcal{H}$ as natural projection and embedding.\footnote{
For the Fock vacuum $| \Omega \rangle \equiv | \mathrm{lc} \rangle \otimes | a^{\pm}, b,c \rangle$, 
we define $\pi : |\Omega \rangle \mapsto  |\mathrm{lc} \rangle$ and $\iota : |\mathrm{lc} \rangle \mapsto |\Omega \rangle $.  
For excitations on these vacua, we define 
$\pi \circ ( p^{\mu } , a_{n}^{I} , c_{0} ; a_{n}^{\pm } , c_{n} , b_{n} ) = ( p^{\mu } , a_{n}^{I} , c_{0} ) \circ \pi$ and $\iota \circ ( p^{\mu } , a_{n}^{I} , c_{0} ) = ( p^{\mu } , a_{n}^{I} , c_{0} ; 0 , 0, 0) \circ \iota $ for $n \not=0$. }
We can take $c_{0} L^{lc}$ as a perturbation to (\ref{quartet}) and get 
\begin{align}
\label{no-ghost} 
\kappa ^{-1} \, 
\rotatebox{-70}{$\circlearrowright $} \, 
\big{(} \, \mathcal{H} , \,  c_{0} L_{0}^{lc} + d \, \big{)} 
\hspace{3mm} 
\overset{\pi }{\underset{\iota }{
\scalebox{2}[1]{$\rightleftarrows $}
}}
\hspace{3mm} 
\big{(} \, \mathcal{H}_{\mathrm{lc} } , \,  c_{0} \, L_{0}^{lc} \, \big{)} 
\, . 
\end{align} 
It describes the no-ghost theorem of covariant strings \cite{Kato:1982im}. 
We can take a further perturbation $\mu _{\mathrm{int} } + \hbar \, \mathfrak{L}$ for (\ref{no-ghost}) because of the $A_{\infty }$ structure $(\mu _{1} + \mu _{\mathrm{int} } + \hbar \, \mathfrak{L} )^{2} = 0$ and obtain  
\begin{align}
\label{lc red}
\mathsf{k}^{-1} \, 
\rotatebox{-70}{$\circlearrowright $} \, 
\big{(} \, \mathcal{T} ( \mathcal{H} ) , \,  \boldsymbol{\mu _{1} }  + \boldsymbol{\mu _{\mathrm{int} } } + \hbar \, \boldsymbol{\mathfrak{L} } 
\, \big{)} 
\hspace{3mm} 
\overset{\mathsf{P} }{\underset{\mathsf{I} }{
\scalebox{2.5}[1]{$\rightleftarrows $}
}}
\hspace{3mm} 
\big{(} \, \mathcal{T} ( \mathcal{H}_{\mathrm{lc} } ) , \,  \boldsymbol{\nu _{1}^{lc}} +
\boldsymbol{\nu _{\mathrm{int}}^{lc} } \, \big{)} 
\, . 
\end{align}
While the left side has the $A_{\infty }$ structure $\boldsymbol{\mu }$ of the covariant string field theory (\ref{cov}), 
the right side provides the transferred $A_{\infty }$ structure $\boldsymbol{\nu ^{lc}}$ of the light-cone string field theory.\footnote{Note that $\hbar $ must be zero unless the covariant theory solves the quantum BV master equation. }  
By using the light-cone string field $\varphi \in \mathcal{H}_{\mathrm{lc}}$ and the light-cone vertices $\boldsymbol{\nu _{\mathrm{int} }^{lc} } \equiv \boldsymbol{\pi } \, \boldsymbol{\mu _{\mathrm{int}} } \, \mathsf{I} $, 
we obtain the light-cone string field theory $S_{\mathrm{lc} } [\varphi ]$ extracted from the covariant theory (\ref{cov}), 
\begin{align}
\label{lc}
S_{\mathrm{lc} } [ \varphi ] 
= \frac{1}{2} \omega \big{(} \varphi , \, c_{0} \, L_{0}^{lc} \, \varphi \big{)} 
+ \sum_{n=2}^{\infty } \frac{1}{n+1}  \omega \big{(}  \varphi , \, \boldsymbol{\nu }^{lc}_{n} (\underbrace{ \varphi , \dots , \varphi }_{n} )  \big{)} \, , 
\end{align}
where we used loose notation $\varphi = \iota (\varphi ) $ and $c_{0} \, L^{lc}_{0} = \pi \, ( c_{0} \, L_{0}^{lc} ) \, \iota = \mu _{1}$ for simplicity. 
Note that the vertices $\boldsymbol{\nu _{\mathrm{int} }^{lc} }$ consists of the original vertices $\boldsymbol{\mu _{\mathrm{int} } }$ (with projections and embeddings) and \textit{effective} vertices $\mu _{\mathrm{eff} }$ including propagators $\boldsymbol{\kappa ^{-1}}$ as follows 
\begin{align*}
\boldsymbol{\nu _{\mathrm{int} }^{lc} } 
= \boldsymbol{\pi } \,
\boldsymbol{\mu _{\mathrm{int} } } 
\, \boldsymbol{\iota } 
+ \boldsymbol{\pi } \, 
\underbrace{ 
\bigg[ 
\sum_{n=1}^{\infty } 
(-)^{n} \boldsymbol{\mu _{\mathrm{int} } } 
\big{(} \boldsymbol{\kappa ^{-1} } \, \boldsymbol{\mu _{\mathrm{int} } } \big{)}^{n} 
+ \sum_{g} \hbar ^{g} \, (\mathrm{g\mathchar`-loop})
\bigg] 
}_{\mu _{\mathrm{eff} } (\varphi , ... , \varphi ) }
\, \boldsymbol{\iota } 
\, . 
\end{align*}  
In this sense, 
the light-cone reduction (\ref{lc red}) can be cast as the form which consists of 
the light-cone kinetic term, 
the original vertices, 
and effective vertices. 
Hence, 
the action (\ref{lc}) has higher interacting terms and takes the different form from the original covariant theory (\ref{cov}) unless all of the effective vertices $\mu _{\mathrm{eff}} (\varphi , ... , \varphi )$ exactly equal to zero. 
See \cite{EM} for further discussions.

\subsection{S-matrix and asymptotic string fields}

When a given (quantum) $A_{\infty }$ structure $\boldsymbol{\mu } = \boldsymbol{\mu }_{2} + \cdots $ has no linear part $\boldsymbol{\mu }_{1}$, 
it is called minimal. 
The $S$-matrix is realized as a minimal model, 
which can be obtained by using the homological perturbation. 
The uniqueness of the minimal $A_{\infty }$ structure is ensured by the minimal model theorem in mathematics. 
In terms of physics, 
it implies that the \textit{on-shell} amplitudes are independent of given gauge-fixing conditions or propagators and thus are unique.

\vspace{2mm} 

The S-matrix is a set of multi-linear forms $\{ \mathcal{A} _{n} \} _{n \geq 3}$ defined on the tensor algebra $\mathcal{T} (\cH_{as} )$ of the state space $\cH_{as}$, 
whose inputs are asymptotic string fields $\Psi _{as} \in \cH _{as }$. 
We consider the free action of asymptotic string fields, 
\begin{align}
\label{asympt}
S_{as} [ \Psi _{as} ] 
= \frac{1}{2} \omega \big{(} \, \Psi _{as} , Q \, \Psi _{as} \, \big{)} \, . 
\end{align} 
The asymptotic string field $\Psi _{as} \in \cH _{as}$ has the linear gauge invariance $\delta \Psi _{as} = Q \, \lambda _{as}$ and the physical states condense on the cohomology of $Q$ acting on $\cH _{as}$. 
We assume that the cohomology $\cH _{as \, \mathrm{phys}}$ of the asymptotic theory is isomorphic to that of the free theory, 
$\cH _{\mathrm{phys}} \equiv I \, ( \mathcal{H}_{as \, \mathrm{phys}} )$.

\vspace{2mm} 

We first solve the free theory and derive a propagator $\kappa ^{-1}$, 
which gives the Hodge decomposition (\ref{solving free theory}). 
Then, 
by defining morphisms $\iota _{as } \equiv \iota \, I$ and $\pi _{as } \equiv I^{-1} \pi $ that satisfy $\pi _{as} \, \kappa ^{-1} = \iota _{as} \, \kappa ^{-1} = 0$, 
we can consider 
\begin{align}
\label{asymptotic}
\boldsymbol{\kappa ^{-1} } \, 
\rotatebox{-70}{$\circlearrowright $} \, 
\big{(} \, \mathcal{T} ( \mathcal{H} )  , \, \boldsymbol{\mu _{1} } \, \big{)} 
\hspace{3mm} 
\overset{\boldsymbol{\pi _{as}} }{\underset{\boldsymbol{\iota _{as}}}{
\scalebox{2}[1]{$\rightleftarrows $}
}}
\hspace{3mm} 
\big{(} \, \mathcal{T} ( \mathcal{H} _{as \, \mathrm{phys} } ) ,  \, \boldsymbol{0} \, \big{)} 
\hspace{3mm} 
\overset{\boldsymbol{I } }{\underset{\boldsymbol{I^{-1} }}{
\scalebox{2}[1]{$\rightleftarrows $}
}}
\hspace{3mm} 
\big{(} \, \mathcal{T} ( \mathcal{H}_{\mathrm{phys} } ) , \, \boldsymbol{0} \, \big{)} 
 \, . 
\end{align} 
Because of $\iota \, \pi = \iota \, (I \, I^{-1} ) \, \pi = \iota _{as } \, \pi _{as} $, 
we quickly find $1 - \iota _{as} \pi _{as} = \mu _{1} \kappa ^{-1} + \kappa ^{-1} \mu _{1}$\,. 
The minimal model is obtained by taking interacting terms $\boldsymbol{\mu _{\mathrm{int}} } $ as the perturbation to (\ref{asymptotic}). 
The (quantum) $A_{\infty }$ structure of the $S$-matrix is given by the right side of 
\begin{align}
\mathsf{K}^{-1}_{as} \, 
\rotatebox{-70}{$\circlearrowright $} \, 
\big{(}  \, \mathcal{T} ( \mathcal{H} )  , \, \boldsymbol{\mu _{1}} + \boldsymbol{\mu _{\mathrm{int} } } \, \big{)} 
\hspace{3mm} 
\overset{ \mathsf{P}_{as} }{\underset{\mathsf{I}_{as} }{
\scalebox{4.5}[1]{$\rightleftarrows $} }
}
\hspace{3mm} 
\big{(} \, \mathcal{T} ( \mathcal{H} _{as \, \mathrm{phys} } ) , \, \boldsymbol{\mu '_{\mathrm{int} } } \, \big{)} 
 \, . 
\end{align} 
This is a minimal model because $\mu '_{1} \equiv Q$ vanishes 
and it has no gauge degree. 
The morphism $\mathsf{P}_{as}$ determines a nonlinear field relation between interacting and asymptotic theories on-shell. 
The $(n+1)$-point amplitude is given by the $\mu '_{n}$ part of the homotopy Maurer-Cartan action 
\begin{align}
\label{S-matrix}
\mathcal{A} [ \Psi ' ] 
= \sum_{n} \frac{1}{n+1} \omega ' \big{(} \, \Psi '_{as} , \, \boldsymbol{\mu }'_{n} ( \Psi '_{as} , ... , \Psi '_{as} ) \, \big{)} \, . 
\end{align}
%
It defines multi-linear maps acting on the on-shell asymptotic string fields. 
As we showed, 
it is the same as the Feynman graph expansion and thus gives the amplitudes correctly. 
In addition, 
as long as it is minimal, 
the $A_{\infty }$ relation $(\mu '_{\mathrm{int} } )^{2} = 0$ implies the BRST identities 
\begin{align}
\label{BRST id}
 \omega ' \big{(} \, Q \, \Psi '_{0} , \, \boldsymbol{\mu }'_{n} ( \Psi '_{1} , ... , \Psi '_{n} ) \, \big{)}  
 + \sum_{k=1}^{n} \omega ' \big{(} \, \Psi '_{0} , \, \boldsymbol{\mu }'_{n} ( \Psi '_{1} , ... , Q \, \Psi '_{k} , ... , \Psi '_{n} ) \, \big{)} = 0 \, , 
\end{align}
which corresponds to the Stokes theorem. 
Hence, 
even if we replace $\mathcal{H}_{as \, \mathrm{phys} }$ by $\mathrm{Ker}[ \, Q \, ]$, 
the amplitudes (\ref{S-matrix}) reproduce the same values because of the BRST identities (\ref{BRST id}). 

\vspace{2mm} 

\subsection*{Open string field theory}

We obtained a generic result (\ref{S-matrix}) which is valid whenever we consider ordinary perturbative calculations, 
in which propagators of $S$-matrix and gauge-fixing conditions should be written in terms of the free theory. 
When we apply it to open string field theory, 
our homological techniques suggest that 
we can consider somewhat unconventional situations where each pieces of $S$-matrix are given by using information of interacting terms: 
formally, 
we may use these \textit{unconventional} gauge-fixing conditions or propagators in the Feynman graph calculations. 

\vspace{2mm} 

Let us consider Witten's open string field theory \cite{Witten:1985cc}, 
which satisfies the classical BV master equation.\footnote{If this open string field theory gives a well-defined quantum theory, 
it solves the quantum BV master equation without any modification. 
Then, 
we can extend these results to the loop amplitudes since it guarantees that the theory gives amplitudes independent of the gauge-fixing condition. 
} 
We can obtain the tree amplitudes on the basis of the classical limit of the homological perturbation \cite{Matsunaga:2019fnc}. 
Since it is a cubic theory, 
the $A_{\infty }$ structure has no higher product $\boldsymbol{\mu }_{n} = 0$ for $n >2$. 
The interacting vertex $\boldsymbol{\mu _{\mathrm{int} }} = \boldsymbol{\mu }_{2}$ is given by the star product 
\begin{align}
\boldsymbol{\mu }_{2} ( A , B ) \equiv (-)^{A} \, A \, \ast  \, B \, . 
\end{align}
We first consider the Siegel gauge and the linear $b$-gauge \cite{Kiermaier:2007jg}, 
which give a standard perturbative calculus and valid results. 
Next, 
we consider formal gauges, 
the dressed $\mathcal{B}_{0}^{-}$ gauge and $A_{T}$ gauge, 
which are singular but reproduce correct on-shell amplitudes under appropriate assumptions. 

\subsubsection*{Siegel gauge}

In the Siegel gauge $b_{0} \Psi = 0$, 
the propagator $\kappa ^{-1}_{Siegel} \equiv b_{0} \, L_{0}^{-1}$ has poles on the kernel of $L_{0}$. 
We can represent the projector onto the physical states as $(\iota \, \pi )_{Siegel} \equiv e^{-\infty L_{0}}$\,.  
Note that the Schwinger representation of the inverse of $L_{0}$ naturally includes $e^{- \infty L_{0} }$ as a boundary term \cite{Sen:2019jpm} 
\begin{align} 
\label{Siegel}
b_{0} \, L_{0}^{-1} \equiv 
b_{0} \int_{0}^{\infty } e^{- t \, L_{0} } \, dt
= \frac{b_{0} }{L_{0}} ( 1 - e^{- \infty L_{0} } ) \, . 
\end{align}
Since $\mu _{1} \equiv Q$ is the BRST operator of open strings, 
we obtain the decomposition 
\begin{align}
1 - e^{-\infty L_{0}} = Q \, ( b_{0} \, L_{0}^{-1} ) + ( b_{0} \, L_{0}^{-1} ) \, Q \, . 
\end{align}
As is known, 
the Siegel gauge is the standard gauge used in perturbative calculations and it provides a conventional propagator.

\subsubsection*{Linear $b$-gauge}

Let us consider a linear combination of the oscillators $b_{n}$, 
which we write $\mathcal{B}_{(g)}$, 
that can be encoded in a vector field $v (z) = \sum_{n \in \mathbb{Z} } v _{n} z^{n+1}$. 
The linear $b$-gauge is given by 
\begin{align} 
\label{linear b}
\mathcal{B}_{(g)} \, \Psi _{g} = 0 
\hspace{5mm} 
\mathrm{with}
\hspace{5mm} 
\mathcal{B }_{(g)} \equiv \sum_{n \in \mathbb{Z} } v_{n} b_{n} = \oint \frac{d z}{2 \pi i } v (z) b (z) 
\, ,
\end{align}
where $g$ denotes the label of the space-time ghost number. 
Note that the BPZ properties $\mathcal{B}_{(-g)} = \mathcal{B}_{(g-1)}^{\ast} $ must be satisfied for the consistency.  
For each $\mathcal{B}_{(g)}$ or $\mathcal{B}_{(g)}^{\ast }$, 
we define a linear combination of the Virasoro generators $\mathcal{L}_{(g)} \equiv Q \, \mathcal{B}_{(g)} + \mathcal{B}_{(g)} \, Q$, 
which appears in propagators. 

\vspace{2mm} 

In general, 
the linear $b$-gauge may not be invariant under the BPZ conjugation $\mathcal{B}_{(g) } \not= \mathcal{B}_{(g)}^{\ast }$ and then we cannot impose the same gauge-fixing condition for all space-time ghost numbers, 
such as $\mathcal{B}_{(g-1) } = \mathcal{B}_{(-g) }^{\ast } \not= \mathcal{B}_{(-g)}$. 
We write $\Psi = \sum \Psi _{g}$, 
$\mathcal{B} \equiv \sum _{g} \mathcal{B}_{(g)}$ and $\mathcal{L}_{0} \equiv \sum \mathcal{L}_{(g)}$ for simplicity. 
The double Schwinger representation of the propagator 
\begin{align}
\label{double Schwinger}
\kappa _{double}^{-1} \equiv 
\big{(} \mathcal{B}^{\ast } \mathcal{L}_{0}^{\ast \, -1} \big{)} \, Q \, \big{(} \mathcal{B} \mathcal{L}_{0}^{-1} \big{)} 
= \frac{\mathcal{B}^{\ast } }{\mathcal{L}^{\ast }_{0} } \, Q \, \frac{\mathcal{B} }{\mathcal{L}_{0} } \big{(} \, 1-e^{- \infty \mathcal{L}_{0} } \, \big{)} \big{(} \, 1- e^{- \infty \mathcal{L}_{0}^{\ast } } \, \big{)}
\end{align}
provides the decomposition (\ref{solving free theory}) with $1- ( \iota \pi )_{double} \equiv 
[1 + Q ( \frac{\mathcal{B} }{\mathcal{L}_{0}} - \frac{\mathcal{B}^{\ast } }{\mathcal{L}_{0}^{\ast } } ) ] 
( 1- e^{- \infty \mathcal{L}_{0} } ) (1- e^{- \infty \mathcal{L}_{0}^{\ast } }) $, 
where we assume that $e^{- \infty \mathcal{L}_{0}^{\ast } }$ commutes with $\mathcal{B}$, $\mathcal{L}_{0}$ and $e^{- \infty \mathcal{L}_{0} }$. 
It gives correct on-shell amplitudes unless the vector field $v(z)$ is singular.\footnote{For singular $v(z)$, 
such as a sliver frame, 
we can obtain correct on-shell tree amplitudes.  
However, 
for loops, 
\cite{Kiermaier:2008jy} suggests a gauge dependent result. 
}  
Calculations of homological perturbation suggest us an interesting but slightly unconventional propagator\footnote{In principle, 
more unconventional propagator $\frac{1}{2} ( \mathcal{B} (\mathcal{L}_{0})^{-1} + \mathcal{B}^{\ast }(\mathcal{L}_{0}^{\ast } )^{-1})$ may be allowed since $(\iota \pi )$ does not have to be a projector to apply the homological perturbation, 
which gives $( \iota \pi ) = \frac{1}{2}( e^{- \infty \mathcal{L}_{0} } + e^{- \infty \mathcal{L}_{0}^{\ast } } )$. 
} 
\begin{align}
\label{unconventional Schwinger}
\kappa ^{-1}_{mean} \equiv 
\frac{1}{2} \Big{(} \mathcal{B} \, ( \mathcal{L}_{0} )^{-1} +\mathcal{B}^{\ast } \, ( \mathcal{L}^{\ast }_{0} )^{-1} \Big{)}  
\big{(} \, 1-e^{- \infty \mathcal{L}_{0} } \, \big{)} \big{(} \, 1- e^{- \infty \mathcal{L}_{0}^{\ast } } \, \big{)}  
\end{align}
with the gauge-fixing condition $( \mathcal{B} + \mathcal{B}^{\ast } ) \Psi = 0$, 
which gives the decomposition (\ref{solving free theory}) with $1 - (\iota \, \pi )_{mean} \equiv (1-e^{- \infty \mathcal{L}_{0} } ) ( 1- e^{- \infty \mathcal{L}_{0}^{\ast } } )$\,. 
Both of (\ref{double Schwinger}) and (\ref{unconventional Schwinger}) reduces to the ordinary propagator $( \mathcal{B} + \mathcal{B}^{\ast } ) ( \mathcal{L}_{0}  + \mathcal{L}_{0}^{\ast } )^{-1}$ with the gauge-fixing condition $(\mathcal{B} + \mathcal{B}^{\ast }) \Psi = 0$ when $\mathcal{B}^{\ast }_{(g)} = \mathcal{B}_{(g)}$ holds.

\subsubsection*{Dressed $\mathcal{B}_{0}^{-}$ gauge}

We consider formal properties of singular gauge fixing conditions on the basis of the homological perturbation. 
Let $z$ be a coordinate of the sliver frame. 
We set $\mathcal{B}^{-}_{0} = \mathcal{B}_{0}  + \mathcal{B}_{0}^{\ast }$ for $\mathcal{B}_{0}$ defined by $v(z) = z$ of (\ref{linear b}). 
Although the $\mathcal{B}_{0}^{-}$ gauge would be understood as a special case of the linear $b$-gauge defined in the sliver frame, 
it may have more unconventional or non-perturbative aspects. 
We can regard it as a gauge-fixing condition based on the star product multiplications \cite{Erler:2009uj}. 
In the sliver frame, 
the conformal stress tensor $T(z)$ naturally defines a state  
\begin{align}
K \equiv \int_{i \infty }^{-i \infty } dz \, T(z) \, \big{|} \, \mathrm{id} \, \ra \, , 
\end{align} 
where $| \mathrm{id }\rangle $ denotes the identity state of the star product. 
By using any functions $F = F(K) $ and $G = G(K)$ of \textit{the string field} $K$, 
where multiplications are given by the star product $\ast $, 
we can consider the operator $\mathcal{B}_{F, G}$ defined by 
\begin{align}
\label{dressed b}
\mathcal{B}_{F, G} \, \Phi \equiv \frac{1}{2} F (K) \ast  \mathcal{B}_{0}^{-} \Big{[} F(K)^{-1} \ast \Phi \ast G(K) \Big{]} \ast G(K)  \, . 
\end{align} 
Since the interactions of open string fields are given by the star product, 
(\ref{dressed b}) gives a gauge-fixing condition $\mathcal{B}_{F, G} \, \Phi  = 0$ written by using information of interacting terms and would be unconventional in a perturbation from the free theory. 
While the linear $b$-gauge is written in terms of the free theory or the world-sheet theory, 
the dressed $\mathcal{B}_{0}^{-}$ gauge needs the star product defining the interacting term and deviates from the free theory. 
In this sense, 
it seems that we cannot use (\ref{dressed b}) within an ordinary perturbation from the free theory. 
It however gives a Hodge decomposition of operators acting on the identity state, 
which implies that we can apply the homological perturbation. 
As long as the gauge-fixing condition $\mathcal{B}_{F, G} \, \Phi  = 0$ is valid, 
which we just assume, 
it gives (\ref{S-matrix}) correctly. 
For any state $\Phi \in \mathcal{H}$, 
the identity state $|\mathrm{id} \rangle $ satisfies 
\begin{align}
\big{|} \, \mathrm{id} \ra  \, \ast \, \Phi = \Phi = \Phi \, \ast \, \big{|} \, \mathrm{id} \ra  \, . 
\end{align}
Recall that we can represent a given state $\Psi $ as a set of operators $\mathcal{O}_{\Psi }$ acting on the conformal vacuum $|0\rangle $. 
Likewise, 
we may represent $\Psi $ as a set of operators $\widehat{\Psi }$ acting on the identity state $| \mathrm{id} \rangle $, 
\begin{align}
\label{op on id}
\mathcal{O}_{\Psi } \, \big{|} \, 0 \, \ra 
= \Psi 
= (\widehat{\Psi })_{L} \, \big{|} \, \mathrm{id} \ra 
= (\widehat{\Psi })_{R} \, \big{|} \mathrm{id} \ra \, , 
\end{align}
where $(\widehat{\Psi } )_{L} \, \Phi = \Psi  \ast \Phi $ and $(\widehat{\Psi } )_{R} \, \Phi = (-)^{\Psi \, \Phi } \Phi \, \ast \, \Psi $ for any state $\Phi \in \mathcal{H}$\,. 
The propagator (\ref{dressed b}) gives a decomposition on the identity state and in this sense reproduces (\ref{S-matrix}).

\subsection*{$A_{T}$ gauge}

In open string field theory, 
in addition to the perturbative vacuum, 
the tachyon vacuum is well studied \cite{Sen:1999xm, Schnabl:2005gv}. 
We consider the tachyon vacuum solution $\Psi _{T}$ of the Witten theory, 
\begin{align}
Q \, \Psi _{T} + \Psi _{T} \ast \Psi _{T} = 0 \, .
\end{align}
As is known, 
the tachyon vacuum has empty cohomology: 
there exist a state $A_{T}$ satisfying  
\begin{align}
\label{homotopy state}
Q_{T} \, A_{T} = \big{|} \, \mathrm{id} \ra  \, , 
\end{align}
which is called a homotopy contracting state \cite{Ellwood:2001ig}. 
The BRST operator around the tachyon vacuum, 
$Q_{T} \equiv Q + (\widehat{\Psi }_{T})_{L} - (\widehat{\Psi }_{T} )_{R}$,  
satisfies $Q_{T} \, | \mathrm{id} \rangle =0$, 
as $Q \, | \mathrm{id} \rangle = 0$.  
We show that this non-perturbative relation (\ref{homotopy state}) gives an interesting Hodge decomposition with helps of a state defined by 
\begin{align} 
W \equiv  \Psi _{T} \, \ast \, A_{T} + A_{T} \, \ast \, \Psi _{T} \, . 
\end{align} 
As we see, 
it provides the tree $S$-matrices based on unconventional propagators,\footnote{For an $A_{\infty }$ type string field theory, 
when its tachyon solution $\Psi _{T}$ and an operator $\widehat{A}_{T}$ satisfying (\ref{tachyon sft}) are given, 
the same computations can be done by setting $W = \sum_{n} \sum_{\mathrm{cyclic} } \mu _{n+1} ( \Psi _{T} , ... , \Psi _{T} , A_{T} )$. } 
whose $4$-point amplitude reproduces that of \cite{Masuda:2019rgv}. 
In the rest of this section, 
we consider formal but interesting properties suggested by $A_{T} \, \phi = 0$ on the basis of the homological perturbation. 
We would like to emphasise that the following discussions are not a proof of the validity of this gauge fixing condition or the correctness of obtained amplitudes.

\vspace{2mm}

Let us consider the free action $S_{T} [\phi _{T} ] = \frac{1}{2} \omega ( \phi _{T} ,  Q_{T} \, \phi _{T} ) $ for string field theory around the tachyon vacuum. 
We write $\mathcal{H}_{\Psi _{T}}$ for the state space of this free theory around $\Psi _{T}$, 
namely, $\phi _{T} \in \mathcal{H}_{\Psi _{T}}$. 
We assume $A_{T} \ast A_{T} =0$\,. 
The relation (\ref{homotopy state}) implies that this theory has no physical state, 
which gives the following deformation retract 
\begin{align}
\label{tachyon sft}
\widehat{A}_{T} \, 
\rotatebox{-70}{$\circlearrowright $} \, 
\big{(} \, \mathcal{H}_{\Psi _{T} } , \, Q_{T} \, \big{)} 
\hspace{3mm} 
\overset{ \pi }{\underset{ \iota }{
\scalebox{2}[1]{$\rightleftarrows $}
}}
\hspace{3mm} 
\big{(} \, 0 , \, 0 \, \big{)} 
\hspace{5mm}
\mathrm{with} 
\hspace{3mm}  
\widehat{1} = Q_{T} \, \widehat{A}_{T} + \widehat{A}_{T} \, Q_{T} 
\, , 
\end{align}
where operators $\widehat{1}$ and $\widehat{A}_{T}$ are defined by $\widehat{1} \, \Phi \equiv | \, \mathrm{id} \rangle \ast \Phi = \Phi \ast | \, \mathrm{id} \rangle $ and $2 \, \widehat{A}_{T} \equiv (\widehat{A})_{L} + (\widehat{A})_{R}$. 
It is helpful to introduce the operator $\widehat{W} \equiv \frac{1}{2} (\widehat{W})_{L} + \frac{1}{2} (\widehat{W})_{R}$, 
which commutes with $Q$ and $\widehat{A}_{T}$ because of $Q \, W = 0$ and $W \ast A_{T} = A_{T} \ast W $. 
The relation (\ref{homotopy state}) or the above Hodge decomposition of (\ref{tachyon sft}) can be cast as 
\begin{align}
\label{A-W}
Q \, \widehat{A}_{T} + \widehat{A}_{T} \, Q + \widehat{W} = \widehat{1}  \, . 
\end{align}  
We would like to transfer the Hodge decomposition (\ref{A-W}) obtained by the non-perturbative relation (\ref{homotopy state}) to that of the perturbation theory.    

\vspace{2mm} 

We consider the free action $S [\phi  ] = \frac{1}{2} \omega ( \phi , Q \, \phi )$ for string field theory around the perturbative vacuum 
and write $\mathcal{H}$ for its state space: 
$\phi \in \mathcal{H}$. 
This theory has physical states and the $Q$-cohomology $\mathcal{H}_{\mathrm{phys} }$ is not empty. 
A gauge fixing gives a decomposition $\mathcal{H} = \mathcal{H}_{\mathrm{phys} } \oplus \mathcal{H}_{\mathrm{unphys} } \oplus \mathcal{H}_{\mathrm{gauge} }$. 
We assume $\mathcal{H} \cap \mathcal{H}_{\Psi _{T}} \not= 0$. 
When we consider a restriction of the nonperturbative relation (\ref{homotopy state}) or (\ref{A-W}) onto the perturbative state space $\mathcal{H}$, 
there would be two possibility: 
the relation holds in the same form as (\ref{A-W}) in $\mathcal{H}$, 
which we expect for most cases, 
or not. 
%
%
When (\ref{A-W}) holds in $\mathcal{H}$, 
%
for any physical state $\phi _{\mathrm{phys}} \in \mathcal{H}_{\mathrm{phys}}$, 
the relation (\ref{A-W}) provides 
\begin{align}
\label{value of W}
\widehat{W} \, \phi _{\mathrm{phys} } = \phi _{\mathrm{phys}} + Q \, \Lambda 
\end{align}
with the gauge parameter $\Lambda \equiv - \widehat{A}_{T} \, \phi _{\mathrm{phys} }$.\footnote{ 
%
Note that for any $n \in \mathbb{N}$, 
we have $\widehat{1}^{n} \, \phi = 1^{n} \, \phi = \phi $, 
where $1 \in \mathbb{R}$ and $\phi \in \mathcal{H}$. 
If $\lim_{n \rightarrow \infty } \widehat{1}^{n}$ exists and equals to $1$,  
a possible choose of $W = e^{K}$ suggests that we may represent a given $\phi _{\mathrm{phys} }$ as  
\begin{align*}
\phi _{\mathrm{phys }} = \Omega _{\infty } \, \phi _{\mathrm{phys} } \,  \Omega _{\infty } + Q \, \lambda \, , 
\end{align*}
where $\Omega _{\infty } = \lim_{n \rightarrow \infty } e^{n \, K}$ denotes the sliver state and $\lambda$ is a gauge parameter. 
This relation implies that we may use $\Omega _{\infty } \, \phi _{\mathrm{phys} } \,  \Omega _{\infty } $ as external lines of the S-matrix, 
which gives the same value as the $S$-matrix whose external lines are $ \phi _{\mathrm{phys} }$ with the $Q$-exact shifts.  
A state interposed between sliver states belongs to the kernel of $(1- \widehat{W})$. } 
Note that the $Q$-exact term of (\ref{value of W}) can be set to zero by imposing $\widehat{A}_{T} \, \phi _{\mathrm{phys}} = 0$
We write $\iota \, \pi$ for a projector onto $\mathrm{Ker}[ 1 - \widehat{W} ]$. 
With $\widehat{A}_{T} \, \phi =0$, 
the relation (\ref{value of W}) implies that $ \iota \, \pi \, \mathcal{H}$ includes $\mathcal{H}_{\mathrm{phys} }$, 
which suggests that we may use 
\begin{align}
\label{unconventional propagator}
\widehat{\kappa }^{-1} \equiv \widehat{A}_{T} \, ( \, \widehat{1} - \widehat{W} \, )^{-1} \, ( \, \widehat{1} - \iota \, \pi \, )
\end{align}
as a propagator: 
these $\iota \, \pi $, 
$\widehat{\kappa }^{-1}$ and $Q$ give a Hodge decomposition of $\mathcal{H}$. 
We assume that they satisfy $\widehat{\kappa }^{-1} \, \iota \, \pi  = \iota \, \pi \, \widehat{\kappa }^{-1} = 0$.\footnote{When $\widehat{\kappa }^{-1} \, \iota \, \pi  = 0$ or $ \iota \, \pi \, \widehat{\kappa }^{-1} = 0$ does not hold, 
instead of (\ref{unconventional propagator}), 
we use $\widehat{\kappa }^{\prime -1} = \widehat{\kappa }^{-1} ( Q \, \widehat{\kappa }^{-1} + \widehat{\kappa }^{-1} Q )$ or $\widehat{\kappa }^{\prime -1} = ( Q \, \widehat{\kappa }^{-1} + \widehat{\kappa }^{-1} Q ) \, \widehat{\kappa }^{-1} $ as a propagator, 
respectively. 
}  
Then, 
we can start with the deformation retract 
\begin{align}
\label{transfer of most cases}
\widehat{\kappa }^{-1} \, 
\rotatebox{-70}{$\circlearrowright $} \, 
\big{(} \, \mathcal{H}, \, Q \, \big{)} 
\hspace{3mm} 
\overset{ \pi }{\underset{ \iota }{
\scalebox{2}[1]{$\rightleftarrows $}
}}
\hspace{3mm} 
\big{(} \, \mathcal{H}_{\mathrm{phys} } 
, \, 0 \, \big{)} 
\hspace{5mm}
\mathrm{with} 
\hspace{3mm}  
\widehat{1} - \iota \, \pi = Q \, \widehat{\kappa }^{-1} + \widehat{\kappa }^{-1} \, Q
\, . 
\end{align} 
The above $\mathcal{H}_{\mathrm{phys} }$ should be relaxed to $\mathcal{H}_{\mathrm{phys} } \oplus \mathcal{H}_{\mathrm{gauge} }$ when we use the left hand side of (\ref{value of W}) as external lines instead of $\phi _{\mathrm{phys}}$. 
As a result of the perturbation $\mu _{2}$ to (\ref{transfer of most cases}), 
we obtain an unconventional $S$-matrix whose propagator is (\ref{unconventional propagator}). 
A gauge fixing is needed to specify the external lines $\phi _{\mathrm{phys}}$ explicitly. 
The relation (\ref{value of W}) suggest that the condition $\widehat{A}_{T} \, \phi = 0$ for $\phi \in \mathcal{H}$ works as a gauge fixing condition, 
although it may take a singular expression.

\vspace{2mm} 

Next, 
we consider the other case that (\ref{A-W}) does not hold in the same form after the restriction onto $\mathcal{H}$. 
We assume that such a perturbative breakdown of (\ref{A-W}) occurs in $\mathcal{H}_{\mathrm{phys} }$ only. 
In this case, 
there exists $\phi _{\mathrm{phys} } \in \mathcal{H}_{\mathrm{phys}}$ such that the equality of the relation (\ref{value of W}) or (\ref{A-W}) does not hold.\footnote{It includes a situation where we cannot define an inner product of $\widehat{W} \phi _{\mathrm{phys}}$ and a given $\psi _{\mathrm{phys}} \in \mathcal{H}_{\mathrm{phys} }$. } 
Then, 
we need to replace the operator $\widehat{W}$ by a projected one $\widehat{W} (\, \widehat{1} - ( \iota \, \pi )_{\mathrm{phys}} )$ to transfer the relation (\ref{tachyon sft}) to that of $\mathcal{H}$, 
where $( \iota \, \pi )_{\mathrm{phys}} : \mathcal{H} \rightarrow \mathcal{H}_{\mathrm{phys} }$ denotes a projector onto the physical space. 
We consider  
\begin{align}
\label{other case}
\widehat{A}_{T} \, 
\rotatebox{-70}{$\circlearrowright $} \, 
\big{(} \, \mathcal{H}, \, Q_{T} \, \big{)} 
\hspace{3mm} 
\overset{ \pi _{\mathrm{phys}} }{\underset{ \iota _{\mathrm{phys}} }{
\scalebox{2.5}[1]{$\rightleftarrows $}
}}
\hspace{3mm} 
\big{(} \, \mathcal{H}_{\mathrm{phys} } , \, 0 \, \big{)} 
\hspace{5mm}
\mathrm{with} 
\hspace{3mm}  
\widehat{1} - ( \iota \, \pi )_{\mathrm{phys}} = Q_{T} \, \widehat{A}_{T} + \widehat{A}_{T} \, Q_{T} 
\, , 
\end{align} 
where we set $\iota _{\mathrm{phys}} =1$ and used $\mathcal{H}_{\mathrm{phys} } = \pi _{\mathrm{phys}} \, \mathcal{H}$ for simplicity. 
As a result of the perturbation $Q - Q_{T}$ to (\ref{other case}), 
we obtain the transferred relation   
\begin{align}
\label{transfer of other case}
\widetilde{\kappa }^{-1} \, 
\rotatebox{-70}{$\circlearrowright $} \, 
\big{(} \, \mathcal{H}, \, Q \, \big{)} 
\hspace{3mm} 
\overset{ \pi _{\mathrm{phys}}}{\underset{ \iota _{\mathrm{phys}}}{
\scalebox{2.5}[1]{$\rightleftarrows $}
}}
\hspace{3mm} 
\big{(} \, \mathcal{H}_{\mathrm{phys} } , \, 0 \, \big{)} 
\hspace{5mm}
\mathrm{with} 
\hspace{3mm}  
\widehat{1} - (\iota \, \pi )_{\mathrm{phys}} = Q \, \widetilde{\kappa } + \widetilde{\kappa } \, Q
\, , 
\end{align} 
 where the perturbed homotopy contracting operator $\widetilde{\kappa }^{-1}$ is given by  
\begin{align} 
\label{formal A_T}
\widetilde{\kappa }^{-1} \equiv \widehat{A}_{T} \, \big{(} \, \widehat{1} - \widehat{W} \, \big{)}^{-1}   \big{(} \, \widehat{1} - (\iota \, \pi )_{\mathrm{phys}} \big{)} \, . 
\end{align}
It solves the Hodge decomposition (\ref{solving free theory}) on $| \mathrm{id} \ra $ and thus provides the on-shell $S$-matrix (\ref{S-matrix}) formally. 
The propagator (\ref{formal A_T}) consists of $A_{T}$, 
$\Psi _{T}$ and the projector $(\iota \, \pi )_{\mathrm{phys}}$. 
The explicit form of $A_{T}$ is determined by specifying the explicit form of $\Psi _{T}$, 
which is free from the gauge choice. 
Note however that we need to impose a gauge fixing condition to determine an explicit form of the projector $(\iota \, \pi )_{\mathrm{phys}}$ or external lines and to compute the on-shell $S$-matrix.  
In this sense, 
unless the choice of external lines are taken into account, 
the $S$-matrix (without external lines) obtained by using (\ref{formal A_T}) is free from a gauge-fixing. 

\vspace{2mm} 

The $S$-matrix obtained from (\ref{transfer of most cases}) has the same form as the $S$-matrix obtained from (\ref{transfer of other case}) except for the external lines and projectors. 
Actually, 
we can check that (\ref{unconventional propagator}) or (\ref{formal A_T}) indeed gives a correct $4$-point amplitude. 
For any state $\phi \in \mathcal{H}$, 
we find 
\begin{align} 
\widehat{\kappa }^{-1} \, \widehat{W} \, \phi 
= - \big{(} \, A_{T} - \widehat{\kappa }^{-1}  \, \big{)} \, \phi  \, . 
\end{align} 
It resembles (\ref{Siegel}) and can be understood as separating the main contribution from the boundary contribution. 
By using the cyclic property, 
the $4$-point amplitude reduces to  
\begin{align}
\label{Vene}
\mathcal{A}_{4} (\phi ' , ... ,\phi ' ) 
= \frac{1}{2} \La ( \widehat{\kappa }^{-1} -\widehat{A}_{T} ) \widehat{\phi }' \, (\widehat{W} \, \widehat{\phi }' )^{3} \Ra _{\mathrm{sliver}} \, , 
\end{align}
where $\langle ... \rangle _{\mathrm{sliver}}$ denotes the correlation function of the conformal field theory on the sliver frame. 
As shown by \cite{Masuda:2019rgv}, 
when external lines $\phi '$ are specified, 
the expression (\ref{Vene}) reproduces the on-shell $4$-point amplitudes of the world-sheet theory, 
where $\widehat{\kappa }^{-1}$ of (\ref{Vene}) is identified with $A_{\Psi }$ of \cite{Masuda:2019rgv}. 
This result also supports the validity of the formal object (\ref{unconventional propagator}) or (\ref{formal A_T}) as a propagator.

\section{Conclusion and Discussions} 

In this paper, 
we explicitly showed that the perturbative path-integral can be performed in terms of both of the BV-BRST cohomology and Lagrangian's homotopy algebraic structure on the basis of the homological perturbation. 
%
%
Although many of these results are already known, 
we believe that an explicit and direct derivation given in this paper includes new points and is worthwhile, 
in particular, 
for physicists.
%
%
%
As we checked explicitly, 
the homological perturbation for Lagrangian's $A_{\infty }$ gives an alternative representation of the perturbative path-integral, 
for every BV-quantizable field theory has own $A_{\infty }$ structure -- each solution of the BV master equation is in one-to-one correspondence with a quantum $A_{\infty }$ structure. 
As a corollary of these results, 
we showed that when the original BV master action includes source terms or interactions with external fields, 
its effective theory must have a weak (quantum) $A_{\infty }$ structure. 
We also discussed that such a homological approach may enable us to use unconventional propagators for calculating $S$-matrix, 
which may provide further applications. 
As long as physicists believe that the path-integral condenses configurations of integrated fields onto the on-shell physical ones, 
our results seem to be a quite natural (or trivial) 
because the BV formalism itself is based on the homological perturbation and determines the physical states from it. 
%

\vspace{2mm}

As we explained, 
the BV master equation and the intrinsic $A_{\infty }$ structure play central roles in perturbative quantum field theory. 
Once we solve the BV master equation, 
we can quickly obtain each quantity given by the perturbative path-integral, 
such as effective theory or scattering amplitude. 
Thus, 
it would be important tasks to try to derive BV master actions for some superstring field theories \cite{Berkovits:2012np, Matsunaga:2016zsu, Erler:2017onq, Matsunaga:2018hlh}. 
It would be worth mentioning that we checked how the perturbative path-integral preserves the $A_{\infty }$ structure, 
although it may be originally a property of the non-perturbative path-integral (\ref{effective BV master action}). 
Our results suggest that the non-perturbative corrections would preserve the $A_{\infty }$ structure we consider. 

\vspace{2mm}

We would like to emphasize that such algebraic approaches to Lagrangian field theory have been exploited since long-time before: 
it is not a new idea. 
However, 
the link between homotopy algebras and the BV formalism have developed recently and minimal models of quantum homotopy algebras are now available \cite{Braun:2017ikg, Doubek:2017naz}. 
We thus believe that it would be worth studying these approaches more explicitly and physicist-friendly in terms of higher algebra.\footnote{Also, 
homotopy algebras would be more accessible to mathematicians, 
rather than the BV formalism. } 
We end this section by mentioning related earlier works. 
%
%
%
The earliest and outstanding work would be \cite{Zwiebach:1992ie}, 
which introduced quantum $L_{\infty }$ algebras and established the link to the BV master equation. 
The geometry and meaning of the classical BV formalism were given by various authors in the early days, 
for example \cite{Schwarz:1992nx, Alexandrov:1995kv}. 
Recently, 
a nice review was given by \cite{Jurco:2018sby}. 
Also \cite{Albert} is very suggestive. 
Application of minimal models of homotopy algebras to field theory was given by \cite{Kajiura:2003ax, Kajiura:2001ng}, 
which pointed out that minimal models give S-matrices. 
Quantum minimal models is given by \cite{Braun:2017ikg, Doubek:2017naz} recently. 
Derivations of S-matrix based on the homological perturbation were given by many authors. 
For example, 
see \cite{Konopka:2015tta, Matsunaga:2019fnc, Macrelli:2019afx} for the tree level and \cite{Plumann, Doubek:2017naz, Jurco:2019yfd} for the loop. 
See also \cite{Arvanitakis:2019ald}. 
The work \cite{Nakatsu:2001da} discussed the classical part of effective theory and renormalization group by using the $A_{\infty }$ structure. 
The work \cite{Costello:2007ei} presented that the BV formalism is very useful to discuss renormalization group flows. 
Also, 
the works \cite{Gwilliam:2012jg, JohnsonFreyd:2012ww} derived Wick's theorem and Feynman rules for finite-dimensional integrals by using BV differentials. 
The link between solutions of BV master equation and homotopy algebras originates from their operadic relations, 
which were studied by \cite{Barannikov:2017, Barannikov:2010np, Doubek:2013dpa, Jurco}. 

\vspace{-3mm} 

\subsection*{Acknowledgments}

H.M. would like to thank Martin Markl and Jan Plumann for helpful discussions of homotopy transfer or homological perturbation. 
Also, 
the authors would like to thank Ted Erler, Carlo Maccaferri and Yuji Okawa for discussions of SFT 
at the GGI workshop ``String Theory from the World-Sheet Perspective'', 
at the YITP workshop ``Strings and Fields 2019'', 
or at Italian or Czech restaurants. 

\vspace{1mm} 

This work was supported by GACR Grant EXPRO 19-28628X. 
T.M. was supported by the GACR grant 20-25775X. 
H.M. was supported by Praemium Academiae RVO\,67985840 at the institute of mathematics of the Czech Academy of Sciences until 2019, 
in which main part of the first version of this manuscript was completed; 
Yukawa Institute for Theoretical Physics, Kyoto University from late 2020 to early 2022; 
JSPS KAKENHI (Grant-in-Aid for Early-Career Scientists) grant number JP22K14038 from 2022, 
whose support completed the final version. 

\appendix

\section{Symplectic basis and Lagrangian's $A_{\infty }$} 

In this paper, 
we start with a given BV master action $S[\psi ]$ for a given quantum field theory, 
which is constructed for a given classical action 
\begin{align}
S_{\mathrm{cl} } [\psi _{\mathrm{cl} } ] 
= - \sum_{n} \frac{1}{n+1} \int dx \, \psi _{\mathrm{cl} } (x) \, \mu ^{\mathrm{cl} }_{n} (\psi _{\mathrm{cl} } (x) , ... , \psi _{\mathrm{cl} } (x) )  
\end{align}
on the basis of the BV formalism. 
While $\psi _{\mathrm{cl}} = \{ \psi _{\mathrm{cl}}^{a} \}_{a}$ denotes given classical fields, 
$\psi = \{ \psi _{a} , \, \psi ^{\ast }_{a} \}_{a}$ denotes all of the fields and antifields. 
The BV master action $S[\psi ]$ has the neutral grading, 
as is the classical action. 
In the BV formalism, 
a pairing that assigns each field $\psi _{a}$ to its antifield $\psi ^{\ast }_{a}$ gives a symplectic form, 
which can be diagonalised by using the basis 
\begin{align}
\psi \equiv 
\begin{pmatrix}
\vec{\psi } \\ 
\vec{\psi }^{\ast } 
\end{pmatrix} 
\, , 
\end{align}
where $\vec{\psi }$ or $\vec{\psi }^{\ast }$ denotes the column vector of fields $\{ \psi _{a} \}_{a}$ or antifields $\{ \psi ^{\ast }_{a} \}_{a}$ respectively. 
For $\psi = \{ \psi _{a} , \, \psi ^{\ast }_{a} \}_{a}$ and $\widetilde{\psi } = \{ \widetilde{\psi }_{a} , \, \widetilde{\psi }^{\ast }_{a} \} _{a}$, 
it takes the form of  
\begin{align}
\la \, \widetilde{\psi } \, , \, \psi  \, \ra 
\equiv \sum_{a} \int dx \, 
\begin{pmatrix}
\widetilde{\psi }_{a} & 
\widetilde{\psi }^{\ast }_{a} 
\end{pmatrix} 
\, 
\begin{pmatrix}
0 & -1 \\
1 & 0 
\end{pmatrix} 
\, 
\begin{pmatrix}
\psi _{a} \\ 
\psi ^{\ast }_{a} 
\end{pmatrix} 
\, . 
\end{align}
We can regard the above basis and thus the pairing $\langle \,\, , \,\, \rangle$ to be degree zero: 
the grading consists of physical ones. 
Notice that the $A_{\infty }$ products $\mu = (-)^{\psi } ( S , \psi )$ can be represented as 
\begin{align}
\sum _{n} \mu _{n} (\psi , ... , \psi ) 
= \sum_{a} 
\begin{pmatrix}
\frac{\partial S}{\partial \psi ^{\ast }_{a} } \\[1mm] 
- \frac{\partial S}{\partial \psi _{a} } 
\end{pmatrix} 
\, , 
\end{align}
so that $\mu _{n}$ also has degree zero. 
In this representation, 
the grading is always given by the sum of the physical gradings carried by fields $\psi $, 
such as the ghost number or Grassmann parity. 
When a given field theory has no gauge degrees, 
we find that its $A_{\infty }$ relations come from the form of 
\begin{align}
\mu _{n} \bigg{(}
\begin{pmatrix}
\psi _{\mathrm{cl}}^{a_{1}} \\ 
\psi ^{\ast }_{a_{1} } 
\end{pmatrix}
\, , \, ... \, , \, 
\begin{pmatrix}
\psi _{\mathrm{cl}}^{a_{n}} \\ 
\psi ^{\ast }_{a_{n} } 
\end{pmatrix}
\bigg{)} 
=  
\begin{pmatrix}
0 \\ 
\mu _{n} (\psi _{\mathrm{cl} }^{a_{1}} , ... , \psi _{\mathrm{cl}}^{a_{n}} )
\end{pmatrix} 
\, , 
\end{align} 
where $\psi ^{\ast }_{a}$ denotes the antifield for a classical field $\psi ^{a}_{\mathrm{cl}}$. 
Using these symplectic basis and multilinear maps acting on them, 
a given BV master action $S[\psi ]$ can be cast as follows 
\begin{align}
S [\psi ] 
= \sum_{n} \frac{1}{n+1} \la \, \psi \, , \, \mu _{n} ( \underbrace{\psi , ... , \psi }_{n} ) \, \ra 
\, . 
\end{align}
Although the grading comes from physical ones, 
sign factors of the $A_{\infty }$ relations of $\{ \mu _{n} \}_{n}$ are not simple: 
the sum $\psi = \sum _{g} \psi _{g} + \psi ^{\ast }_{g}$ includes odd and even grading fields since the $g$-th ghost fields $\psi _{g}$ have space-time ghost number $g$ and their antifields $\psi ^{\ast }_{g}$ has $-1-g$ respectively. 

\vspace{3mm}

In contrast to the above physical grading, 
the simplest $A_{\infty }$ grading can be realized by assigning unphysical grading, 
which we call basis ghost number, 
to each symplectic basis. 
Note that there exist a set of bases $\{ \hat{e}_{-g} , \, \hat{e}^{\ast }_{-g} \}_{g}$ for a given set of fields-antifields $\{ \psi _{g} , \psi ^{\ast }_{g} \}_{g}$\,. 
We assign\footnote{We omit the label $a$ distinguishing species of fields $(\psi _{a})_{g}$ having the same space-time ghost number for brevity. 
The state (\ref{total state}) should be understood as $\hat{e}_{-g} \otimes \psi _{g} \equiv \sum_{a} (\hat{e}_{a})_{-g} \otimes (\psi _{a})_{g}$\,. } 
degree $-g$ to the bases $\hat{e}_{-g}$ corresponding to the fields $\psi _{g}$ having space-time ghost number $g$, 
so that the following state is degree zero 
\begin{align}
\label{total state}
\Psi  \equiv \sum \hat{e}_{-g}  \otimes \psi _{g}+ \sum \hat{e}^{\, \ast }_{-g} \otimes \psi ^{\ast }_{g}\, , 
\end{align}
as is the string field. 
The sum of the space-time and basis ghost numbers gives the simplest $A_{\infty }$ degree. 
For $g \geq 0$, 
as $\psi _{-1-g} \equiv \psi ^{\ast }_{g}$, 
we often write $\hat{e}_{1+g} \equiv (-)^{g} \hat{e}^{\, \ast }_{-g}$ and $\Psi _{-1-g} \equiv (-)^{g} \Psi ^{\ast }_{g}$ for brevity. 
%
%
Let $\hat{\mathcal{H} } \equiv E \otimes \mathcal{H}$ be the state space of (\ref{total state}), 
where $E$ denotes the space of basis. 
The set of bases $\{ \hat{e}_{-g} , \hat{e}^{\ast }_{-g} \}_{g}$ lead a degree $-1$ symplectic form $\hat{\omega } : E^{\otimes 2} \rightarrow \mathbb{C}$ with 
\begin{align}
\label{symplec}
\hat{\omega } \big{(} \, \hat{e}_{g} , \, \hat{e}^{\ast }_{g'} \, \big{)} 
= - \hat{\omega } \big{(} \, \hat{e}^{\ast }_{g'} , \, \hat{e}_{g} \, \big{)} 
= \delta _{g, g'} \, , 
\end{align}
or a degree $-3$ inner product on $E$ by $\langle \hat{e}_{-g} , \hat{e}^{\ast }_{-g'} \rangle _{E} \equiv (-)^{\hat{e}_{g}} \hat{\omega } ( \hat{e}_{-g} , \hat{e}^{\ast }_{g'} )$. 
Then, 
we can define a degree $-1$ symplectic form $\omega : \hat{\mathcal{H} }^{\otimes 2} \rightarrow \mathbb{C}$ in the BV formalism, 
\begin{align}
\omega ( \hat{e}_{p_{1} } \otimes \psi _{g_{1} }  , \,  \hat{e}^{\ast }_{p_{2}} \otimes \psi _{g_{2} }  ) 
\equiv (-)^{\hat{e}^{\ast }_{p_{2}} \psi _{g_{1}} } \hat{\omega }( \hat{e}_{p_{1} } , \, \hat{e}^{\ast }_{p_{2} } ) \, 
\la \psi _{g_{1} } , \, \psi _{g_{2}} \ra \, , 
\end{align} 
which is graded antisymmetric $\omega (A ,B) = - (-)^{AB} \omega (B,A)$ as usual, 
and a set of degree $1$ multilinear maps $\boldsymbol{\mu }_{n} : \hat{\mathcal{H} }^{\otimes n } \rightarrow \hat{\mathcal{H} }$,  
\begin{align}
\label{adjusted A}
\sum _{g \in \mathbb{Z} } 
\boldsymbol{\mu}_{n} ( \Psi _{g_{1}} , ... \, , \Psi _{g_{n} } ) \big{|}_{g } 
\equiv 
\sum_{g>0} \hat{e}_{1-g} 
\otimes \mu _{n} ( \psi _{g_{1}} , ... , \psi _{g_{n}}  ) \big{|}_{g}
+ \sum_{g \geq 0} \hat{e}^{\, \ast }_{-g} 
\otimes \mu _{n} ( \psi _{g_{1}} , ... , \psi _{g_{n}}  ) \big{|}_{-g}
\, ,
\end{align}
where $g \equiv \sum _{k=1}^{n} g_{k}$ denotes the sum of the ghost numbers of fields $\{ \psi _{g_{k}} \}_{k=1}^{n}$ and $A |_{g}$ denotes the restriction of $A$ to the states having space-time ghost number $g$. 
Notice that the grading of the linear map $\boldsymbol{\mu } = \sum_{n} \boldsymbol{\mu }_{n}$ acting on $\mathcal{T}(\hat{\mathcal{H}} )$ is indeed $1$ since $\{ \mu _{n} \}_{n}$ have no ghost number, 
bases $e_{1+g}$ have basis ghost number $1+g$ and fields have ghost number $g$ respectively. 
Then, 
the BV master action $S[\psi ]$ takes the homotopy Maurer-Cartan form 
\begin{align} 
\label{hMC S}
\hat{S} [\Psi ] = \sum_{n} \frac{1}{n+1} \omega \big{(}  \, \Psi , \, \boldsymbol{\mu }_{n} ( \, \underbrace{\Psi ,  . . . \, , \Psi }_{n} \, ) \,  \big{)} 
\end{align}
which, 
of course, 
equals to the above representations of the BV master action $\hat{S}[\Psi ] = S[\psi ]$.\footnote{Notice that a different pair $\mathsf{a} = (\psi _{a_{1} } )_{g_{1} } \otimes \cdots \otimes (\psi _{a_{n} })_{g_{n}}$ gives different base $(e _{\mathsf{a} } )^{\ast }_{g_{1} + \cdots + g_{n} } $, 
which can be read from a given BV master action explicitly. 
We suppress this $\mathsf{a}$-label distinguishing pairs of fields.} 
The action $\hat{S} [\Psi ]$ consists of fields $\Psi$ of degree $0$, 
the (quantum) $A_{\infty }$ structure $\boldsymbol{\mu }$ of degree $1$, 
and the graded symplectic form $\omega$ of degree $-1$, 
which gives the simplest $A_{\infty }$ degrees. 
In this representation, 
the $A_{\infty }$ structure can be simply read from 
\begin{align}
\sum_{n} \boldsymbol{\mu }_{n} ( \Psi , ... , \Psi ) 
= \hbar \, \Delta _{S[\psi ] } \, \Psi 
= \sum_{g} (-)^{g} \bigg{[} 
\frac{\partial \hat{S} [\Psi ] }{\partial \Psi _{g} } + \frac{\partial \hat{S} [\Psi ] }{\partial \Psi ^{\ast }_{g} }
\bigg{]} 
\end{align} 
and the $\Psi$-derivatives can be computed in the same way as the string-field derivatives \cite{Matsunaga:2018hlh}. 
The $A_{\infty }$ relations of $\boldsymbol{\mu } = \sum_{n} \boldsymbol{\mu }_{n}$ have simple sign factors (\ref{def of A-relations}) or (\ref{def of quantum A-relations}), 
which reduce to the $A_{\infty }$ relations of $\mu = \{ \mu _{n} \} _{n}$ with the sign factors of (\ref{A in sec2}) or (\ref{quantum A}) by removing the bases $\{ \hat{e}_{g} \}_{g}$ having unphysical degrees. 
The above set of string-field-inspired bases $\{ e_{-g} , e^{\ast }_{-g} \}_{g}$ having unphysical grading naturally provides a set of bases of a degree $-1$ symplectic form in the BV formalism.


\footnotesize 

\begin{thebibliography}{99}



\bibitem{Zwiebach:1992ie}
  B.~Zwiebach,
  ``Closed string field theory: Quantum action and the B-V master equation,''
  Nucl.\ Phys.\  B {\bf 390}, 33 (1993)
  [arXiv:hep-th/9206084].


\bibitem{Batalin:1981jr}
  I.~A.~Batalin and G.~A.~Vilkovisky,
  ``Gauge Algebra and Quantization,''
  Phys.\ Lett.\  {\bf 102B} (1981) 27.
\bibitem{Batalin:1984jr}
  I.~A.~Batalin and G.~A.~Vilkovisky,
  ``Quantization of Gauge Theories with Linearly Dependent Generators,''
  Phys.\ Rev.\ D {\bf 28} (1983) 2567
   Erratum: [Phys.\ Rev.\ D {\bf 30} (1984) 508].
\bibitem{Henneaux} 
  M.~Henneaux and C.~Teitelboim, 
  ``Quantization of gauge systems,'' 
  (Princeton University Press), 1992. 
  
\bibitem{Barannikov:2017}
  “Modular Operads and Batalin-Vilkovisky Geometry”, 
  International Mathematics Research Notices, Vol. 2007, rnm075. 
  [arXiv:1710.08442 [math.QA]]
   
\bibitem{Barannikov:2010np}
  S.~Barannikov,
  ``Solving the noncommutative Batalin-Vilkovisky equation,''
  Lett.\ Math.\ Phys.\  {\bf 103} (2013) 605
  [arXiv:1004.2253 [math.QA]]. 
  
  
\bibitem{Doubek:2013dpa}
  M.~Doubek, B.~Jurco and K.~Munster,
  ``Modular operads and the quantum open-closed homotopy algebra,''
  JHEP {\bf 1512} (2015) 158
  [arXiv:1308.3223 [math.AT]].


\bibitem{Schwarz:1992nx}
  A.~S.~Schwarz,
  ``Geometry of Batalin-Vilkovisky quantization,''
  Commun.\ Math.\ Phys.\  {\bf 155} (1993) 249
  [hep-th/9205088].

\bibitem{Alexandrov:1995kv}
  M.~Alexandrov, A.~Schwarz, O.~Zaboronsky and M.~Kontsevich,
  ``The Geometry of the master equation and topological quantum field theory,''
  Int.\ J.\ Mod.\ Phys.\ A {\bf 12} (1997) 1405
  [hep-th/9502010].

\bibitem{Jurco:2018sby}
  B.~Jurco, L.~Raspollini, C.~Samann and M.~Wolf,
  ``$L_\infty$-Algebras of Classical Field Theories and the Batalin-Vilkovisky Formalism,''
  Fortsch.\ Phys.\  {\bf 67} (2019) no.7,  1900025
  [arXiv:1809.09899 [hep-th]].

\bibitem{Macrelli:2019afx}
  T.~Macrelli, C.~Samann and M.~Wolf,
  ``Scattering amplitude recursion relations in Batalin-Vilkovisky-quantizable theories,''
  Phys.\ Rev.\ D {\bf 100} (2019) no.4,  045017
  [arXiv:1903.05713 [hep-th]].

\bibitem{Jurco:2019yfd}
  B.~Jurco, T.~Macrelli, C.~Samann and M.~Wolf,
  ``Loop Amplitudes and Quantum Homotopy Algebras,''
  JHEP {\bf 07} (2020) 003 [arXiv:1912.06695 [hep-th]]. 

\bibitem{Sen:2016qap}
  A.~Sen,
  ``Wilsonian Effective Action of Superstring Theory,''
  JHEP {\bf 1701} (2017) 108
  [arXiv:1609.00459 [hep-th]].
\bibitem{Matsunaga:2019fnc}
  H.~Matsunaga,
  ``Light-cone reduction of Witten’s open string field theory,''
  JHEP {\bf 1904} (2019) 143
  [arXiv:1901.08555 [hep-th]]. 
\bibitem{EM}
  T.~Erler and H.~Matsunaga,
  ``Mapping between Witten and Lightcone String Field Theories,''
  JHEP \textbf{11} (2021), 208
  [arXiv:2012.09521 [hep-th]]. 
\bibitem{Masuda:2019rgv}
  T.~Masuda and H.~Matsunaga,
  ``Deriving on-shell open string field amplitudes without using Feynman rules,''
  PTEP \textbf{2022} (2022) no.1, 013B06
  [arXiv:1908.09784 [hep-th]].


\bibitem{Markl:1997bj}
  M.~Markl,
  ``Loop homotopy algebras in closed string field theory,''
  Commun.\ Math.\ Phys.\  {\bf 221} (2001) 367
  [hep-th/9711045]. 

\bibitem{Herbst:2006kt}
  M.~Herbst,
  ``Quantum A-infinity structures for open-closed topological strings,''
  hep-th/0602018.

\bibitem{Munster:2011ij}
  K.~Munster and I.~Sachs,
  ``Quantum Open-Closed Homotopy Algebra and String Field Theory,''
  Commun.\ Math.\ Phys.\  {\bf 321} (2013) 769
  [arXiv:1109.4101 [hep-th]].

  

\bibitem{Kajiura:2003ax}
  H.~Kajiura,
  ``Noncommutative homotopy algebras associated with open strings,''
  Rev.\ Math.\ Phys.\  {\bf 19} (2007) 1.
  [math/0306332 [math-qa]].
\bibitem{Kajiura:2001ng}
  H.~Kajiura,
  ``Homotopy algebra morphism and geometry of classical string field theory,''
  Nucl.\ Phys.\ B {\bf 630} (2002) 361.
  [hep-th/0112228]. 

\bibitem{Crainic}
  M.~Crainic, 
  ``On the perturbation lemma, and deformations,'' 
  arXiv:math/0403266 [math.AT]. 

\bibitem{Doubek:2017naz}
  M.~Doubek, B.~Jurco and J.~Pulmann,
  ``Quantum $L_{\infty }$ Algebras and the Homological Perturbation Lemma,''
  Commun.\ Math.\ Phys.\ {\bf 367} (2019) 1, 215-240 [arXiv:1712.02696 [math-ph]].

\bibitem{Albert} 
  C.~Albert, 
  ``Batalin-Vilkovisky Gauge-Fixing via Homological Perturbation Theory,'' 
  \url{
  https://math.unice.fr/~patras/CargeseConference/ACQFT09_CarloALBERT.pdf}   

\bibitem{Plumann} 
  J.~Pulmann, 
  ``S-matrix and homological perturbation lemma,'' 
  Master Thesis, 2016, Mathematical Institute of Charles University. 

\bibitem{Munster:2012gy}
  K.~Munster and I.~Sachs,
  ``Homotopy Classification of Bosonic String Field Theory,''
  Commun.\ Math.\ Phys.\  {\bf 330} (2014) 1227
  [arXiv:1208.5626 [hep-th]].



\bibitem{Aisaka:2004ga}
  Y.~Aisaka and Y.~Kazama,
  ``Relating Green-Schwarz and extended pure spinor formalisms by similarity transformation,''
  JHEP {\bf 0404} (2004) 070
  [hep-th/0404141]. 

\bibitem{Kato:1982im}
  M.~Kato and K.~Ogawa,
  ``Covariant Quantization of String Based on BRS Invariance,''
  Nucl.\ Phys.\ B {\bf 212} (1983) 443.


\bibitem{Witten:1985cc}
  E.~Witten,
  ``Noncommutative Geometry and String Field Theory,''
  Nucl.\ Phys.\  B {\bf 268}, 253 (1986).

\bibitem{Sen:2019jpm}
  A.~Sen,
  ``String Field Theory as World-sheet UV Regulator,''
  JHEP {\bf 1910} (2019) 119
  [arXiv:1902.00263 [hep-th]].

\bibitem{Kiermaier:2007jg}
  M.~Kiermaier, A.~Sen and B.~Zwiebach,
  ``Linear b-Gauges for Open String Fields,''
  JHEP {\bf 0803} (2008) 050
  [arXiv:0712.0627 [hep-th]].

\bibitem{Kiermaier:2008jy}
  M.~Kiermaier and B.~Zwiebach,
  ``One-Loop Riemann Surfaces in Schnabl Gauge,''
  JHEP {\bf 0807} (2008) 063
  [arXiv:0805.3701 [hep-th]].

\bibitem{Erler:2009uj}
  T.~Erler and M.~Schnabl,
  ``A Simple Analytic Solution for Tachyon Condensation,''
  JHEP {\bf 0910} (2009) 066
  [arXiv:0906.0979 [hep-th]].

\bibitem{Sen:1999xm}
  A.~Sen,
  ``Universality of the tachyon potential,''
  JHEP {\bf 9912} (1999) 027
  [hep-th/9911116].

\bibitem{Schnabl:2005gv}
  M.~Schnabl,
  ``Analytic solution for tachyon condensation in open string field theory,''
  Adv.\ Theor.\ Math.\ Phys.\  {\bf 10} (2006) no.4,  433
  [hep-th/0511286].

\bibitem{Ellwood:2001ig}
  I.~Ellwood, B.~Feng, Y.~H.~He and N.~Moeller,
  ``The Identity string field and the tachyon vacuum,''
  JHEP {\bf 0107} (2001) 016
  [hep-th/0105024].


\bibitem{Berkovits:2012np}
  N.~Berkovits,
  ``Constrained BV Description of String Field Theory,''
  JHEP {\bf 1203} (2012) 012
  [arXiv:1201.1769 [hep-th]].
\bibitem{Matsunaga:2016zsu}
  H.~Matsunaga,
  ``Notes on the Wess-Zumino-Witten-like structure: L$_{\infty }$ triplet and NS-NS superstring field theory,''
  JHEP {\bf 1705} (2017) 095
  [arXiv:1612.08827 [hep-th]].
\bibitem{Erler:2017onq}
  T.~Erler,
  ``Superstring Field Theory and the Wess-Zumino-Witten Action,''
  JHEP {\bf 1710} (2017) 057
  [arXiv:1706.02629 [hep-th]].  
\bibitem{Matsunaga:2018hlh}
  H.~Matsunaga and M.~Nomura,
  ``On the BV formalism of open superstring field theory in the large Hilbert space,''
  JHEP {\bf 1805} (2018) 020
  [arXiv:1802.04171 [hep-th]].


\bibitem{Braun:2017ikg}
  C.~Braun and J.~Maunder,
  ``Minimal models of quantum homotopy Lie algebras via the BV-formalism,''
  J.\ Math.\ Phys.\  {\bf 59} (2018) no.6,  063512
  [arXiv:1703.00082 [math.QA]].

  
\bibitem{Konopka:2015tta}
  S.~Konopka,
  ``The S-Matrix of superstring field theory,''
  JHEP {\bf 1511} (2015) 187
  [arXiv:1507.08250 [hep-th]]. 

\bibitem{Arvanitakis:2019ald}
A.~S.~Arvanitakis,
``The L$_\infty$-algebra of the S-matrix,''
JHEP \textbf{07} (2019), 115
[arXiv:1903.05643 [hep-th]].

\bibitem{Nakatsu:2001da}
  T.~Nakatsu,
  ``Classical open string field theory: A(infinity) algebra, renormalization group and boundary states,''
  Nucl.\ Phys.\ B {\bf 642} (2002) 13
  [hep-th/0105272].

\bibitem{Costello:2007ei}
  K.~J.~Costello,
  ``Renormalisation and the Batalin-Vilkovisky formalism,''
  arXiv:0706.1533 [math.QA].

\bibitem{Gwilliam:2012jg}
  O.~Gwilliam and T.~Johnson-Freyd,
  ``How to derive Feynman diagrams for finite-dimensional integrals directly from the BV formalism,''
  Topology and quantum theory in interaction, 175-185, Contemp.
  Math., 718, Amer. Math. Soc., Providence, RI, 2018
  [arXiv:1202.1554 [math-ph]].
  
\bibitem{JohnsonFreyd:2012ww}
  T.~Johnson-Freyd,
  ``Homological perturbation theory for nonperturbative integrals,''
  Lett.\ Math.\ Phys.\  {\bf 105} (2015) no.11,  1605
  [arXiv:1206.5319 [math-ph]].

\bibitem{Jurco} 
   B.~Jurco, 
   Presentation at the Solvay workshop on ``Higher Spin Gauge Theories, Topological Field Theory and Deformation Quantization'', Brussels, February 17-21, 2020. 



\end{thebibliography}
\end{document}